\RequirePackage{tikz}
\documentclass{interact}

\usepackage{epstopdf}
\usepackage[caption=false]{subfig}

\usepackage[numbers,sort&compress]{natbib}
\bibpunct[, ]{[}{]}{,}{n}{,}{,}

\usepackage[utf8]{inputenc}
\usepackage{comment,enumitem, amsmath, mathrsfs,  dsfont, booktabs,lipsum, parskip, multirow,amsfonts, amssymb, makecell, siunitx, booktabs, tikz, fix-cm, mathtools, multicol, bigints,  float, mathtools, parskip, tabstackengine, tabulary, tabularx, adjustbox, booktabs, longtable, lscape, graphics, lipsum, xr-hyper, color, colortbl,  threeparttable}

\usepackage[myheadings]{fullpage}
\usepackage[draft=false, citecolor=blue]{hyperref}
\hypersetup{
    colorlinks=true,
    linkcolor=blue,
    filecolor=magenta,      
    urlcolor=cyan,
    pdftitle={Overleaf Example},
    pdfpagemode=FullScreen,
    }
    
\numberwithin{equation}{section}
\newlist{enumthm}{enumerate}{1}
\setlist[enumthm]{label=(\alph*)}

\newtheorem{Theorem}{Theorem}[section]

\newtheorem{Remark}{Remark}[section]

\counterwithin{table}{section}
\counterwithin{figure}{section}

\newcounter{desccount}
\newcommand{\descitem}[1]{%
  \item[#1] \refstepcounter{desccount}\label{#1}
}
\newcommand{\descref}[1]{\hyperref[#1]{#1}}

\begin{document}


\title{Robust and Efficient Estimation in Ordinal Response Models using the Density Power Divergence}

\author{
\name{Arijit Pyne\thanks{Arijit Pyne. Email: arijitp\_r@isical.ac.in}, 
Subhrajyoty Roy\thanks{Subhrajyoty Roy. Email: roysubhra98@gmail.com},
Abhik Ghosh\thanks{Abhik Ghosh. Email: abhik.ghosh@isical.ac.in} \and
Ayanendranath Basu\thanks{Ayanendranath Basu. Email: ayanbasu@isical.ac.in}
}
\affil{{Interdisciplinary Statistical Research Unit, Indian Statistical Institute, Kolkata, India}}
}

\maketitle

\begin{abstract}
\label{abstract}
In real life we frequently come across data sets that involve some independent explanatory variables, generating a set of ordinal responses. These ordinal responses may be thought to depend on a continuous latent variable and a set of unknown cut-off points. The latent variable is further assumed to be linearly related to the explanatory variables which in turn drive the ordinal responses. One way of estimating the unknown parameters (i.e., the regression coefficients and the cut-offs), that comes through modelling the data, is to find its maximum likelihood estimates. Maximum likelihood estimates are noted for being fully asymptotically efficient at the true model. However, a small proportion of outlying observations, e.g., responses incoherent to the categories or unbounded covariate(s) may destabilize the likelihood function to a great extent. Therefore, the reliability of the maximum likelihood estimates is strongly affected. Minimum distance methods are widely used in the literature when robustness is a concern. In the present paper, we use the density power divergence to estimate the parameters. The roles of different link functions are analyzed through the lens of the density power divergence. Asymptotic properties of the minimum density power divergence estimator are discussed theoretically. Its robustness is investigated through the influence function analysis. Analytically, we have shown that the proposed class of estimates has a very high asymptotic breakdown point against data contamination. Numerically it is further demonstrated that the proposed method yields slope estimates that never implode towards zero. The finite-sample performance of the minimum density power divergence estimators is investigated through extensive numerical experiments, either at the model or when data contamination occurs across different link functions. It is shown to outperform the maximum likelihood estimators, producing more stable results in the face of data contamination. Moreover, our estimators are very competitive with the other robust alternatives. Finally, we wrap up this article with an application on a real data example.  

\end{abstract}


\section{Introduction}
\label{Introduction}

In recent years, the analysis of ordinal response data has become a popular topic in mainstream research. Such data arise naturally in many areas of scientific studies, for instance in psychology, sociology, economics, medicine, political science, and in several other disciplines where the final response of a subject belongs to a finite number of ordered categories based on the values of several explanatory variables in a way described later in this article. One such example may be the qualitative customer review of a particular vehicle where its price, mileage, carbon emission properties, etc., are to be taken into account to arrive at a qualitative review on an ordinal scale. While these ratings summarize many important explanatory variables and are primarily useful to a new customer, these customer feedbacks often turn out to be equally important to the manufacturer as the latter might want to learn about the statistical relationship between the ordinal response and its covariates to improve their product, or for post-manufacturing surveys to fix things.

A pioneering work in this field has been done by McCullagh \cite{mccullagh1980regression} who advocated the use of an underlying continuous latent variable that drives ordinal responses based on some unknown cut-offs. This method has become popular as it enables us to view the ordinal response model within a unified framework of the generalized linear model (GLM); see, e.g., McCullagh \cite{mccullagh2019generalized} and Nelder et al. \cite{nelder1972generalized}. Moustaki \cite{moustaki2000latent} uses the maximum likelihood (ML) method to fit a multi-dimensional latent variable model to a set of observed ordinal variables and also discusses the related goodness-of-fit problem. Also, see Moustaki \cite{moustaki2003general} for related discussions. Piccolo \cite{piccolo2003moments} and Iannario et al. \cite{iannario2016generalized} suggest a different approach which uses the response variable as a combination of a discrete mixture of uniform and a shifted binomial (CUB) random variables.  

Although the area of robust statistics has a very rich and developed body of literature, applications in the direction of ordinal response data are rather limited. An early reference is Hampel \cite{hampel1968contributions}, where, in addition to the development of a classical infinitesimal approach to robustness, some pointers about robustness in the case of binomial model fitting are discussed. Robust estimators have been developed by Victoria-Feser and Ronchetti \cite{victoria1997robust} for grouped data. Ruckstuhl and Welsh \cite{ruckstuhl2001robust} have considered different classes of estimators in the context of fitting a robust binomial model to a data set. Moustaki and Victoria-Feser \cite{moustaki2004bounded, moustaki2006bounded} have developed bounded-bias and bounded-influence robust estimators for the generalized linear latent (GLL) variable. Lack of robustness in the maximum likelihood estimates for the logistic regression model has been extensively studied in the literature (Croux et al. \cite{croux2002breakdown}; M$\ddot{\rm u}$ller and Neykov \cite{muller2003breakdown}). Croux et al. \cite{croux2013robust} have proposed a weighted maximum likelihood (WML) estimation method for the logit link function. Iannario et al. \cite{iannario2016robustness} dealt with robustness for the class of CUB models. More recently, Iannario et al. \cite{iannario2017robust} used a weighted likelihood function where weights vary depending on the choice of link functions. Unlike the approaches of Croux et al. \cite{croux2013robust}, who propose to use weights as functions of robust Mahalanobis type distances, Iannario et al. \cite{iannario2017robust} considered Huber's weight functions that combine both the generalized residuals and robust Mahalanobis distance or the normalized MAD as appropriate for different link functions. The primary objective aims at controlling the influential observations with respect to the parametric model. Recently, Scalera et al. \cite{scalera2021robust} analyzed the role of different link functions towards robustness in this context.

In this article, we propose to use the density power divergence (DPD) measure, proposed by Basu et al. \cite{basu1998robust}, to obtain robust and (asymptotically highly) efficient estimates in ordinal response data under the same setup as Iannario et al. \cite{iannario2017robust}. The independent but non-homogeneous version of the DPD-based inference (Ghosh and Basu, \cite{ghosh2013robust}) is best suited for this application.

The key highlights of this paper are the following. 
\begin{itemize}
     \item[(a)] To study the role of different link functions in estimating the slope parameter, we plot a DPD-version of the generalized residuals. As it turns out, the DPD-version of the generalized residuals stays bounded even for the commonly used links that produce unbounded generalized residuals in the MLE. This gives a clear insight into why DPD should produce robust slope estimates.     

    \item [(b)] Asymptotic properties of the proposed minimum DPD estimator (MDPDE) are discussed in the context of the ordinal response regression problem.    

    \item[(c)] Robustness of the minimum density power divergence functional is investigated through the influence function analysis. As expected, we find the effect of infinitesimal data contamination to be limited, as compared to the MLE, whenever the tuning parameter $\alpha$ is strictly positive.  

    \item[(d)] We have shown that under suitable assumptions, the asymptotic breakdown point of the MDPDE is $\frac{1}{2}$, the maximum possible value, compared to the very low breakdown of MLE.

     \item[(e)] We have empirically shown that the implosive breakdown point of the MDPDE of the regression parameter is very high. This means that even when the sample contains a high proportion of outlying observations, we can still obtain a stable MDPDE of the slope parameter.
     
     \item[(f)] In continuation of the earlier point, these estimates are also used to find the prediction misclassification rate. When $\alpha>0$, the misclassification rate incurred by the MDPDE becomes much lower than the MLE.  

    \item[(g)] Through extensive simulation studies, we show that our proposed estimator outperforms the MLE in the presence of outlying observations at higher values of the tuning parameter $\alpha$. Also, it is almost as good as the MLE in the true model when $\alpha$ is relatively small. Our estimator also performs better than the weighted likelihood estimator proposed by Croux et al. \cite{croux2013robust}, and it is very competitive to the M-estimator proposed by Iannario et al. \cite{iannario2017robust}. Computationally our method is less expensive than Iannario et al. \cite{iannario2017robust} ( See Figure \textbf{S.Fig. 4.} in the supplementary material). 

    \item[(h)] To apply this method in real-life data examples, we make use of a tuning parameter selection algorithm, as proposed by \cite{warwick2005choosing}. The performance of this algorithm is validated through simulation studies.    
    
    \item[(i)] Finally, we analyze a real data example with the proposed robust estimator, where we choose the optimal robustness parameter using the above tuning parameter selection algorithm. The prediction from the resulting estimator achieves no lesser (higher in some cases) accuracy than the MLE.  

\end{itemize}

The rest of this article is organized as follows. In Section \ref{parametric model and mle} we state the problem under study, and briefly review the maximum likelihood estimation in the present setup. Next, we introduce the method of minimum density power divergence estimation in Section \ref{The density power divergence and estimating eq}, and further discuss the role of generalized residuals under different error distributions in view of this divergence measure. Asymptotic properties of the minimum density power divergence estimators (i.e., consistency and asymptotic normality) are discussed in Section \ref{asymptotic properties}. The robustness of the proposed estimator is investigated in Section \ref{Robustness Studies}. In particular, we plot the gross error sensitivity in Subsection \ref{IF analysis}, and present an asymptotic breakdown point result in Subsection \ref{Breakdown point}. We also study the implosion resistance property of the slope estimates in Subsection \ref{implosion Breakdown of the slope estimates}. Finite-sample performance of the proposed estimator is compared with the MLE, and the estimators proposed by Croux et al. \cite{croux2013robust} and Iannario et al. \cite{iannario2017robust} in Section \ref{Numerical Studies}. Next, we briefly discuss a data-driven strategy for the tuning parameter selection in Section \ref{tuning parameter selection}. This algorithm is validated through a simulation study. Further, we apply this algorithm to a real data example in Section \ref{real data analysis}. Concluding remarks have been made in Section \ref{conclusions}. The proofs of all the results and some additional tables and figures are provided in the Supplementary material. 

\section{Parametric Model and the Maximum Likelihood Estimation}
\label{parametric model and mle}

Consider a random sample $\{( x_{i}, Y_{i}): i=1,2, \ldots , n\}$ of size $n$. The $i$-th explanatory vector, denoted by $x_{i}=(x_{i1}, x_{i2}, \ldots , x_{ip})^{T}$, is assumed to be non-stochastic in $\mathds{R}^{p}$. Here $Y_{i}$ is the realization of the response variable $Y$ when conditioned on $x_{i}$. Further, $Y$ is supported on a finite set $\chi=\{1,2, \ldots ,m\}$. Following McCullagh \cite{mccullagh1980regression} we presume that there exists a continuous latent random variable $Y^{*}$ such that it is related to $Y$ as 
\begin{equation}
\label{latent relationship}
Y=j   \iff \gamma_{j-1} < Y^{*} \le \gamma_{j} \mbox{ for } j \in \chi,
\end{equation}
where $-\infty=\gamma_{0} <\gamma_{1}< \gamma_{2} < \cdots < \gamma_{m-1} < \gamma_{m}=+\infty$ are the unknown cut-off points (thresholds) in the continuous support of the latent variable. Moreover, $Y_{i}$ depends on the explanatory variable $x_{i}$ as   
\begin{equation}
\label{linear latent regression model}
 Y^{*}_{i}= x_{i1}\beta_1+x_{i2}\beta_2+\cdots+x_{ip}\beta_{p}+e_{i}
 =x_{i}^{T}\beta+e_{i} \mbox{ for all } i=1, 2, \ldots,n.
 \end{equation}
 Here $\beta=(\beta_{1},\beta_{2},\ldots,\beta_{p})^{T}$ is a vector of regression coefficients in the latent linear (LL) regression model with $e_{i}$ being a random error term. The $e_{i}$s are assumed to be identically and independently distributed according to a known probability distribution function $F$. The inverse of $F$ is called the link function. We assume that $F$ admits a probability density function $f$, and further denote $\gamma=(\gamma_{1}, \ldots, \gamma_{m-1})^{T}$. Using (\ref{linear latent regression model}) we find that  
 \begin{equation}
\label{model probability}
p_{\theta,i}(j)=Pr(Y=j|x_{i})=F(\gamma_{j} -x_{i}^{T}\beta)-F(\gamma_{j-1} -x_{i}^{T}\beta)
 \mbox{ where } \theta=(\gamma,\beta),
\end{equation}
for $i=1, 2, \ldots ,n$ and $j=1, 2, \ldots ,m$. It is obvious that
$p_{\theta, i}(1)=F(\gamma_{1}-x_{i}^{T}\beta)$ and $p_{\theta, i}(m)=1-F(\gamma_{m-1}-x_{i}^{T}\beta)$. Here the parameter space is denoted by $\Theta \subseteq \mathds{R}^{m+p-1}$. Later on, we shall interchangeably use the term ``slope parameters" for ``regression parameters". 

Now we wish to find an estimate of $\theta$. A traditional way of doing that is to find its maximum likelihood estimate $\hat{\theta}_{ML}$ which maximizes the log-likelihood function   
\begin{equation}
\label{log likelihood}
\sum_{i=1}^{n}\ell(\theta; x_{i}, Y_{i})
\mbox{ where }
\ell(\theta; x_{i}, Y_{i})=\sum_{j=1}^{m} \delta_{i}(j)\ln p_{\theta,i}(j).
\end{equation}
Here $\delta_{i}(j)=\mathds{1}(Y =j| x_{i})$ where $\mathds {1}(\cdot|x_{i})$ symbolizes the indicator of the set $\{Y=j\}$ given at $x_{i}$, and $\ln(\cdot)$ represents the natural logarithm. The log-likelihood function for the entire sample turns out to be the sum of individual log-likelihood functions $\ell(\theta; x_{i}, Y_{i})$ evaluated at each data point $(x_{i}, Y_{i})$, $i=1, 2, \ldots,n$. Under appropriate regularity conditions $\hat{\theta}_{ML}$ consistently estimates true $\theta$. It is well known that the MLE is asymptotically the most efficient among the class of consistent and uniformly asymptotically normal (CUAN) estimators. However, in practice, we hardly come across a data set that truly follows an assumed model. Often a model is deemed a good fit to a data set if the majority of the data points follow that model, leaving out only a small proportion (maybe $5\%$ - $10\%$) of observations that are inconsistent with the model. Observations, that defy the assumed probability distribution, are deemed outliers with regard to that model. The reliability of the MLE becomes questionable even in the presence of a single outlier. This is a problem with the MLE whenever robustness is a concern. Often robustness comes at the cost of asymptotic efficiency, e.g., trimmed MLE, median, etc. To overcome this, it is therefore required to resort to alternative robust methodologies that would make use of the same model but yield estimators balancing between the extreme situations-- asymptotic efficiency and robustness. 

In the vast literature of robust statistics, methods often focus on two primary aspects to eliminate or limit the influence of outlying observations in the estimation process. The most intuitive way is to multiply the log-likelihood by a suitable weight function. Many different types of weights may be suggested along the way. Croux et al. \cite{croux2013robust} propose one such weight function that uses the robust Mahalanobis distance in the space of explanatory variables. In the second approach, a weighted average of the likelihood score is set to zero, and solved for the unknown parameter. Iannario et al. \cite{iannario2017robust} take this second approach. These are expected to lead to robust estimates. Methods that deal with the parameter estimation in ordinal response models are primarily limited to these two methods. In this paper, we use the density power divergence (Basu et al. \cite{basu1998robust}) to find robust parameter estimates with high asymptotic efficiency.

\section{The Density Power Divergence and Estimating Equations}
\label{The density power divergence and estimating eq}

The density power divergence (DPD) between two probability density functions $g$ and $q$ (with respect to a common dominating measure), indexed by a tuning parameter $\alpha$, is defined as 
\begin{equation}
\label{DPD}
d_{\alpha}(g,q)=\bigintssss_{\mathcal{S}}\Bigg\{q^{1+\alpha}-\Bigg(1+\frac{1}{\alpha}\Bigg)q^{\alpha}g+\frac{1}{\alpha}g^{1+\alpha}\Bigg\} \mbox{ for } \alpha > 0.
\end{equation}
Here $\mathcal{S}$ denotes the support common to both $q$ and $g$. In the discrete case, this divergence may be accordingly modified by replacing the integration with the summation. Although the divergence would be undefined when we simply substitute $\alpha=0$ in (\ref{DPD}), its limit is well-defined when $\alpha \downarrow 0+$. The latter is defined to be the value of $d_{0}(g,q)$. Some routine algebraic manipulation shows that
\begin{equation}
\label{KLD}
d_{0}(g,q)=\int_{\mathcal{S}} g \ln{\frac{g}{q}}.   
\end{equation}
This is a version of the Kullback-Leibler divergence.

Now consider the problem of minimum distance estimation. Let a data set, generated by the probability distribution function $G$, be modelled with $\mathcal{Q}=\{Q_{\theta}; \theta \in \Theta\}$. $G$ and $Q_{\theta}$ are supposed to admit probability density functions $g$ and $q_{\theta}$ respectively. Then the minimum density power divergence (MDPD) functional $T_{\alpha}(G)$ is defined as a minimizer of $d_{\alpha}(g, q_{\theta})$ over $\Theta$. When $\alpha=0$ it becomes the maximum likelihood functional $T_{0}(G)$, and we get the minimum $L_{2}$ distance functional at $\alpha=1$. Since the third term in the divergence does not involve $\theta$, the essential objective function may be given by   
\begin{equation}
\label{obj function}
\int_{\mathcal{S}} \Bigg\{ q_{\theta}^{1+\alpha}-\Bigg(1+\frac{1}{\alpha}\Bigg) q_{\theta}^{\alpha}dG\Bigg\}
\mbox{ for } \alpha >0.
\end{equation}   
In practice, the true distribution function $G$ is unknown to us. This may be estimated using the empirical distribution function $G_{n}$ based on an iid random sample. Thus the minimum density power divergence estimator is given by 
\begin{equation}
\label{empirical obj function}
\hat{\theta}_{\alpha}  :=  \arg\min_{\theta\in\Theta}
\int_{\mathcal{S}}
\Bigg\{ q_{\theta}^{1+\alpha}-\Bigg(1+\frac{1}{\alpha}\Bigg)
q^{\alpha}_{\theta}dG_{n}\Bigg\}.
\end{equation}
Alternatively, we also use the notation $T_{\alpha}(G_{n})$ for $\hat{\theta}_{\alpha}$. Notice that the dependency of the model on data appears linearly in (\ref{empirical obj function}). Therefore, it does not require the use of any nonparametric kernel smoothing even for the continuous models, which is unavoidably necessary for any other $\phi$-divergences. See Basu et al. \cite{basu1998robust} for more details. The minimum density power divergence estimator belongs to the class of M-estimators. An M-estimator $\hat{\theta}_{M}$ solves 
\begin{align}
    \int \psi_{\theta} dG_{n}=0 
\end{align}
where $\psi_{\theta}$ is a real-valued function. For the MDPDE, we have $\psi_{\theta}=(1+\alpha)\Big\{\int q^{\alpha}_{\theta}\nabla q_{\theta} - q^{\alpha-1}_{\theta}\nabla q_{\theta}\Big\} $. Here $\nabla$ denotes the partial derivative operator with respect to $\theta$
    
When the random observations are independent but not necessarily identically distributed, the DPD may be generalized in a variety of different ways. However, we shall follow the approach of Ghosh and Basu \cite{ghosh2013robust}. Now, we consider our problem setup. Let $G_{i}$ be the probability distribution function that generates ordinal responses $Y_{i}$; $i=1,2, \ldots ,n$. The sample observations $Y_{i}$s are independent but distributed according to possibly different distribution functions $G_{i}$s. It is further assumed that $G_{i}$s admit probability density functions $g_{i}$s with respect to a common dominating measure for $i=1,2, \ldots,n$. For each $i$, the true density $g_{i}$ is modelled by  $p_{\theta, i}$. For the $i$-th data point, the DPD is given by
\begin{equation}
\label{i th dpd}
    d_{\alpha}(g_{i}, p_{\theta, i})
    =\begin{cases}
    \sum_{j=1}^{m}\Big[p_{\theta, i}(j)^{1+\alpha}- \Big(1+\frac{1}{\alpha} \Big) p_{\theta, i}(j)^{\alpha}g_{i}(j)+\frac{1}{\alpha} g_{i}(j)^{1+\alpha} \Big]
    &\mbox{ when } \alpha > 0 \\ \\
    \sum_{j=1}^{m}g_{i}(j) \ln \frac{g_{i}(j)}{p_{\theta,i}(j)}
    &\mbox{ when } \alpha=0. 
    \end{cases}
\end{equation}
The overall divergence is defined as the arithmetic mean of individual divergence measures, i.e.,  
\begin{align}
\label{DPD, INH setup}
    \frac{1}{n}\sum_{i=1}^{n}d_{\alpha}(g_{i}, p_{\theta, i}).
\end{align}
The minimum density power divergence functional $\theta_{\alpha}$ minimizes (\ref{DPD, INH setup}). Notice that $\theta_{\alpha}$ depends on the true distributions-- $G_{1}, \ldots , G_{n}$. Essentially, $\theta_{\alpha}$ minimizes the following objective function 
\begin{align}
\label{population INH obj fun}
    H(\theta)=\frac{1}{n}\sum_{i=1}^{n}H^{(i)}(\theta),
\end{align}
where $H^{(i)}(\theta)$ is obtained from (\ref{DPD, INH setup}) excluding its terms independent of $\theta$. To find the minimum density power divergence estimate, we substitute $\delta_{i}(j)$ for $g_{i}(j)$ in the above expressions. The empirical version of $H^{(i)}$ is given by 
\begin{align}
 \label{V_i theta}
 V_{i}(x_{i}, Y_{i}, \theta)=
 \begin{cases}
   \sum_{j=1}^{m}
 p_{\theta,i}(j)^{1+\alpha}- \Big(1+\frac{1}{\alpha}\Big)
 p_{\theta, i}(Y_{i})^{\alpha}
& \mbox{ when } \alpha >0 \\ \\
-\ln p_{\theta, i}(Y_{i})
& \mbox{ when } \alpha=0.
 \end{cases} 
\end{align}
Consequently, $\hat{\theta}_{\alpha}$ minimizes 
\begin{align}
    H_{n}(\theta)=\frac{1}{n}\sum_{i=1}^{n}V_{i}(x_{i}, Y_{i}, \theta)
\end{align}
over the parameter space $\Theta$. When the error distribution $F$ as in (\ref{model probability}) is differentiable, the minimum density power divergence estimator $\hat{\theta}_{\alpha}$ can be obtained by solving the estimating equation
\begin{equation}
\label{DPD, INH, estimating eq}
\nabla H_{n}( \theta)=\frac{(1+\alpha)}{n} \Bigg\{\sum_{j=1}^{m} p_{\theta,i}(j)^{1+\alpha} u_{\theta,i}(j)
-p_{\theta,i}(Y_{i})^{\alpha} u_{\theta,i}(Y_{i})\Bigg\}=0, 
\end{equation}
where $u_{\theta, i}(j)=\frac{\nabla p_{\theta, i}(j)}{p_{\theta, i}(j)}$ is the likelihood score function. Notice that $\nabla p_{\theta,i}(j)$ is a vector-valued function which is given by 
\begin{equation}
\label{score function}
 \nabla p_{\theta,i}(j)
   =\Big( \frac{\partial }{\partial \gamma^{T}}   p_{\theta, i}(j),  \frac{\partial }{\partial \beta^{T}}  p_{\theta, i}(j)\Big)^{T} \in \mathds{R}^{m+p-1}.
\end{equation}
A simple calculation shows that
\begin{align}
\label{first deriv of p w.r.t gamma}
    \frac{\partial}{\partial \gamma_{s}} p_{\theta, i}(j)
    =\begin{cases}
    f(\gamma_{s}-x^{T}_{i}\beta)  &\mbox{ when } j=s \\
    -f(\gamma_{s}-x^{T}_{i}\beta) &\mbox{ when } j=s+1 \\
    0      &\mbox{ otherwise}, 
   \end{cases} 
   \end{align}
 \begin{align}
\label{first deriv of p w.r.t beta}
\frac{\partial}{\partial \beta_{k}}  p_{\theta, i}(j)
   &=\Big\{ f(\gamma_{j-1}- x^{T}_{i}\beta)-f(\gamma_{j}-x^{T}_{i}\beta ) \Big\}x_{ik}
\end{align}
for $s=1,2,\ldots ,(m-1)$ and $k=1, \ldots,p$. Equations (\ref{first deriv of p w.r.t gamma}) and (\ref{first deriv of p w.r.t beta}) would together imply  
\begin{align}
\label{first deriv of V w.r.t gamma}
  \frac{1}{1+\alpha}\cdot\frac{\partial }{\partial \gamma_{s}} V_{i}(x_{i}, Y_{i}, \theta)
  &=\Bigg\{p_{\theta, i}(s)^{\alpha}f(\gamma_{s}-x_{i}^{T}\beta)-p_{\theta, i}(Y_{i})^{\alpha-1} f(\gamma_{Y_{i}}-x_{i}^{T}\beta) \mathds{1}(Y_{i}=s|x_{i})\Bigg\} \nonumber \\
  &-\Bigg\{p_{\theta, i}(s+1)^{\alpha}f(\gamma_{s}-x_{i}^{T}\beta)-p_{\theta, i}(Y_{i})^{\alpha-1} f(\gamma_{Y_{i}-1}-x_{i}^{T}\beta) \mathds{1}(Y_{i}=s+1|x_{i})\Bigg\}, \\
\label{first deriv of V w.r.t beta}  
\frac{1}{1+\alpha}\cdot\frac{\partial}{\partial \beta_{k}}V_{i}(x_{i}, Y_{i}, \theta)
&=x_{ik} \Bigg[ \sum_{j=1}^{m}
p_{\theta, i}(j)^{\alpha} \Big\{f(\gamma_{j-1}- x^{T}_{i}\beta)-f(\gamma_{j}-x^{T}_{i}\beta )
 \Big\}  \nonumber \\
 &-p_{\theta, i}(Y_{i})^{\alpha-1} \Big\{f(\gamma_{Y_{i}-1}- x^{T}_{i}\beta)-f(\gamma_{Y_{i}}-x^{T}_{i}\beta )
 \Big\} \Bigg]
\end{align}
for $s=1, 2, \ldots,(m-1)$ and $k=1, 2, \ldots,p$. Using (\ref{first deriv of V w.r.t gamma}) and (\ref{first deriv of V w.r.t beta}) the estimating equations in (\ref{DPD, INH, estimating eq}) may be further simplified. Observe that the quantity in (\ref{DPD, INH, estimating eq}) is unbiased when the true densities belong to the model families, i.e., $g_{i}(j)= p_{\theta_{*}, i}(j)$ for all $j=1,\ldots,m$ and $i=1, \ldots,n$. In that case, the minimum density power divergence functional is Fisher consistent, i.e., $\theta_{\alpha}=\theta_{*}$ for true $\theta_{*}$.

Another justification for adapting the general theory of the non-homogeneous DPD over the usual one in this context is the following. Assume that $(X, Y)$ is jointly distributed according to a probability distribution that involves all the parameters of our interest. Then a single DPD can still be constructed between the data and a model using the original formulation of Basu et al. \cite{basu1998robust}. Consequently, all the parameters may be estimated by minimizing the divergence albeit with computational complexity that may arise due to modelling both the covariates and response variables in a higher dimension. This may be completely avoided or at least reduced to a great extent if we take the conditioning approach on the explanatory variables keeping the usual flavour of regression analysis as widely used in most of its applications. This, in a way, gives a reason for using the non-homogeneous version of the DPD in the context of the present situation. Not only that, the loss of efficiency incurred by the ordinary MDPDE which usually occurs in higher dimensions at a fixed $\alpha$, may be completely circumvented in this approach.

\section{DPD-version of the Generalized Residuals}
\label{DPD-version of the Generalized Residuals}

\begin{figure}[!ht]
\begin{multicols}{2}
    \includegraphics[scale=0.4]{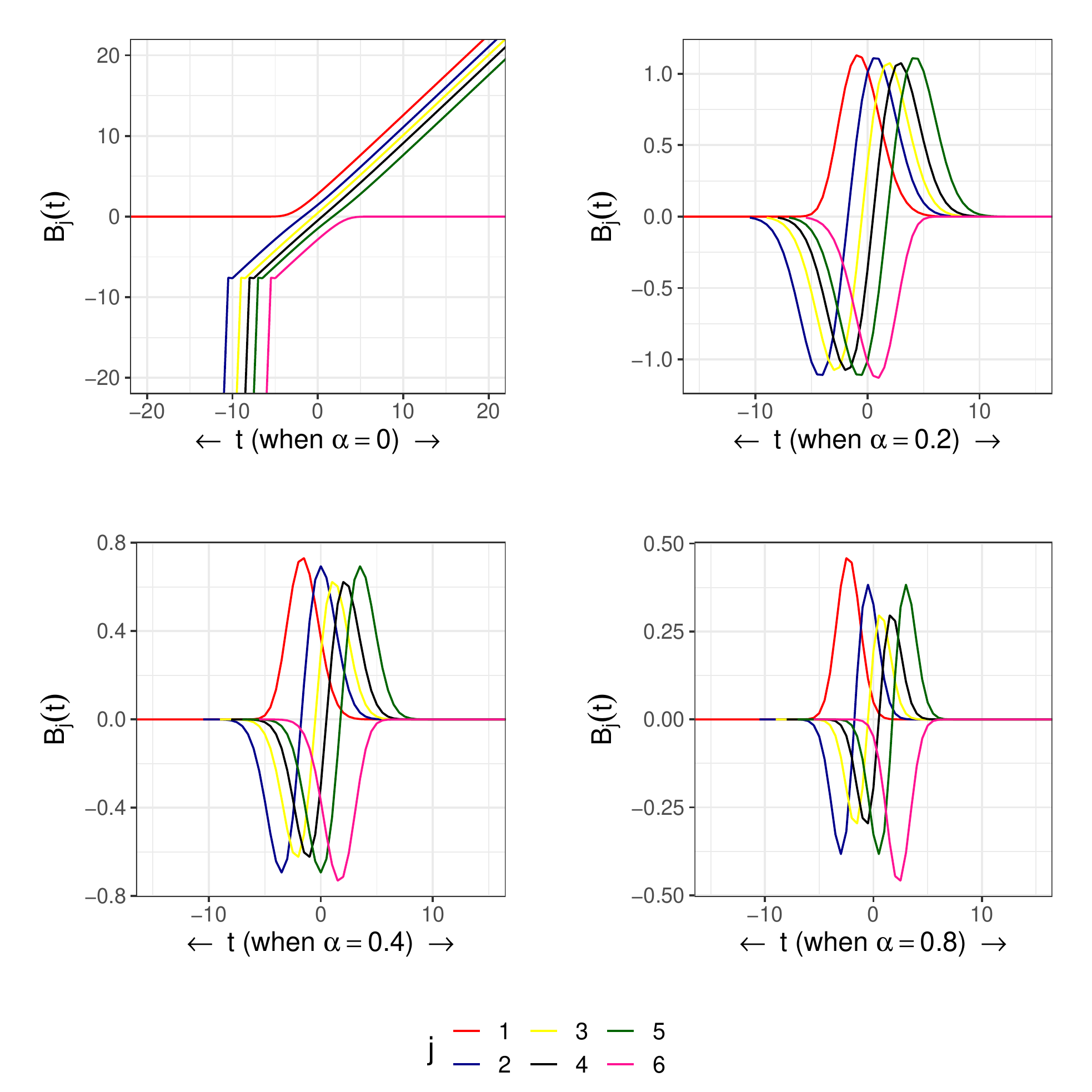} 
    \caption{Generalized residuals for the probit link with thresholds $\gamma=(-2.5, -1, 0, 1, 2.5)^{T}$.} 
   \label{fig:gen res probit}
    \par 
    \includegraphics[scale=0.4]{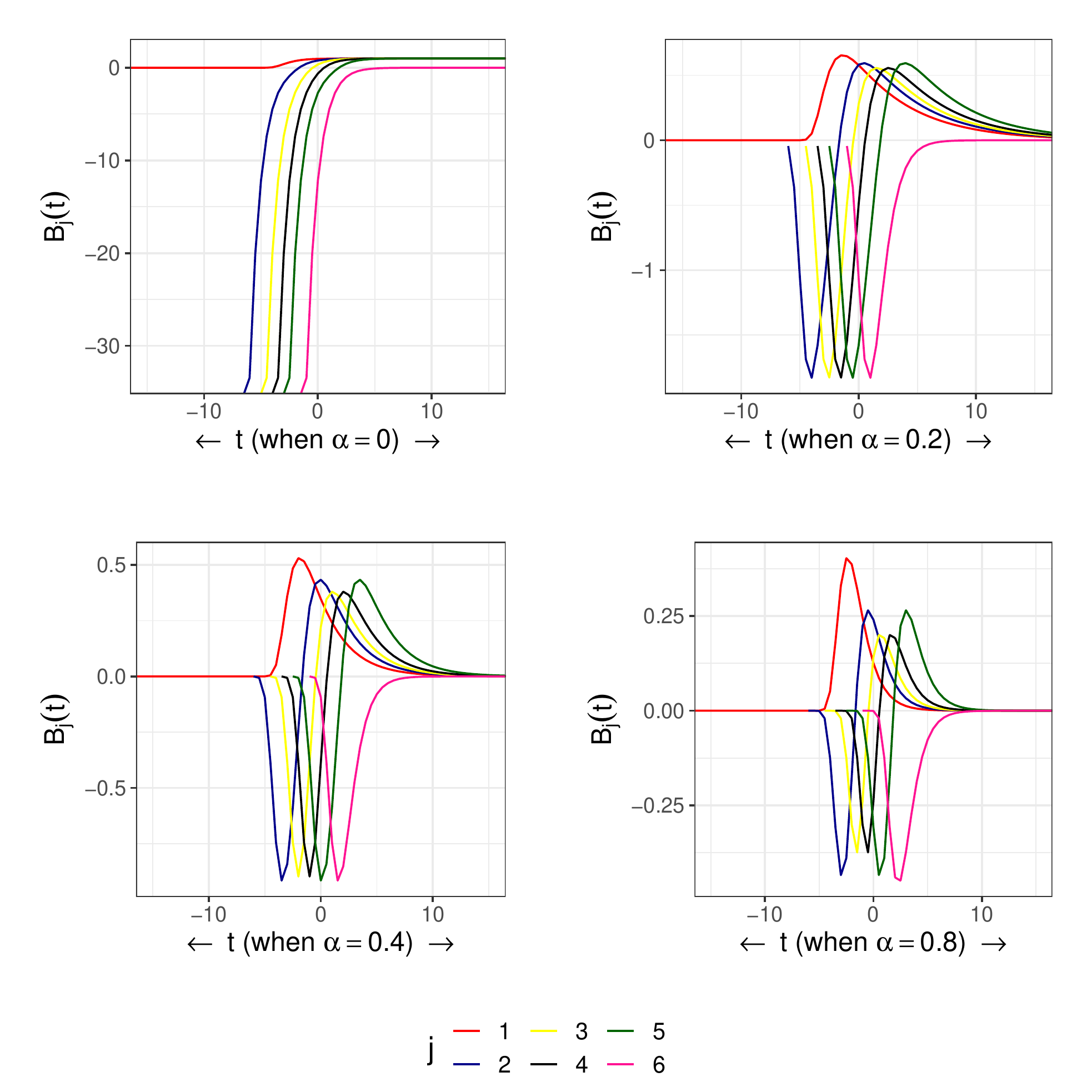} 
   \caption{Generalized residuals for the complementary log-log link with thresholds $\gamma=(-2.5, -1, 0, 1, 2.5)^{T}$.} 
\label{fig:gen res complementary log-log}
    \par 
    \end{multicols}
\end{figure}

Next, we discuss the role of generalized residuals as introduced by Iannario et al. \cite{iannario2017robust}. We know that $-\frac{1}{1+\alpha}\sum_{i=1}^{n}V_{i}(x_{i}, Y_{i}, \theta)$ is akin to the log-likelihood function. It is called the $\beta$-likelihood function (cf. Fujisawa and Eguchi \cite{fujisawa2006robust}). To find the DPD-version of the generalized residuals, we express 
\begin{align}
-\frac{1}{1+\alpha}\cdot \sum_{i=1}^{n}
\frac{\partial}{\partial \beta_{k}}V_{i}(x_{i}, Y_{i}, \theta)
=\sum_{i=1}^{n} \sum_{j=1}^{m} \mathcal{E}_{ij}(\theta, \alpha) x_{ik}
\mbox{ for all } k, 
\end{align}
where $\mathcal{E}_{ij}(\theta, \alpha)=\Big[ p_{\theta, i}(j)
- \delta_{i}(j) \Big] e_{ij}(\theta) p_{\theta, i}(j)^{\alpha}$ with $e_{ij}(\theta)$ being the generalized residuals as in (6) of Iannario et al. \cite{iannario2017robust}. Here $\mathcal{E}_{ij}(\theta, \alpha)$ plays the same role in the estimation of $\hat{\beta}_{\alpha}$ as $\delta_{i}(j)e_{ij}(\theta)$ does for $\hat{\beta}_{ML}$, the likelihood estimates of the slope parameters. Excluding the term $[p_{\theta,i}(j)-\delta_{i}(j)]$, which is bounded by $2$ anyway, we may consider $e_{ij}(\theta)p_{\theta,i}(j)^{\alpha}$ as a simple analogue for {\it generalized residuals} in the context of the DPD. To study its behaviour, denote 
\begin{align}
\label{Bj}
    B_{j}(t)=A_{j}(j)\Big[ F(\gamma_{j}-t)-F(\gamma_{j-1}-t)\Big]^{\alpha}    
    \mbox{ for } j=1, 2, \ldots , m
\end{align}
where $t=x^{T}_{i}\beta$ and 
\begin{align}
\label{Aj}
  A_{j}(t)=\frac{ f(\gamma_{j}-t)-f(\gamma_{j-1}-t)}{ F(\gamma_{j}-t)-F(\gamma_{j-1}-t)}. 
\end{align}
Here $A_{j}(t)$ comes from Equation (9) in Iannario et al. \cite{iannario2017robust}. As in Figure 1 - Figure 4 of Iannario et al. \cite{iannario2017robust}, we plot $B_{j}(t)$ in Figure \ref{fig:gen res probit} and Figure \ref{fig:gen res complementary log-log} in a panel for different values of the tuning parameter for the probit and complementary log-log link functions. Figures for the Cauchy and logit link are moved to the supplementary material. In all these graphs, we find that when $\alpha$ increases from $0$ to $1$, the magnitude of the DPD-version of the generalized residuals is significantly dampened. An outlying 
observation can make the generalized residuals unbounded when the probit and the complementary log-log link are used in the likelihood-based procedure. However, these outliers only have a limited impact on the DPD-based procedure. This also explains why the minimum density power divergence estimate of the slope parameter in the ordinal response models is more stable than the MLE when robustness is a concern. 

\section{Asymptotic Properties}
\label{asymptotic properties}

Now, we shall present the weak consistency and asymptotic normality results for $\hat{\theta}_{\alpha}$. These follow from Ghosh and Basu \cite{ghosh2013robust} with some suitable modifications. Let us introduce the quantities $J^{(i)}(\alpha)$ and $\xi_i(\alpha)$ as 
\begin{equation}
\label{J(i)}
    J^{(i)}(\alpha)
    =\frac{1}{1+\alpha} \mathds{E}_{g_{i}}\Big[\nabla^{2} V_{i}(x_{i}, Y_{i}, \theta_{\alpha})\Big]
    \mbox{, }
    \xi_{i}(\alpha)=\sum_{j=1}^{m}u_{\theta, i}(j)p^{\alpha}_{\theta, i}(j)g_{i}(j).
\end{equation}
The matrix $J^{(i)}(\alpha)$ is assumed to be positive definite for $i=1,2, \ldots,n$. Also, define
\begin{gather}
\label{psi_n}
\Psi_{n}(\alpha)=\frac{1}{n}\sum_{i=1}^{n}J^{(i)}(\alpha), \\
\label{omega_n}
\Omega_{n}(\alpha)=\frac{1}{n}\sum_{i=1}^{n} \mathds{V}ar_{g_{i}}\Big[\nabla V_{i}(x_{i}, Y_{i}, \theta_{\alpha})\Big].
\end{gather}
We now present the following assumptions that will be necessary to apply the consistency and asymptotic normality result of Ghosh and Basu \cite{ghosh2013robust}. 
\begin{description}
\descitem{(A1)} The best-fitting parameter $\theta_{\alpha}$ is an interior point of $\Theta$.

\descitem{(A2)} The error distribution function $F$ is thrice continuously differentiable with respect to its argument having bounded derivatives. 

\descitem{(A3)} The matrices $J^{(i)}(\alpha)$s are positive definite for all $i$, and 
\begin{equation}
\label{A2}
   \lambda_{0}:=\underset{n}{\inf}\Big[\min \mbox{ eigenvalue of } \Psi_{n}(\alpha)\Big]>0. 
\end{equation}

\descitem{(A4)} The vector $x_{i}=(x_{i1}, \ldots , x_{ip})^{T}$ is such that the following conditions are true: 
    \begin{align}
      \frac{1}{n}\sum_{i=1}^{n}|x_{ij}x_{ij'}x_{ij^{*}}|=\mathcal{O}(1)  
      \mbox{ , }      
      \sup_{n} \max_{1 \le i \le n}|x_{ij}|=\mathcal{O}(1)
      \mbox{ and }
      \sup_{n}\max_{1 \le i \le n}|x_{ij}x_{ij'}|=\mathcal{O}(1)
    \end{align}
 for all $j, j', j^{*} =1, 2, \ldots ,p$.   
\end{description}
\begin{Remark}
\label{remark: bounded derivs of f}
It may be easily checked that 
\begin{align}
    f'(x)=
    \begin{cases}
    \frac{e^{-x}(e^{-x}-1)}{(e^{-x}+1)^{3}}    &\mbox{ when } X \sim Logistic(0,1),\\
    -\frac{1}{\sqrt{2\pi}} e^{-\frac{x^{2}}{2}} x    &\mbox{ when } X \sim \mathcal{N}(0,1),\\
    -\frac{1}{\pi} \cdot \frac{2x}{(1+x^{2})^{2}}
    &\mbox{ when } X \sim \mathcal{C}(0, 1) ,\\
   e^{x-e^{x}}(1-e^{x})
    &\mbox{ when } F(x)=1-e^{-e^{x}}.
    \end{cases}
\end{align}
See that $f'$s are bounded for all these above links as long as we assume $0 \times  \infty=0$. In all these cases, density functions $f$ and $f''$ are also bounded. This boundedness ensures that Assumption (A2) is satisfied for these link functions. Assumption \descref{(A3)} refers to the condition that the smallest eigen root of $\Psi_{n}(\alpha)$ should stay positive in limit. Also, Assumption \descref{(A4)} requires that the cross-products between the components of the non-stochastic covariates $x_{i}$s be bounded.

\end{Remark}

\begin{Theorem}
\label{Theorem: Consistency and CLT}
Suppose the Assumptions \descref{(A1)} to \descref{(A4)} are true. Then the following holds: 
\begin{description}
\descitem{(a)} $\hat{\theta}_{\alpha} \overset{\mathds{P}}{\longrightarrow}\theta_{\alpha}$  as n $\to +\infty$, 
\descitem{(b)} $\sqrt{n}\Omega_{n}^{-\frac{1}{2}}(\alpha)\Psi_{n}(\alpha)\Big(\hat{\theta}_{\alpha}-\theta_{\alpha}\Big)\overset{\mathcal{L}}{\longrightarrow} \mathcal{N}\Big(0,I_{m+p-1}\Big)$ as $n \to +\infty$. 
\end{description}
\end{Theorem}

\begin{Remark}
Let the true distributions belong to the model families, i.e., $g_{i}(j)=p_{\theta_{*}, i}(j)$ for all $j=1, 2, \ldots,m$ and $i=1, 2, \ldots,n$. In this case, we get $\xi_{i}(0)=0$ and $\Psi_{n}(0)=\Omega_{n}(0)=I(\theta_{*})$ where $I(\theta_{*})$ is the Fisher information. So the asymptotic covariance matrix of the MLE becomes $I^{-1}(\theta_{*})$.     
\end{Remark}

\begin{Remark}
 In simulation studies, one may compute the MSE to compare the performance of $\hat{\theta}_{\alpha}$ with the MLE. In such cases, the observed efficiency (Eff) of $\hat{\theta}_{\alpha}$ may be defined as $Eff=\frac{MSE_{ML}}{MSE_{\hat{\theta}_{\alpha}}}$. Because the estimators are consistent when the true distributions belong to the model family, it approximately becomes   
\begin{equation}
\label{efficiency}
Eff \approx \frac{tr\Big(\Psi^{-1}_{n}(0)\Omega_{n}(0) \Psi^{-1}_{n}(0)\Big)} {tr\Big(\Psi^{-1}_{n}(\alpha)\Omega_{n}(\alpha) \Psi^{-1}_{n}(\alpha) \Big)}
\end{equation}
for sufficiently large $n$, where $tr(A)$ denotes the trace of a matrix $A$. When true distributions belong to the model family, smaller values of $\alpha$ should perform almost as good as the MLE. If there are some outliers in a data set, the performance of the MLE may become very unstable depending on the amount of anomaly in the data. As the MDPDE naturally downweights those outlying observations with respect to the model, we expect that the MDPDE will have superior performance over the MLE in the presence of outliers, particularly for relatively large values of $\alpha$.
\end{Remark}

\begin{Remark}
\label{remark:asymp var}
The asymptotic covariance matrix of $\sqrt{n}\hat{\theta}_{\alpha}$ is given by $(\Psi^{-1}_{n}(\alpha)\Omega_{n}(\alpha)\Psi^{-1}_{n}(\alpha))$. This needs to be estimated in real data analysis. A consistent estimator of $\Psi_{n}(\alpha)$ is obtained by plugging in $\hat{\theta}_{\alpha}$ and $\delta_{i}$ respectively for $\theta_{\alpha}$ and $g_{i}$ in (\ref{psi_n}). This gives  
\begin{align}
\label{est psi_n}
\hat{\Psi}_{n}(\alpha) 
=&\frac{1}{n} \sum_{i=1}^{n} \Bigg[\sum_{j=1}^{m}\Big\{\nabla u_{\hat{\theta}_{\alpha},i}(j)+(1+\alpha)u_{\hat{\theta}_{\alpha},i}(j) u_{\hat{\theta}_{ \alpha},i}(j)^{T}\Big\}p_{\hat{\theta}_{\alpha},i}(j)^{1+\alpha}  \nonumber \\
&-\Big\{\nabla u_{\hat{\theta}_{\alpha},i}(Y_{i})+\alpha u_{\hat{\theta}_{\alpha},i}(Y_{i}) u_{\hat{\theta}_{ \alpha},i}(Y_{i})^{T}\Big\}p_{\hat{\theta}_{\alpha},i}(Y_{i})^{\alpha}  \Bigg].
\end{align}
However $\Omega_{n}(\alpha)$ cannot be estimated using only a single observation $Y_{i}$ that comes from $g_{i}, i=1,2 \ldots,n$. To tackle this case, therefore, we make use of the model densities as proxies for true densities in (\ref{omega_n}) substituting $\hat{\theta}_{\alpha}$ for $\theta_{\alpha}$. Thus we obtain 
\begin{align}
 \label{est omega_n}
\hat{\Omega}_{n}(\alpha) &=\frac{1}{n}\sum_{i=1}^{n}\Bigg\{\sum_{j=1}^{m} u_{\hat{\theta}_{\alpha},i}(j)u_{\hat{\theta}_{ \alpha},i}(j)^{T}p_{ \hat{\theta}_{\alpha},i}(j)^{2\alpha+1} -\hat{\xi}_{i}(\alpha)\hat{\xi}_{i}(\alpha)^{T}\Bigg\}, \\
\label{est xi_i}
\hat{\xi}_{i}(\alpha) 
&=\sum_{j=1}^{m}u_{\hat{\theta}_{ \alpha},i}(j)p_{\hat{\theta}_{ \alpha},i}(j)^{1+\alpha}.
\end{align}
\end{Remark}
These estimates will be made use of while doing the tuning parameter selection.

\section{Robustness Studies}
\label{Robustness Studies}

In this section, we will study the robustness of the minimum distance functional. This is, in fact, the main theme of this paper. This section contains the following three parts -- influence function analysis, asymptotic breakdown point analysis, and a discussion of the implosion resistance property of the slope estimates.  

\subsection{Influence Function Analysis}
\label{IF analysis}  

The influence function (IF) is one of the most popular measures of robustness in studying the impact of infinitesimal data contamination on a statistical functional. Essentially, an estimate having a bounded influence function exhibits stable behaviour in the presence of very extreme outlying observations. Here, we shall present the influence function of the MDPD functional $\theta_{\alpha}$ that minimizes $H(\theta)$. Given our setup, where $x_{i}$s are fixed carriers, an outlier may only occur in the vertical direction (i.e., in the $Y$-space). Therefore, having contamination at the $i$-direction in the vertical space only perturbs the distribution of the total mass over the set of ordinal responses given at fixed $x_{i}$. This may be described as the true distribution $G_{i}$ being contaminated at a point $t_{i}$ resulting in the distribution $G_{i, \epsilon}=(1-\epsilon)G_{i}+\epsilon \Lambda_{t_{i}}$, where $\Lambda_{t_{i}}$ is the distribution function degenerate at $t_{i}=1, 2, \ldots ,m$. 

Let the true distribution $G_{i_{j}}$ be contaminated as $G_{i_{j}, \epsilon}$ for $j=1, \ldots, k$ where $k \le n$. Through a straight-forward differentiation, the influence function of $\theta_{\alpha}$ when a subset of the data set is contaminated, is obtained as
\begin{equation}
\label{IF MDPD}
    \mathcal{IF}(\theta_{\alpha},G_{i_{1}},\cdots,G_{i_{k}}, t_{i_{1}},\cdots,t_{i_{k}})
    =\sum_{j=1}^{k} \frac{1}{n}\Psi_{n}^{-1}(\alpha)
    \Big\{p_{\theta_{\alpha},i_{j}}(t_{i_{j}})^{\alpha} u_{\theta_{\alpha},i_{j}}(t_{i_{j}})-\xi_{i_{j}}(\alpha)\Big\}.
\end{equation}
Observe that the $i_{j}$-th summand in (\ref{IF MDPD}) is exactly the influence function when contamination is present only at the $i_{j}$-th distribution $G_{i_{j}}$.  

It is evident in (\ref{IF MDPD}) that the MDPD functional $\theta_{\alpha}$ downweights the influence of the data points that are inconsistent with the model with weights being chosen as model densities raised to the power of $\alpha\in [0,1]$. In the following discussion, we consider $k=n$. In this case, the influence function is a vector-valued function depending on $(t_{1}, t_{2}, \ldots,t_{n})$ where $t_{i} \in \{1, 2, \ldots,m\}$ for all $i$. The influence function also depends on the fixed carriers $(x_{1}, x_{2}, \ldots ,x_{n})$ through the models. Since the number of levels is finite, it may be appropriate to plot the gross error sensitivity (GES) rather than the influence function itself. The GES using (\ref{IF MDPD}) with $k=n$ is given by 
\begin{align}
    \label{IF unstandardized}
    GES(\theta_{\alpha})
    =\max_{t_{1}, \ldots , t_{n}}
    \big|\big|\mathcal{IF}
    (\theta_{\alpha}, G_{1}, \ldots, G_{n}, t_{1}, t_{2}, \ldots , t_{n}) \big|\big|, 
\end{align} 
where $||\cdot||$ is the Euclidean norm. The gross error sensitivity in (\ref{IF unstandardized}) may be standardized using the asymptotic variance of $\hat{\theta}_{\alpha}$. See Ghosh and Basu \cite{ghosh2013robust} for more details. Let the components of the best-fitting parameter $\theta_{\alpha}$ be denoted as $\gamma_{\alpha}=(\gamma_{1, \alpha}, \ldots , \gamma_{(m-1), \alpha})$ and $\beta_{\alpha}=(\beta_{1,\alpha}, \ldots , \beta_{p, \alpha})$. The gross error sensitivity of each component may be similarly obtained using the specified component of the IF given in (\ref{IF MDPD}). These are respectively denoted by $GES(\gamma_{1,\alpha}), \ldots , GES(\gamma_{(m-1), \alpha}), GES(\beta_{1, \alpha}), \ldots , GES(\beta_{p, \alpha})$. Dependence on the true distributions is kept implicit in these notations. 

\begin{figure}[!ht]
\begin{multicols}{2}
    \includegraphics[scale=0.4]{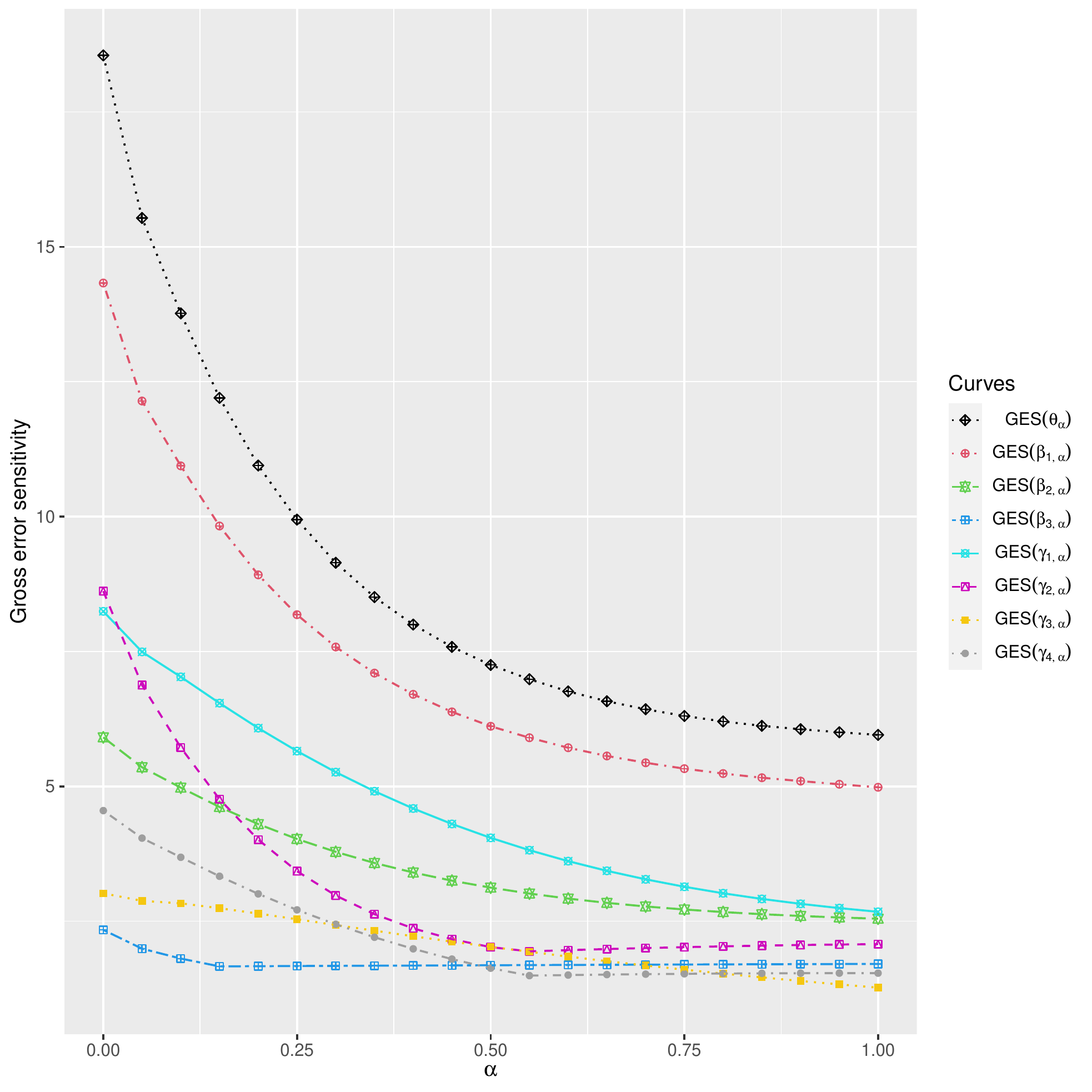}
    \caption{GES of the best-fitting parameters related to \descref{Model 1} with the probit link.} 
   \label{fig:ges of model1 with probit}
    \par 
    \includegraphics[scale=0.4]{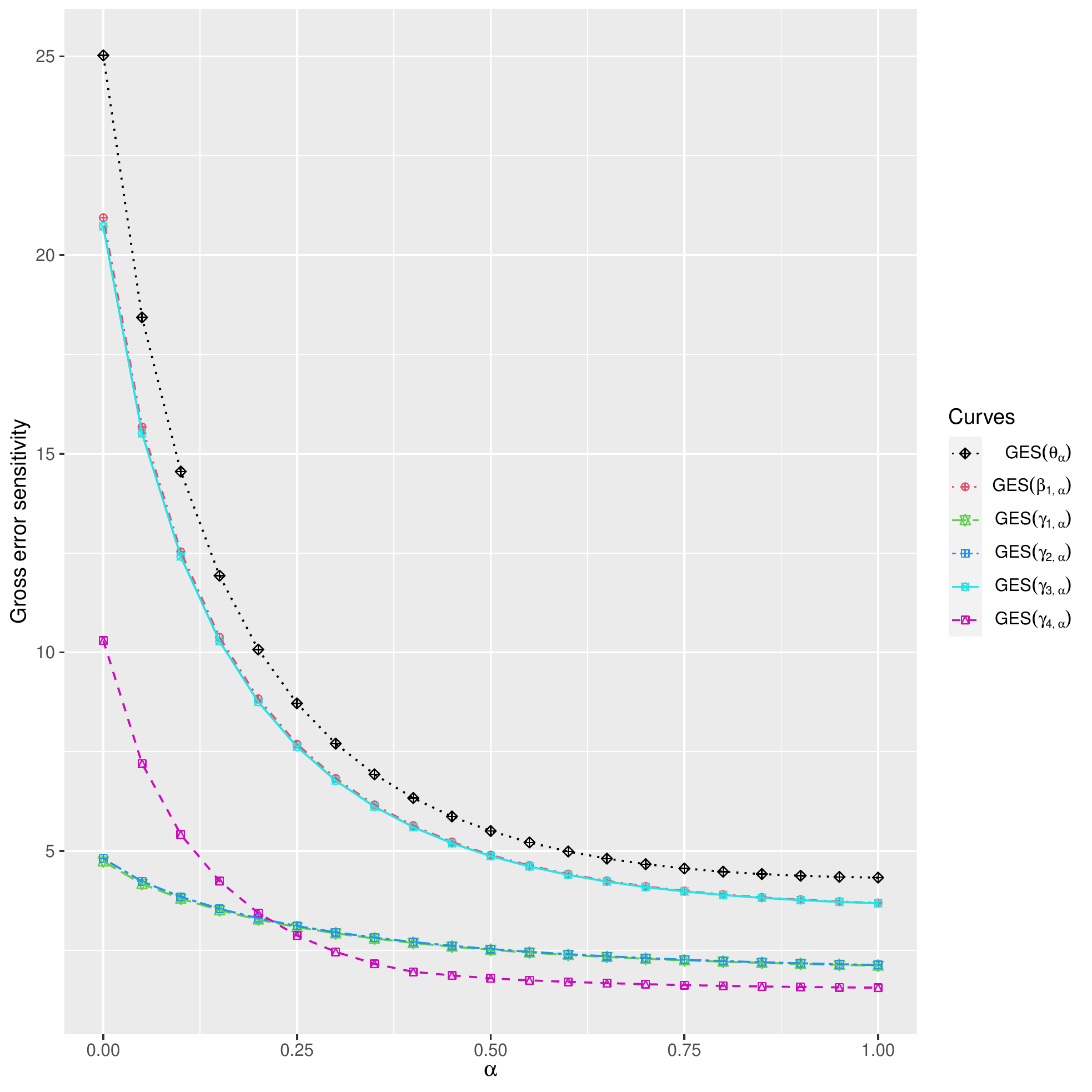} 
   \caption{GES of the best-fitting parameters of \descref{Model 2} with the logit link.} 
\label{fig:ges of model2 with logit}
    \par 
    \end{multicols}
\end{figure}

Next, we plot the GES of different components of $\theta_{\alpha}$ related to \descref{Model 1} and \descref{Model 2} (described in Section \ref{Numerical Studies}) respectively in Figure \ref{fig:ges of model1 with probit} and Figure \ref{fig:ges of model2 with logit}. It is clear from these graphs that $\alpha=0$ represents the case for which GES attains its highest value. It decreases steadily as $\alpha$ increases from $0$ towards $1$. This gives a piece of strong evidence that when $\alpha$ is chosen close to zero the MDPD functional may tend to produce higher absolute bias at misspecified models; this indeed includes the case of maximum likelihood functional even for the logit link. As $\alpha$ increases, the MDPD functionals achieve better stability against model misspecification, particularly at larger values of $\alpha$. 

\subsection{Asymptotic Breakdown Point}
\label{Breakdown point}

We know that the influence function is a local measure of robustness that must be complemented with a global measure of stability. Now, the (asymptotic) breakdown point of $\theta_{\alpha}$ is computed for fixed $\alpha>0$ under appropriate conditions. 

Suppose that the $i$-th true distribution $G_{i}$ is contaminated at $\epsilon$-proportion with a sequence of contaminating distributions $\{K_{i, M}\}_{M=1}^{\infty}$ as $H_{i,\epsilon,M}\equiv (1-\epsilon)G_{i}+\epsilon K_{i,M}$ for fixed $n,m$. The corresponding probability density functions are denoted by $h_{i,\epsilon,M}$, $g_{i}$ and $k_{i,M}$. We assume that the true densities $\{g_{i}\}$, the models $\{p_{\theta,i}: \theta \in \Theta\}$ and the contaminating sequence of densities $\{k_{i,M}\}$ are all supported on $\chi$. Let $\theta_{\alpha}^{h_{\epsilon,M}}$ be the MDPD functional when all $G_{i}$s are $\epsilon$-contaminated as above, i.e.,
\begin{align}
    \theta_{\alpha}^{h_{\epsilon, M}}:=\arg \min_{\theta} 
    \frac{1}{n}\sum_{i=1}^{n} d_{\alpha}(h_{i, \epsilon, M}, p_{\theta,i}).
\end{align}
We declare that the breakdown of $\theta_{\alpha}$ occurs at $\epsilon$-contamination if $||\theta_{\alpha}-\theta_{\alpha}^{h_{\epsilon,M}}|| \to +\infty$ when $M \to +\infty$ for fixed $n,m$ (Simpson \cite{simpson1989hellinger}; Park and Basu \cite{park2004minimum}). 
The asymptotic breakdown point of $\theta_{\alpha}$ is defined as the least proportion of contamination $\epsilon \in [0, \frac{1}{2}]$ such that $||\theta_{\alpha}-\theta_{\alpha}^{h_{\epsilon,M}}|| \to +\infty$ when $M \to +\infty$ for fixed $n,m$. Note that, in consistent with the literature of robustness, the definition restricts the asymptotic breakdown to be at most $\frac{1}{2}$ since we would intuitively expect the majority of the data to be uncontaminated. Define
  \begin{equation}
  D_{\alpha}\Big( g_{i}(j),p_{\theta, i}(j)\Big)=\Bigg\{p_{\theta,i}^{1+\alpha}(j)-\Big(1+\frac{1}{\alpha}\Big) p_{\theta,i}^{\alpha}(j)g_i(j)+\frac{1}{\alpha}g_i(j)^{1+\alpha}\Bigg\} \mbox{ for }  j \in \chi, 
\end{equation}
and $M_{i}^{\alpha}=\sum_{j}p_{\theta_{\alpha},i}(j)^{1+\alpha}$ for all $i$. Let us make the following assumptions to find the asymptotic breakdown point for $\theta_{\alpha}$. Throughout this section, $m$ is assumed to be fixed.
\begin{description}
\descitem{(B1)} 
$g_{i}$ and $p_{\theta,i}$ belong to a common family $\mathscr{F}_{\theta,i}, i=1,2, \ldots ,n$.
\descitem{(B2)} 
There exists a set $B \subset \chi$ and a positive $\delta^{*}_{1}(j)$ depending on $j \in B$, such that 
\begin{equation}
  \big\{k_{i, M}(j)- p_{\theta,i}(j)\big\} \longrightarrow \delta^{*}_{1}(j) \ge 0 
  \mbox{ and } \sum_{j \in B}k_{i, M}(j) \longrightarrow 1
  \mbox{ as } M \to +\infty \mbox{ for all } i
\end{equation}
uniformly for $||\theta||< +\infty$. This condition means that on a set $B \subset \chi$, the contaminating sequence of densities $k_{i, M}$ asymptotically dominate $p_{\theta, i}$ when the parameter $\theta$ is uniformly bounded; moreover, $B^{c}$ asymptotically becomes a $K_{i, M}$-null set as $M$ increases to infinity for each $i$. 
\descitem{(B3)}
There exists a set $C \subset \chi$ and a positive $\delta^{*}_{2}(j)$ depending on $j \in C$, such that 
\begin{align}
 \big\{p_{\theta_{M},i}(j)- p_{\theta_{\alpha},i}(j)\big\} \longrightarrow \delta^{*}_{2}(j) \ge 0 
 \mbox{ and } \sum_{j\in C} p_{\theta_{M},i}(j) \longrightarrow 1
 \mbox{ as } M \to +\infty \mbox{ for all } i,
\end{align}
for any sequence $\{\theta_{M}\}$ such that $||\theta_{M}|| \to +\infty$ as $M \to +\infty$. This means that as $ ||\theta_{M}||$ diverges, the associated sequence of models $p_{\theta_{M},i}$ tend to dominate the true density $p_{\theta_{\alpha},i}$ on a set $C$, and the sequence $p_{\theta_{M},i}$ remain concentrated on $C$ for $i=1,2, \ldots ,n$.
\descitem{(B4)} \hfill 
\begin{itemize}
    \item [(a)]Assume that for any density $q_{i}$ supported on $\chi$, we have
\begin{equation}
\label{B4:eq1}
d_{\alpha}(\epsilon q_{i}, p_{\theta,i}) \ge d_{\alpha}(\epsilon p_{\theta_{\alpha},i}, p_{\theta_{\alpha},i}) \mbox{ for all } \theta,i \mbox{ and } 0 < \epsilon < 1.
\end{equation}
This means that $d_{\alpha}(\epsilon q_{i}, p_{\theta,i})$ will be minimized at $\theta=\theta_{\alpha}$ and $q_{i}=p_{\theta_{\alpha},i}$ for all $i$.

\item[(b)] Further assume that 
\begin{equation}
\label{B4:eq2}
\underset{M \to +\infty}{\limsup} \Big( k_{i,M}(j) \Big) 
\le p_{\theta_{\alpha},i}(j) \mbox{ for all } i,j, \mbox{ and } 
M^{\alpha}=\frac{1}{n} \sum_{i=1}^{n}M^{\alpha}_{i} < +\infty 
\end{equation}  
for fixed $n,m$ and for all $\alpha>0$.
\end{itemize}
\end{description}

\begin{Remark}

Since we know that $p_{\theta,i}(j)=F(\gamma_{j}-x_{i}^{T}\beta)- F(\gamma_{j-1}-x_{i}^{T}\beta)$, Assumption \descref{(B3)} can be verified in the following situations. 
\begin{description}
\descitem{(S1)} Any particular $\gamma_{j}$ decreases to $-\infty$ or increases to $+ \infty$, but $\beta$s remain bounded.  

\descitem{(S2)} Let $\gamma_{j_{1}} \to -\infty$ and $\gamma_{j_{2}} \to +\infty$ for any $1 \le j_{1} < j_{2} \le m-1$, but $\beta$s remain bounded. 

\descitem{(S3)} $\gamma_{j}$s remain bounded but $\beta$s diverge to $\pm \infty$. 
\end{description}
Consider the scenario \descref{(S1)}. Suppose $\gamma_{j} \to -\infty$. Then $\gamma_{r} \to -\infty$ for $r=1, 2, \ldots , j$ because $\gamma_{1} < \gamma_{2} < \cdots \gamma_{j-1}< \gamma_{j}$. Therefore $p_{\theta, i}(r) \to 0$ for $r=1,2, \ldots , j$, and $\sum_{r=j+1}^{m}p_{\theta, i}(r) \to 1$. So the probability $p_{\theta, i}(\cdot)$ will become concentrated on the set $C^{j}(-\infty):=\{j+1, \ldots , m\}$. If we assume that $\gamma_{j} \to +\infty$. Then $\gamma_{r} \to +\infty$ for $r=j, \ldots , m$. This gives $p_{\theta, i}(r) \to 0$ for $r=j+1, \ldots , m$. In this case, the probability $p_{\theta, i}(\cdot)$ gets concentrated on $C^{j}(+\infty):=\{1, 2, \ldots ,j\}$. 

Next we consider \descref{(S2)}. Now $\gamma_{j_{1}} \to -\infty$ and $\gamma_{j_{2}} \to +\infty$ such that $\gamma_{j_{1}} < \gamma_{j_{2}}$. Using the above argument we see that the probability will get concentrated on the set $C^{j_{1}}(-\infty) \cap C^{j_{2}}(+\infty)=\{j_{1}+1, \ldots, j_{2}\}$. As $\chi=C^{0}(-\infty) \cap C^{m}(+\infty)$, both the sets $C^{j}(-\infty), C^{j}(+\infty)$ are proper subsets of $\chi$ for $j=1, \ldots , m-1$. 

In \descref{(S3)}, it will depend on the sign of $x_{i}^{T}\beta$. If $x_{i}^{T}\beta \to +\infty$, then all the terms $(\gamma_{j}-x_{i}^{T}\beta)$ goes to $-\infty$. In that case, the last term $p_{\theta,i}(m)=1-F(\gamma_{m-1}-x_{i}^{T}\beta) \to 1$. Thus the mass gets concentrated on the singleton set $\{m\}$. On the other hand, if $x_{i}^{T}\beta \to -\infty$, then the probability mass gets concentrated on the set $\{1\}$. In all these above cases $||\theta|| \to \infty$. These proper subsets can be taken as $B$ and $C$ as mentioned in Assumptions \descref{(B2)} and \descref{(B3)}. 

Assumption \descref{(B4)}(a) ensures that the divergence in the LHS of (\ref{B4:eq1}) attains its minimum value at the models when $\theta$ being chosen as the best-fitting parameter. In \descref{(B4)}(b) we state the extremity of contamination up to which the true distribution may be contaminated, but still produce reasonable MDPD estimates. 
\end{Remark}

\begin{Theorem}
\label{Theorem: Asymptotic Breakdown point}
Suppose $g_{i}, p_{\theta,i}$ and the contaminating sequence of densities $\{k_{i,M}\}_{M=1}^{\infty}$ are supported on $\chi$, $i=1, \ldots,n$. Then under the Assumptions \descref{(B1)} - \descref{(B4)}, the asymptotic breakdown point $\epsilon^{*}$ of the MDPD functional $\theta_{\alpha}$ is $\frac{1}{2}$ at the model for $\alpha>0$.
\end{Theorem}

\subsection{Implosion Breakdown of the Slope Estimates}
\label{implosion Breakdown of the slope estimates}

The notion of the breakdown discussed earlier may be called the explosive breakdown because it makes an estimator explode towards infinity. Following Croux et al. \cite{croux2013robust} we know that the lack of robustness of the MLE under the present model set-up arises also due to the implosion of the slope estimator $\hat{\beta}_{ML}$ towards zero. Let the initial sample be $Z_{n}=\{z_{1}, \ldots , z_{n}\}$ where $z_{i}=(x_{i}, Y_{i}): i=1, \ldots ,n$. Upon adding $m$ outliers, the initial random sample becomes $Z'_{n+m}=\{z_{1}, \ldots , z_{n}, z_{n+1}, \ldots , z_{n+m}\}$. Then the implosion breakdown point of $\hat{\beta}$ is defined as $\epsilon^{-}=\frac{m^{-}}{m^{-}+n}$
where 
\begin{align}
 m^{-}=\Bigg\{m \in \mathds{N}_{0}: \inf_{z_{n+1}, \ldots, z_{n+m}}||\hat{\beta}(Z^{'}_{n+m})||=0\Bigg\}.
\end{align}
It is difficult to calculate the implosion breakdown point theoretically; we shall empirically demonstrate that the minimum density power divergence estimates of the slope parameter have a very high implosion breakdown point for $\alpha>0$ in situations which lead to implosion breakdown for the MLE.    

\begin{figure}[ht]
\begin{multicols}{2}
    \includegraphics[scale=0.4]{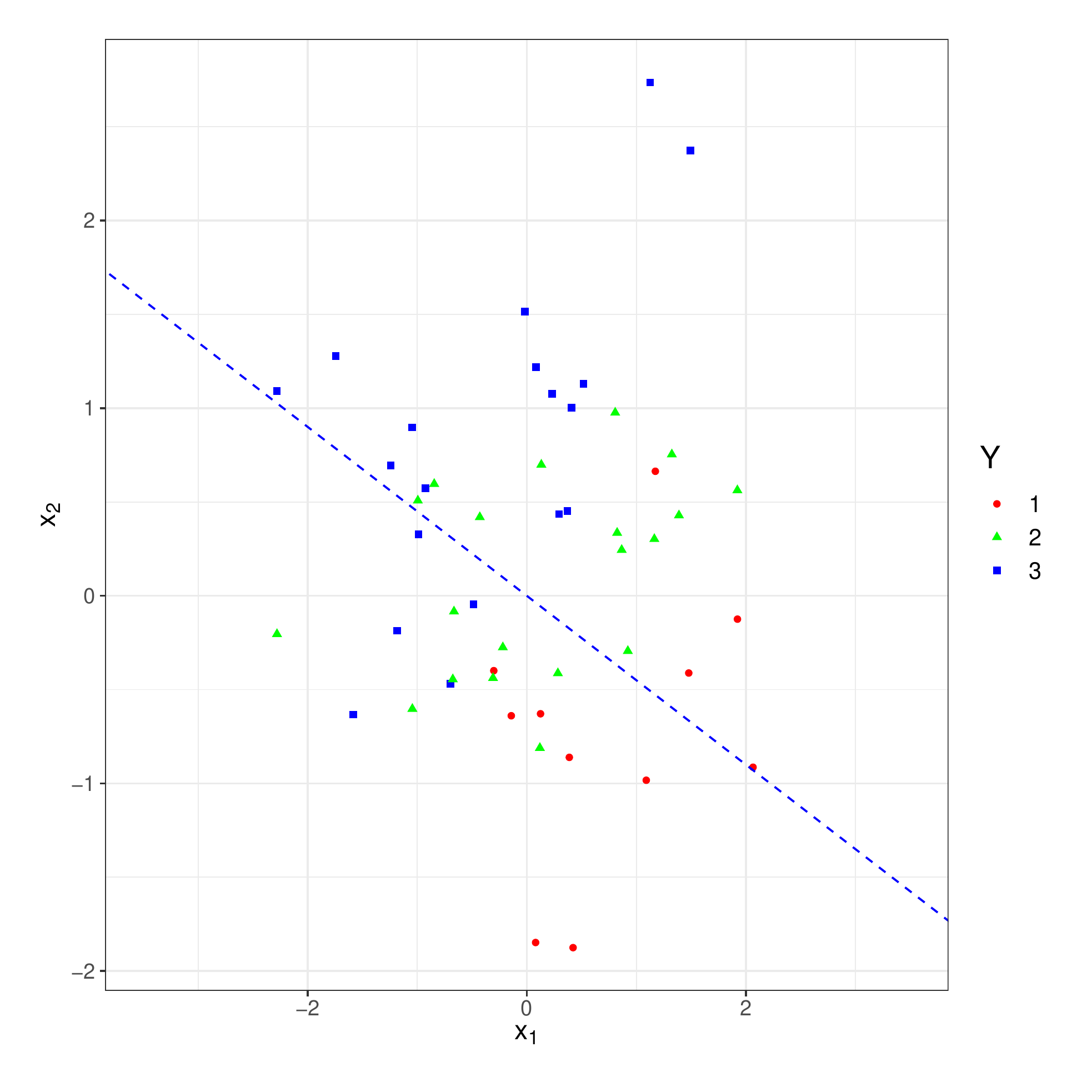} 
    \caption{Scatter plot of a simulated data set. Outliers $((s, -s), 3)$ will be added along the diagonal line. } 
   \label{fig:scatter plots}
    \par 
    \includegraphics[scale=0.4]{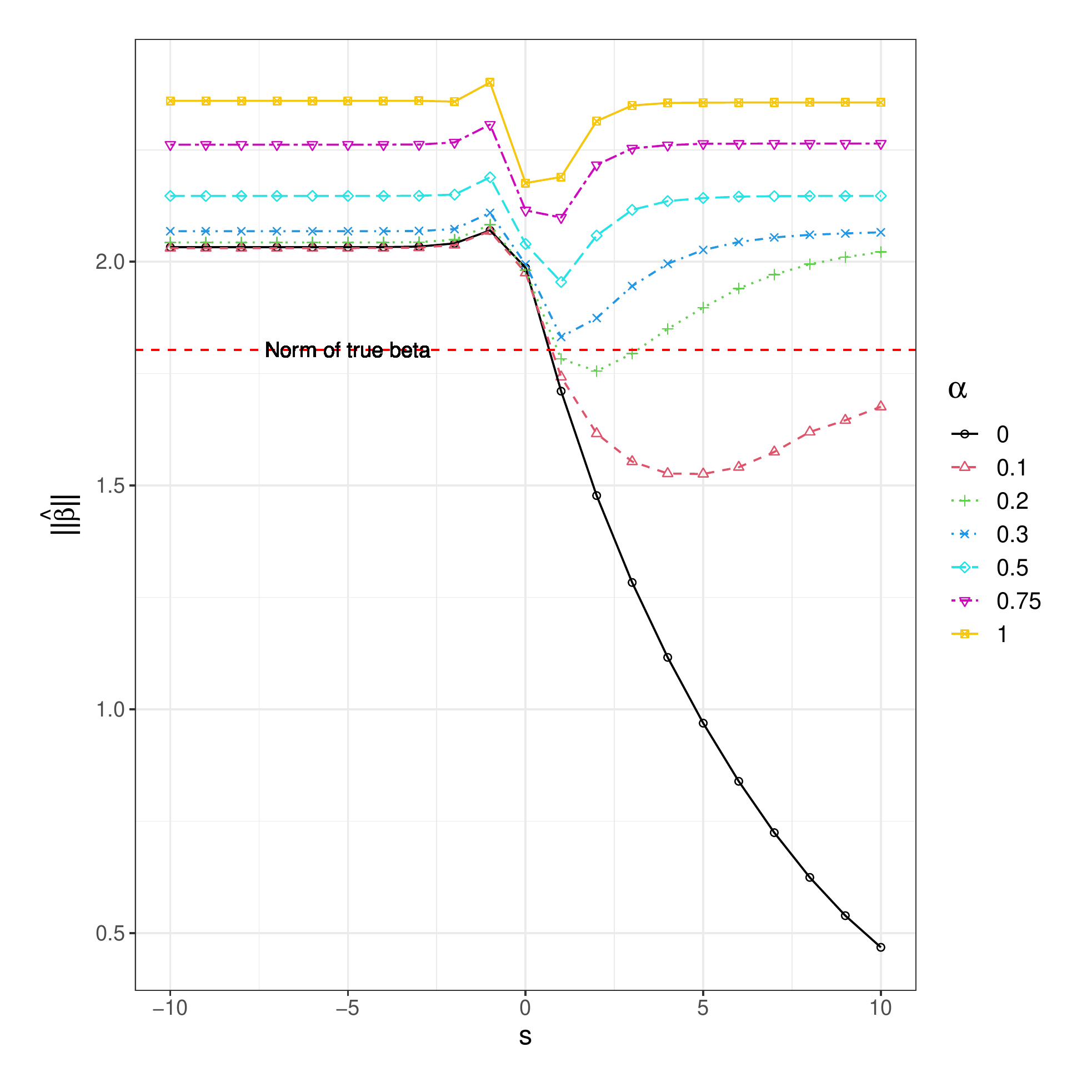} 
\caption{Norm of $\hat{\beta}$ when a single outlier is added along the diagonal line in Figure \ref{fig:scatter plots}.} 
\label{fig:norm of beta}
    \par 
    \end{multicols}
\end{figure}

For illustration, let us exhibit the plot of a random sample of size $50$ as in Figure \ref{fig:scatter plots}. This data set contains non-stochastic covariates $x=(x_{1}, x_{2})^{T}$ whose components are fixed at the values generated by two independent standard normal distributions. The associated error terms are distributed as the standard logistic distribution. Ordinal responses, that classify the observations into $3$ categories, are generated through (\ref{linear latent regression model}) using $\beta=(-1, 1.5)^{T}, \gamma=(-1, 1)^{T}$. The norm of the true slope parameter is $||\beta||=1.803$. At the initial sample, the MLE of slope yields $||\hat{\beta}_{ML}||=2.20$ with the misclassification rate equal to $36\%$. 

\begin{figure}[!ht]
\begin{center}
 \includegraphics[scale=0.4]{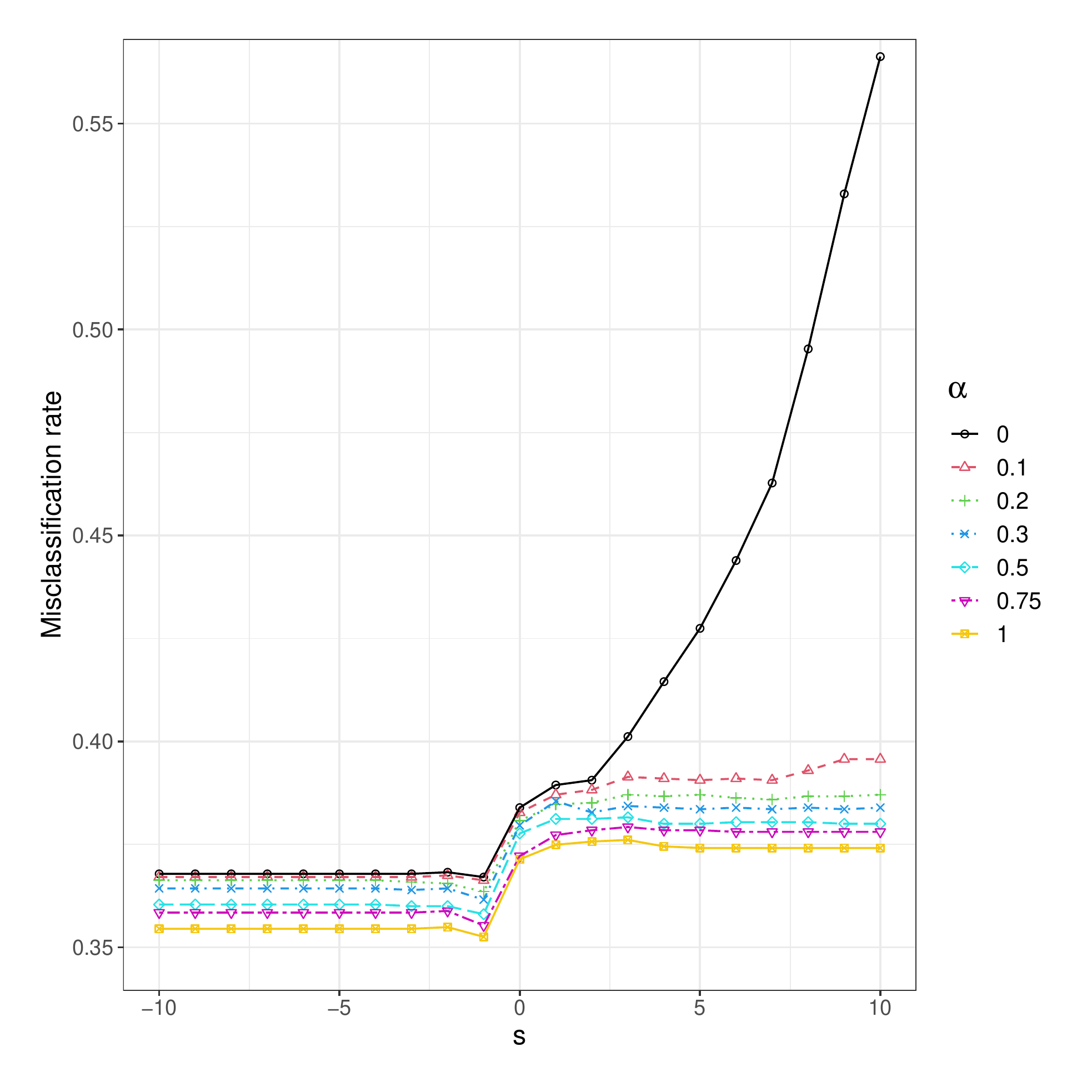} 
\caption{Misclassification rate associated with the slope estimates.} 
\label{fig:misclass prob}
\end{center}
\end{figure}

However $\hat{\beta}_{ML}$ may be completely perturbed when an outlier in the form of $((s, -s), 3)$ is added to the initial sample along the diagonal line as in Figure \ref{fig:scatter plots}. Here $x=(s,-s)$ is the projection of the outlying observation in the $x_{1}x_{2}$-space. Notice that as soon as $s$ goes outside $(-2, 2)$, the additional observation becomes outlying. However, the outlying region in the $x_{1}x_{2}$-space looks more deserted in terms of the presence of fewer $Y$-values whenever $s$ is positive. For each value of $s$, the parameters of the ordinal regression model are estimated by the MDPD methods and the corresponding misclassification rate is computed. The norm of the slope estimate and the rate of misclassification are reported, respectively, in Figure \ref{fig:norm of beta} and Figure \ref{fig:misclass prob} as functions of $s$. As expected, negative values of $s$ do not incur much bias in the slope estimates. On the other hand, as soon as $s$ gets positive, the impact of the added observation becomes more apparent in the ML estimation and gets even more extreme as $s$ increases. Not only does the norm of the slope estimator go to zero, but the misclassification rate also reaches its maximum value which is about $57\%$. When the tuning parameter is increased by a small margin, we notice that MDPDE of $\beta$ starts becoming resistant against implosion; and it becomes the most resistant for the minimum $L_{2}$ distance estimate. Consequently, the misclassification rate decreases steadily along the way as the tuning parameter $\alpha$ increases from $0$ towards $1$. 

In Table \ref{tab:implosion outlier at 8} and Table \ref{tab:implosion outlier at 10} we report the minimized norm of the slope estimates when the set of outliers-- $I_{s}=\{z_{50+i}=(s, -s, 3): i=1, 2, \ldots,50\}$ are added to the initial data with $s=8, 10$. Also, the proportion of outliers corresponding to the lowest $||\hat{\beta}||$ is reported. 
\begin{table}[!ht]
    \parbox{.5\linewidth}{
       \centering
        \caption{Minimum of $||\hat{\beta}||$ over $I_{8}$ }
         \label{tab:implosion outlier at 8}
           \begin{tabular}{cccc}
            \hline
            $\alpha$ & $\min||\hat{\beta}||$ & Prop. of outliers \\\hline
            0 &    0.323 & 0.0566  \\ \hline
            0.1 &  0.374 & 0.0566 \\ \hline
            0.2 &  0.468 & 0.438 \\ \hline
            0.3 &  1.070 & 0.479 \\ \hline
            0.5 &  2.080 & 0.495  \\ \hline
            0.75 & 2.250 & 0.390 \\ \hline
            1 & 2.340 & 0.306   \\ \hline           
        \end{tabular}
    }
    \hfill
    \parbox{.5\linewidth}{
        \centering
          \caption{Minimum of $||\hat{\beta}||$ over $I_{10}$}
          \label{tab:implosion outlier at 10}
           \begin{tabular}{ccc}
            \hline
            $\alpha$ & $\min||\hat{\beta}||$ & Prop. of outliers \\\hline
            0 &   0.319 & 0.0385  \\ \hline
            0.1 & 0.379 & 0.180 \\ \hline
            0.2 & 0.613 & 0.390 \\ \hline
            0.3 & 1.720 & 0.474 \\ \hline
            0.5 & 2.120 & 0.500  \\ \hline
            0.75 & 2.250 & 0.390 \\ \hline
            1 & 2.340 & 0.306   \\ \hline
        \end{tabular}}
\end{table}
The third column in these tables sort of gives an idea about which value of the tuning parameter $\alpha$ adds more resistance to the slope estimates such that it safeguards against implosion. It may be thought of as a finite sample analogue of the implosion breakdown point. Notice that $||\hat{\beta}||$ is close towards zero when $\alpha \downarrow 0+$, resulting into the lowest tolerance of outliers, i.e., $5.7\%$ for $s=8$, and $3.8\%$ for $s=10$. Moreover, as $s$ increases more in the positive direction, this tolerance level of MLE decreases. In this case, the MLE of $\beta$ can accommodate a smaller proportion of outliers before it starts imploding. However, as the tuning parameter $\alpha$ increases, the tolerance level improves significantly. Also, the decreasing trend of tolerance with the increment of $s$ is not observed for the MDPDE. In this numerical study, we see that the implosion breakdown point of the MDPDE of $\beta$ becomes very high for $\alpha>0$ when compared to the MLE. In some cases, this may become as high as $50\%$ for some positive values of $\alpha$.

\section{Numerical Studies}
\label{Numerical Studies}

Samples of sizes $n=150,200$ are drawn from each of the following models described in Subsection \ref{Simulation Studies: Pure Models}, and it is repeated over $1000$ (say, $B$) experiments. For any given method, let $\hat{\theta}^{(b)}=(\hat{\gamma}^{(b)}, \hat{\beta}^{(b)})$ be the estimate of $\theta$ obtained in the $b$-th experiment, $b=1, 2, \ldots, B$. The simulated mean of the estimate is obtained as $\hat{\theta}=\frac{1} {B}\sum_{b=1}^{B}\hat{\theta}^{(b)}$. The squared biases of $\hat{\gamma}$ and $\hat{\beta}$ are respectively given by $||\hat{\gamma}-\gamma||^{2}$ and $||\hat{\beta}-\beta||^{2}$ where $||\cdot||$ denotes the standard Euclidean norm. The mean squared error (MSE) of $\hat{\gamma}$ is defined as $MSE(\hat{\gamma})=\frac{1}{B}\sum_{b=1}^{B}(\hat{\gamma}^{(b)}-\gamma)^{T}(\hat{\gamma}^{(b)}-\gamma)$, and similarly for $\hat{\beta}$. We obtain the MSE of $\hat{\theta}$ simply by adding together the MSEs incurred in each of its components. When $\hat{\theta}$ is consistent its MSE consistently estimates the trace of the asymptotic covariance matrix. 

In the above setup, we will numerically compare the performances of the minimum density power divergence estimates to those of the MLE, the robust alternatives proposed by Croux et al. \cite{croux2013robust} and those by Iannario et al. \cite{iannario2017robust}. For the sake of completeness, we briefly mention here the particular choices of weight functions and tuning constants that are used in the last two methods. 

In Croux et al. \cite{croux2013robust}, a weighted log-likelihood function is constructed by multiplying the usual log-likelihood function with the weight function $w_{i}=\frac{p+3}{d_{i}+3}$ at the $i$-th data point, $i=1, \ldots,n$. Here $d_{i}$ denotes the robust Mahalanobis distance of the $i$-th covariate $x_{i}\in \mathds{R}^{p}$ computed in the space of explanatory variables. The weighted maximum likelihood (WML) estimator is then obtained by maximizing the weighted log-likelihood function. Iannario et al. \cite{iannario2017robust}, on the other hand, suggested the multiplication of  propose to multiply the usual score function by one of the following three types of weight functions 
\begin{gather}
    \label{weights of Iannario}
    w_{1}(Y_{i}, x_{i}, \theta)=\min\Bigg\{1, \frac{c}{\sum_{j=1}^{m}\delta_{i}(j)|e_{ij}(\theta)|}\Bigg\},
    \mbox{  }
    w_{2}(Y_{i}, x_{i}, \theta)=\min\Bigg\{1, \frac{c}{\sum_{j=1}^{m}\delta_{i}(j)|e_{ij}(\theta)|\cdot||x_{i}||}\Bigg\},
    \\
    w_{3}(Y_{i}, x_{i}, \theta)=\min\Bigg\{1, \frac{c}{||x_{i}||}\Bigg\}
\end{gather}
where $c>0$, $||x_{i}||$ and $e_{ij}(\theta)$ respectively denote the tuning constant, norm of $x_{i}$ and indicated generalized residuals. For more details about the computations of $||x_{i}||$ and the definition of the generalized residuals, the readers are referred to Iannario et al. \cite{iannario2017robust}. Given a particular example, a weight function is chosen as the link function is chosen. Based on their numerical studies, Iannario et al. \cite{iannario2017robust} suggest using $ c \in [1.1, 1.7]$ for the probit link, and $c \in [0.6, 1]$ (see Table 2 for Trace criterion in Iannario et al. \cite{iannario2017robust}) for the logit link to keep the efficiency loss below $5\%$. When the complementary log-log (or, simply log-log) and the Cauchy link function are added to the simulation studies, we choose the same values of $c$ as the probit and logit link, since no further suggestion is made in Iannario et al. \cite{iannario2017robust}.   

\subsection{Simulation Studies: Pure Models}
\label{Simulation Studies: Pure Models}

We consider the following models in the simulation studies. 

\begin{description}

\descitem{Model 1}: The response variable $Y$, generated by ({\ref{latent relationship}}), assumes the values $1, \ldots, 5$. $Y$ depends on three dichotomous $0-1$ variables $X_{1}, X_{2}, X_{3}$ such that at most one of them can take the value 1. The cut-off points and the regression coefficients are, respectively, given by $\gamma=(-0.7,0,1.5,2.9)^{T}$ and $\beta=(2.5,1.2,0.5)^{T}$ for both the probit and the complementary log-log link functions. 

\descitem{Model 2}: The response variable $Y$ is generated through the latent variable $Y^{*}=1.5 X + e$ where the regressor $X$ is assumed to have come from $\mathcal{N}(0,1)$. The categories of $Y$ are determined by the cut-off points $\gamma=(-1.7,-0.5,0.5,1.7)^{T}$ and $\gamma=(-2.1,-0.6,0.6,2.1)^{T}$, respectively, when the error component follows $\mathcal{N}(0,1)$ and the logistic distribution with mean $0$ and variance $\frac{\pi^{2}}{3}$. Further, the Cauchy link is used with the same cut-offs as the logit link.

\descitem{Model 3}:  The response variable $Y$ assumes 4 categories. It depends on two regressors $X_{1} \sim \mathcal{N}(0,1)$ and $X_{2} \sim \mathcal{N}(0,4)$ with $ Cov(X_1,X_2)=1.2$. The regression coefficients are given by $\beta=(1.5, 0.7)^{T}$. We use the cut-off points $\gamma=(-2.3,0,2.3)^{T}$ for the probit link and $\gamma=(-2.6,0,2.6)^{T}$ for the logit link.

\descitem{Model 4}: The response variable $Y$, that takes $3$ categories, depends upon three regressors $X_{1} \sim \mathcal{N}(0,1)$, $X_{2} \sim \mathcal{N}(0,4)$ and $X_{3} \sim \mathcal{N}(0,9)$ such that $ Cov(X_{1},X_{2})=1.5$, $Cov(X_{1},X_{3})=0.8$ and $Cov(X_{1},X_{3})=2.5$. The regression parameters are given by $\beta=(2.5,1.2, 0.7)^{T}$. The cut-off points are chosen as $\gamma=(-3.8,3.8)^{T}$ and $\gamma=(-4,4)^{T}$, respectively, for the probit and the logit links. 

\descitem{Model 5}: Here the response variable $Y$ is generated through $Y^{*}=2.5D+1.2X+0.7XD+e$ where $D \sim Bernoulli(\frac{1}{2})$ and $X \sim \mathcal{N}(0,1)$. Here $\gamma=(-1,1,3)^{T}$ for the probit and the complementary log-log links; and $\gamma=(-1.4,1.1,3.4)^{T}$ for the logit link.
\end{description}

Unless otherwise stated, all the links refer to the standard ones.
To be able to apply the theory developed earlier, $X$ must be non-stochastic. A justification is therefore required to bring the above models, where random covariates are involved, into the realm of our proposed theory. In all such models, we assume that the values of $X$ are fixed at the values generated by the aforesaid distributions, and the values of $Y$ are conditional on the generated (fixed) values of $X$. 

We find that squared biases (with one exception at $\alpha=0$; see Table \ref{Table:pure model1 probit} for $||\hat{\beta}-\beta||^{2}$) and MSEs decrease as the sample size increases. Now, we make the following remarks based on the simulation studies of the pure models. 
\begin{itemize}

\item  Highest efficiency is achieved at pure models when $\alpha=0$ (MLE) for all these link functions. As $\alpha$ increases, there is a drop in the efficiencies of the MDPD estimators. However, when $\alpha=0.1$, the loss of efficiency for using the MDPDE is roughly within $10\%$ for \descref{Model 1} - \descref{Model 3} and \descref{Model 5}. However, it drops around $50\%$ in \descref{Model 4}. These observations are clear in Figure \ref{fig:efficiency, pure, model1 and model2} to Figure \ref{fig:efficiency, pure, model5, n=200}. 

\item Generally the logit link gives better efficiency than the probit link which in turn produces more efficient estimates than the complementary log-log link at a fixed value of $\alpha$. Also, the probit link yields better efficiency than the Cauchy link. However in \descref{Model 4}, we notice that the probit link dominates (in the sense of better efficiency) the logit link up to the point around $\alpha=0.37$ ($n=150$) and $\alpha=0.45$ ($n=200$). Beyond that, the usual pattern follows. This is seen in Figure \ref{fig:efficiency, pure, model1 and model2} and Figure \ref{fig:efficiency, pure, model3 and model4}. Also Figure \ref{fig:efficiency, pure, model5, n=200} shows that the probit link dominates the complementary log-log link that in turn dominates the logit link up to a point close to $\alpha=0.45$, and thereafter the order changes when the sample size is $200$. A similar thing is also noticed in Figure \ref{fig:efficiency, pure, model5, n=150} when the sample size is $150$ with different order of domination.    

\item MDPDE performs better (in terms of lower MSE) than both the Croux et al. \cite{croux2013robust} and the Iannario et al. \cite{iannario2017robust} estimators in \descref{Model 1}, \descref{Model 2}, \descref{Model 3} (only with probit link) and \descref{Model 5}. However, Iannario et al. \cite{iannario2017robust} performs slightly better than the MDPDE in \descref{Model 3} (only with logit link) and \descref{Model 4} (only with probit, logit links). See Tables \ref{Table:pure model1 probit} - \ref{Table:pure model5 loglog} and the additional tables in the supplementary material. 

\end{itemize}



\begin{figure}
\begin{center}
 \includegraphics[scale=0.7]{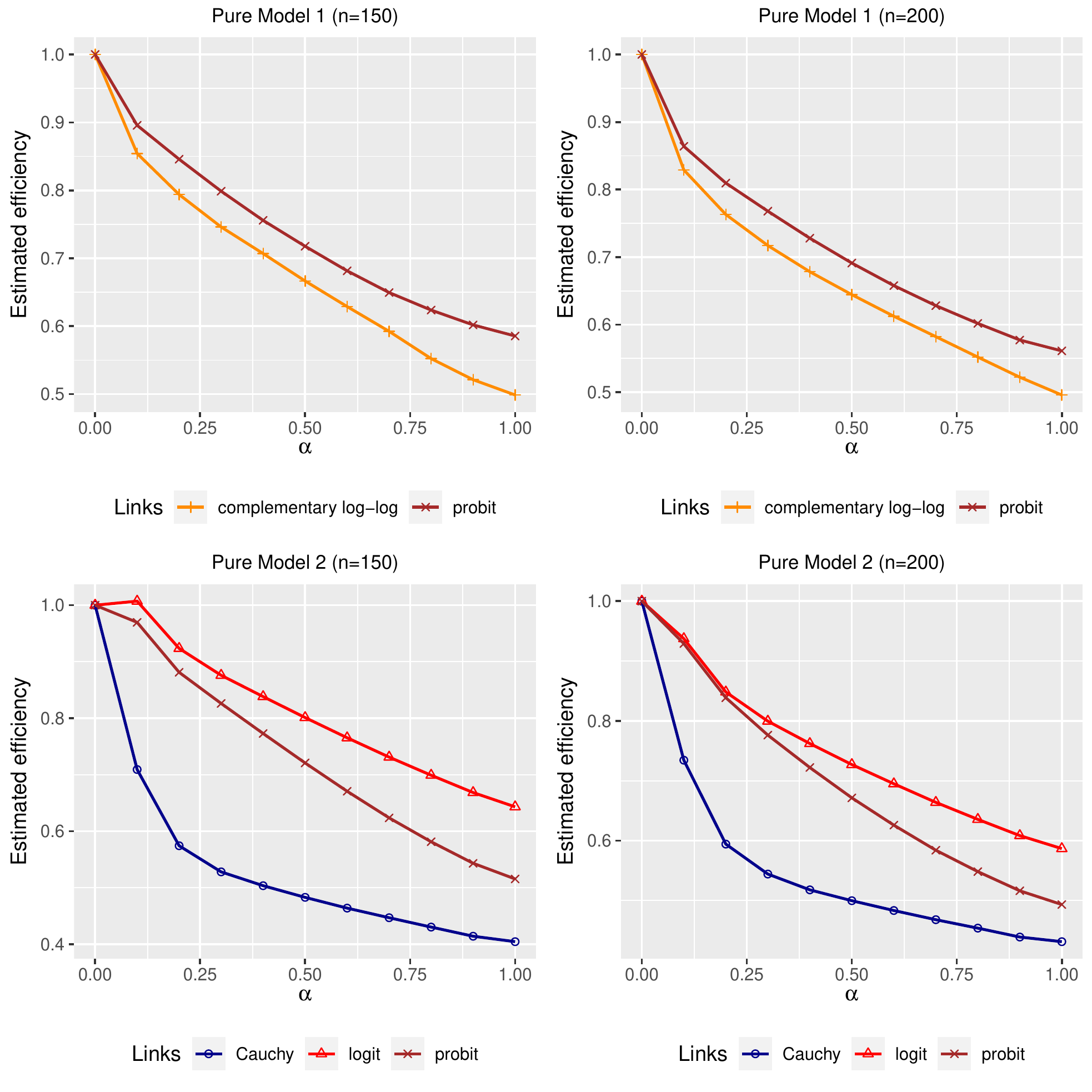} 
\caption{Graphs of efficiency at pure \descref{Model 1} and \descref{Model 2} for different sample sizes.} 
\label{fig:efficiency, pure, model1 and model2}
\end{center}
\end{figure} 

\begin{figure}
\begin{center}
 \includegraphics[scale=0.7]{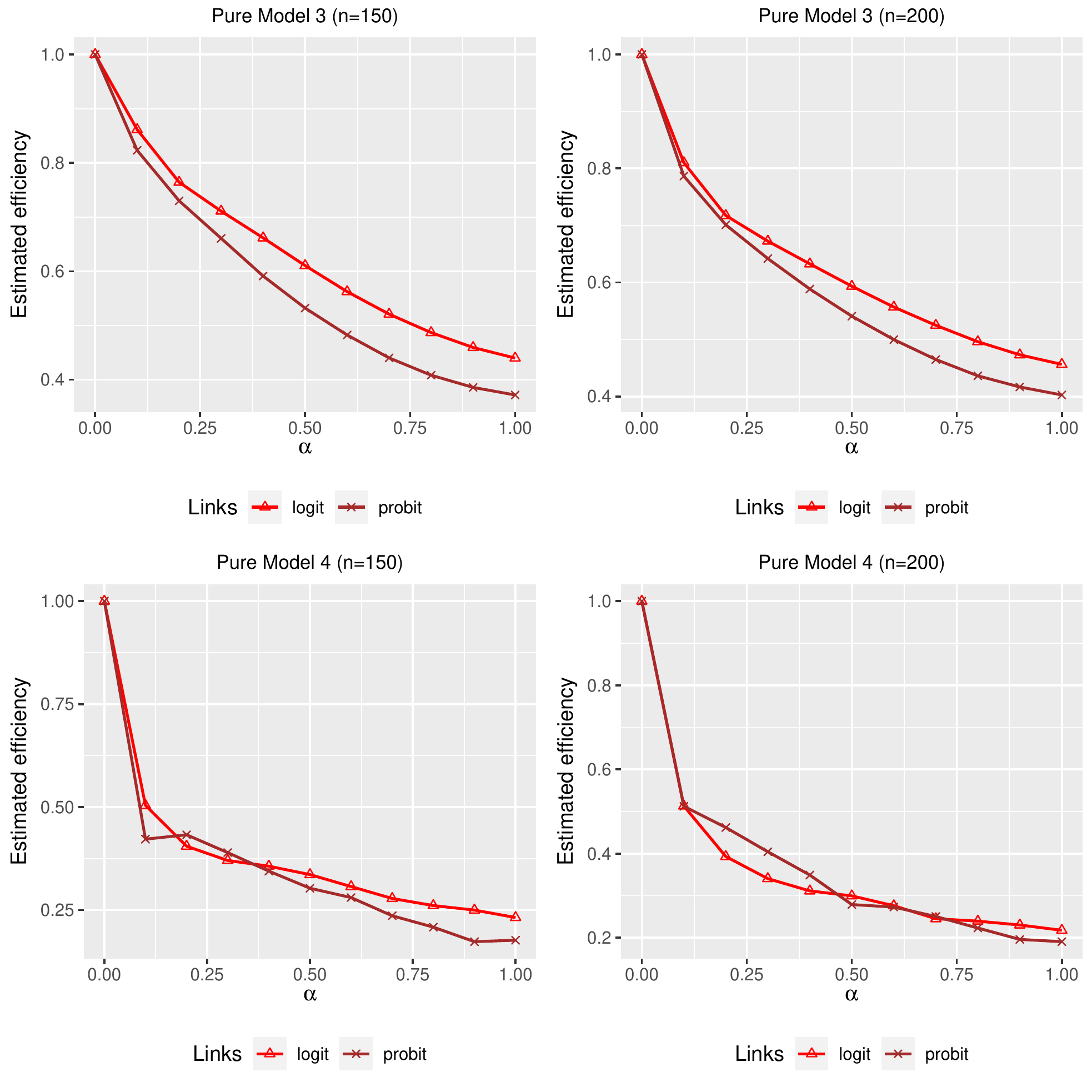} 
\caption{Graphs of efficiency at pure \descref{Model 3} and \descref{Model 4} for different sample sizes.} 
\label{fig:efficiency, pure, model3 and model4}
\end{center}
\end{figure} 

\begin{figure}
\begin{multicols}{2}
    \includegraphics[scale=0.4]{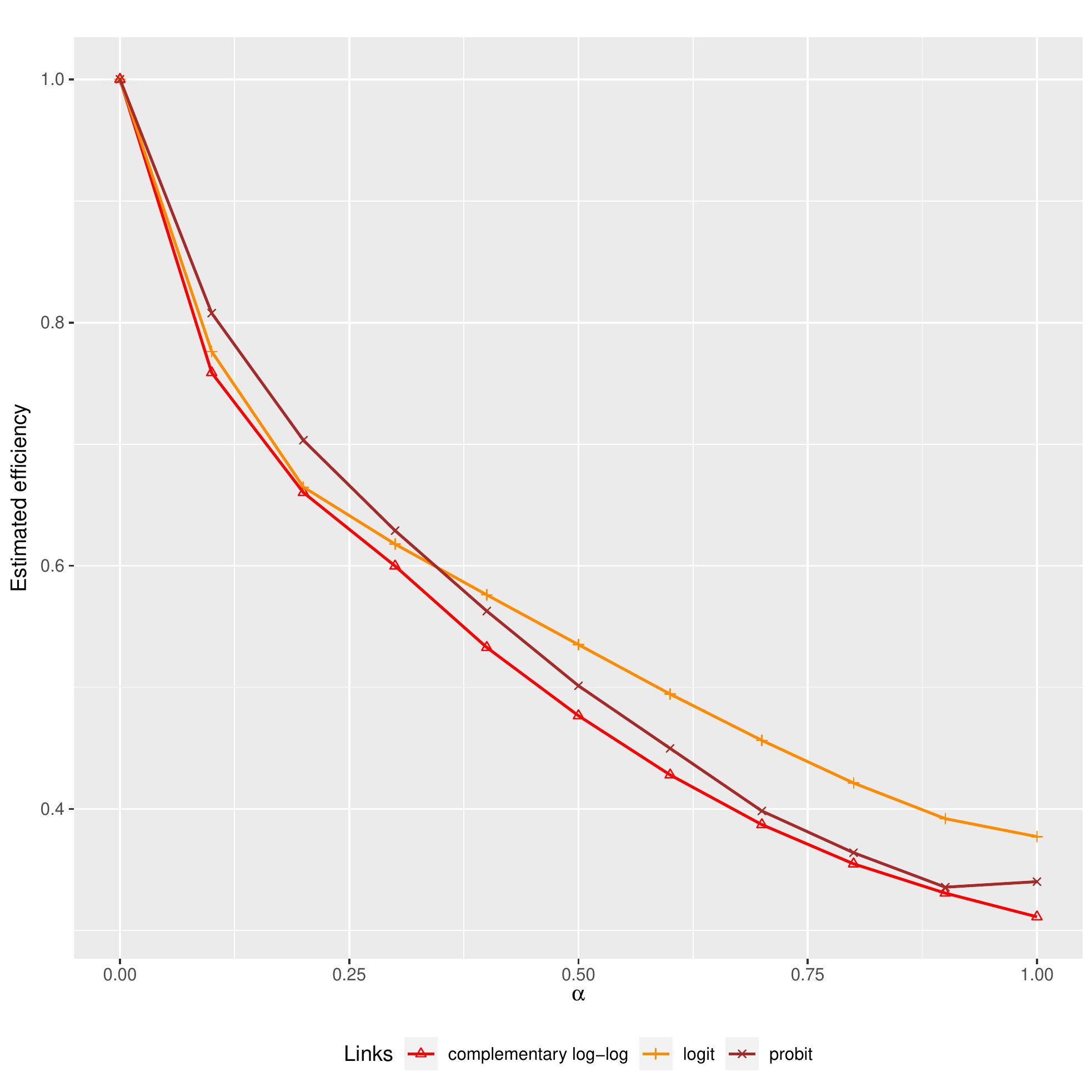} 
\caption{Graphs of efficiency at pure \descref{Model 5} with sample size $n=150$. }
\label{fig:efficiency, pure, model5, n=150}
    \par 
    \includegraphics[scale=0.4]{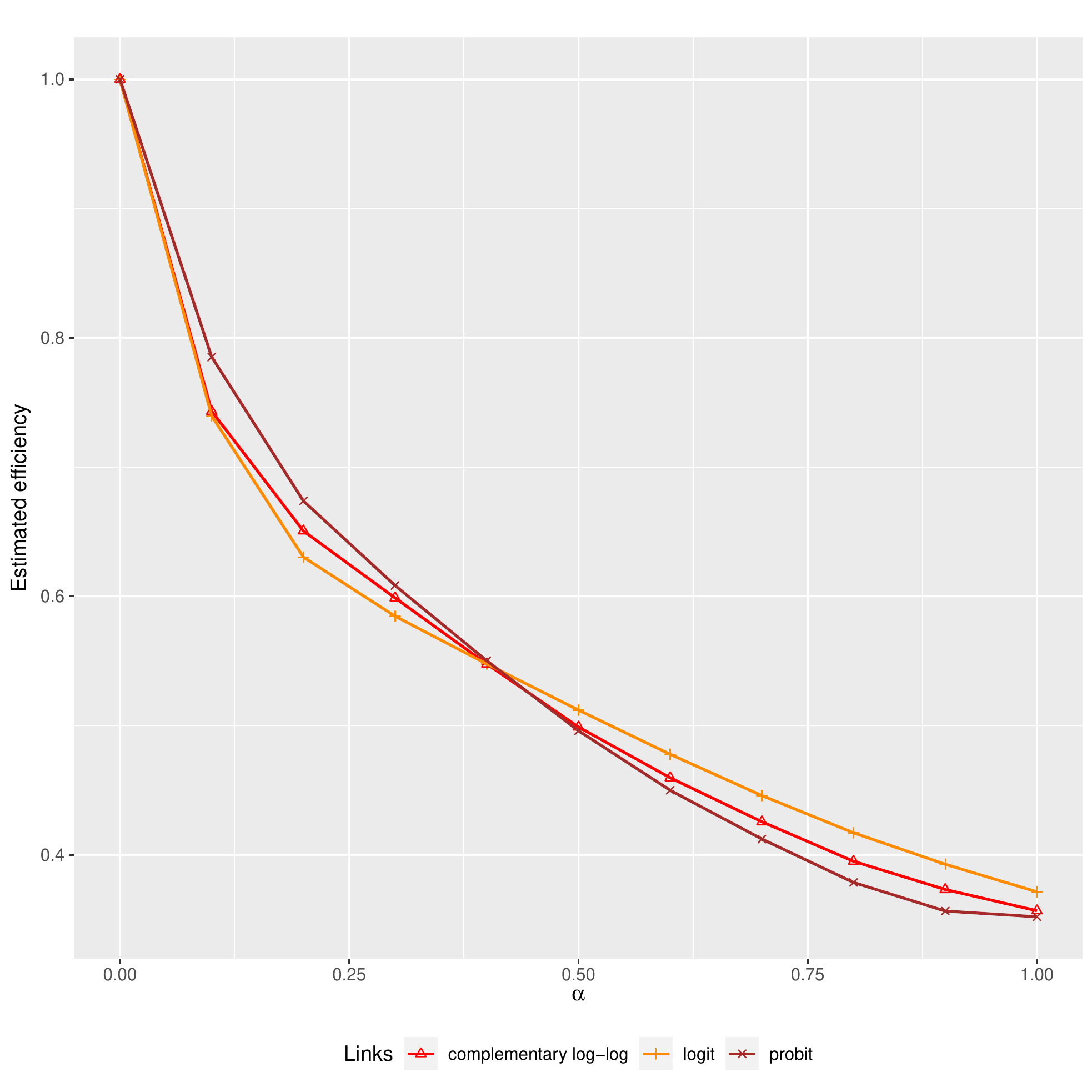} 
\caption{Graphs of efficiency at pure \descref{Model 5} with sample size $n=200$. }
\label{fig:efficiency, pure, model5, n=200}
    \par 
    \end{multicols}
\end{figure} 

\begin{table}
\centering
 \caption{Squared bias and MSE of the estimates at \descref{Model 1} with the probit link}
  \begin{adjustbox}{width=0.7\linewidth}
\begin{tabular}{c c c c c c c }
   \hline
Sample size & Link  & Method & $||\hat{\gamma}-\gamma||^{2}$ & $||\hat{\beta}-\beta||^{2}$ & $MSE(\hat{\gamma})$ & $MSE(\hat{\beta})$ 
\\ 
 \hline
$150$ \textcolor{blue}{$(200)$} & probit & MLE  & $0.00059$ &  $0.00007$ &   $0.01810$ &   $0.02113$ \\ \cline{4-7} 
& & & \textcolor{blue}{$(0.00027)$} & \textcolor{blue}{$(0.00008)$} &  
      \textcolor{blue}{$(0.01232)$} & \textcolor{blue}{$(0.01464)$} 
\\ \cline{3-7}
\rowcolor{yellow}
& & DPD $(\alpha)$ &  &  &  & \\ 
& & $0.1$ &  $0.00050$ & $0.00020$ &  $0.02044$ & $0.02335$
\\ \cline{4-7}
& & & \textcolor{blue}{$(0.00034)$} & \textcolor{blue}{$(0.00009)$} & \textcolor{blue}{$(0.01437)$} & \textcolor{blue}{$(0.01683)$} 
\\ \cline{3-7}          
& &  $0.2$ &  $0.00060$ &  $0.00025$ &  $0.02182$ & $0.02456$
\\ \cline{4-7}
& & & \textcolor{blue}{$(0.00044)$} & \textcolor{blue}{$(0.00012)$} & 
      \textcolor{blue}{$(0.01555)$} & \textcolor{blue}{$(0.01775)$} 
\\ \cline{3-7}     
& & $0.3$ &  $0.00069$ &  $0.00030$ &  $0.02316$ & $0.02595$
\\ \cline{4-7}
& & & \textcolor{blue}{$(0.00050)$} & \textcolor{blue}{$(0.00015)$} & 
      \textcolor{blue}{$(0.01650)$} & \textcolor{blue}{$(0.01861)$} 
\\ \cline{3-7}         
         
& & $0.5$ &  $0.00094$ &  $0.00042$ &  $0.02572$ & $0.02894$ 
\\ \cline{4-7}
& & & \textcolor{blue}{$(0.00067)$} & \textcolor{blue}{$(0.00023)$} & 
      \textcolor{blue}{$(0.01837)$} & \textcolor{blue}{$(0.02064)$} 
\\ \cline{3-7}             
       
& & $0.8$ &  $0.00137$ &  $0.00064$ &  $0.02920$ & $0.03369$ 
\\ \cline{4-7}
& & & \textcolor{blue}{$(0.00098)$} & \textcolor{blue}{$(0.00040)$} & 
      \textcolor{blue}{$(0.02091)$} & \textcolor{blue}{$(0.02389)$} 
\\ \cline{3-7}           
      
& & $1.0$ &  $0.00161$ &  $0.00078$ &  $0.03057$ & $0.03642$ 
\\ \cline{4-7}
& & & \textcolor{blue}{$(0.00116)$} & \textcolor{blue}{$(0.00053)$} & 
      \textcolor{blue}{$(0.02210)$} & \textcolor{blue}{$(0.02594)$} 
\\ \cline{3-7} 
\rowcolor{yellow}
& & Iannario $(c)$ &  &  &  & \\ 
& & $1.1$  &  $0.91683$ &  $0.39137$ &  $1.11269$ & $0.50400$
\\ \cline{4-7}
& & & \textcolor{blue}{$(0.88153)$} & \textcolor{blue}{$(0.38396)$} & 
      \textcolor{blue}{$(0.94061)$} & \textcolor{blue}{$(0.44379)$} 
\\ \cline{3-7}   
& & $1.4$  &  $0.50530$ &  $0.31753$ &  $0.53996$ & $0.35084$
\\ \cline{4-7}
& & & \textcolor{blue}{$(0.49168)$} & \textcolor{blue}{$(0.32074)$} & 
      \textcolor{blue}{$(0.51512)$} & \textcolor{blue}{$(0.34237)$} 
\\ \cline{3-7}  
& & $1.5$  &  $0.35871$ &  $0.16981$ &  $0.41747$ & $0.26115$
\\ \cline{4-7}
& & & \textcolor{blue}{$(0.36809)$} & \textcolor{blue}{$(0.18241)$} & 
      \textcolor{blue}{$(0.43070)$} & \textcolor{blue}{$(0.30426)$} 
\\ \cline{3-7}  
& & $1.7$  &  $0.11496$ &  $0.04636$ &  $0.18342$ & $0.13085$
\\ \cline{4-7}
& & & \textcolor{blue}{$(0.10749)$} & \textcolor{blue}{$(0.04001)$} & 
      \textcolor{blue}{$(0.16445)$} & \textcolor{blue}{$(0.10902)$} 
\\ \cline{3-7}          
& & Croux  &  $0.00083$ &  $0.00030$ &  $0.02712$ & $0.02527$ 
\\ \cline{4-7}
& & & \textcolor{blue}{($0.00048$)} & \textcolor{blue}{$(0.00013)$} & 
      \textcolor{blue}{$(0.01582)$} & \textcolor{blue}{$(0.01791)$} 
\\ \hline 
   \end{tabular}
   \end{adjustbox}
   \label{Table:pure model1 probit}
\end{table}

\begin{table}
\centering
 \caption{Squared bias and MSE of the estimates at \descref{Model 2} with the Cauchy link}
  \begin{adjustbox}{width=0.7\linewidth}
\begin{tabular}{c c c c c c c }
   \hline
Sample size & Link  & Method & $||\hat{\gamma}-\gamma||^{2}$ & $||\hat{\beta}-\beta||^{2}$ & $MSE(\hat{\gamma})$ & $MSE(\hat{\beta})$ 
\\ 
 \hline
          
$150$ \textcolor{blue}{$(200)$} & Cauchy & MLE  & $0.00025$ &  $0.00044$ &   $0.04725$ &  $0.04025$ \\ \cline{4-7}           
& & & \textcolor{blue}{$(0.00010)$} & \textcolor{blue}{$(0.00006)$} &  
      \textcolor{blue}{$(0.03468)$} & \textcolor{blue}{$(0.02549)$} 
\\ \cline{3-7}         
\rowcolor{yellow}
& & DPD $(\alpha)$ &  &  &  & \\           

& & $0.1$ &  $0.00118$ & $0.00118$ &  $0.06698$ & $0.05642$
\\ \cline{4-7}
& & & \textcolor{blue}{$(0.00071)$} & \textcolor{blue}{$(0.00052)$} & \textcolor{blue}{$(0.04670)$} & \textcolor{blue}{$(0.03524)$} 
\\ \cline{3-7}                     
          
& &  $0.2$ &  $0.00236$ &  $0.00252$ &  $0.08108$ & $0.07130$
\\ \cline{4-7}
& & & \textcolor{blue}{$(0.00138)$} & \textcolor{blue}{$(0.00111)$} & 
      \textcolor{blue}{$(0.05670)$} & \textcolor{blue}{$(0.04457)$} 
\\ \cline{3-7}            
                   
& & $0.3$ &  $0.00312$ &  $0.00343$ &  $0.08731$ & $0.07842$
\\ \cline{4-7}
& & & \textcolor{blue}{$(0.00180)$} & \textcolor{blue}{$(0.00155)$} & 
      \textcolor{blue}{$(0.06129)$} & \textcolor{blue}{$(0.04929)$} 
\\ \cline{3-7}              
           
& & $0.5$ &  $0.00450$ &  $0.00517$ &  $0.09456$ & $0.08664$ 
\\ \cline{4-7}
& & & \textcolor{blue}{$(0.00257)$} & \textcolor{blue}{$(0.00235)$} & 
      \textcolor{blue}{$(0.06628)$} & \textcolor{blue}{$(0.05414)$} 
\\ \cline{3-7}                    
          
& & $0.8$ &  $0.00676$ &  $0.00793$ &  $0.10515$ & $0.09820$ 
\\ \cline{4-7}
& & & \textcolor{blue}{$(0.00374)$} & \textcolor{blue}{$(0.00356)$} & 
      \textcolor{blue}{$(0.07282)$} & \textcolor{blue}{$(0.05975)$} 
\\ \cline{3-7}                                                 
          
& & $1.0$ &  $0.00824$ &  $0.00962$ &  $0.11165$ & $0.10468$ 
\\ \cline{4-7}
& & & \textcolor{blue}{$(0.00451)$} & \textcolor{blue}{$(0.00431)$} & 
      \textcolor{blue}{$(0.07680)$} & \textcolor{blue}{$(0.06272)$} 
\\ \cline{3-7} 
\rowcolor{yellow}
& & Iannario $(c)$ &  &  &  & \\           
                    
& & $0.6$  &  $0.00316$ &  $0.00350$ &  $0.09661$ & $0.09002$
\\ \cline{4-7}
& & & \textcolor{blue}{$(0.00187)$} & \textcolor{blue}{$(0.00175)$} & 
      \textcolor{blue}{$(0.06960)$} & \textcolor{blue}{$(0.06131)$} 
\\ \cline{3-7}                                   
           
& & $0.8$  &  $0.00311$ &  $0.00339$ &  $0.09288$ & $0.08628$
\\ \cline{4-7}
& & & \textcolor{blue}{$(0.00183)$} & \textcolor{blue}{$(0.00165)$} & 
      \textcolor{blue}{$(0.06642)$} & \textcolor{blue}{$(0.05806)$} 
\\ \cline{3-7}                      
          
& & $0.9$  &  $0.00308$ &  $0.00331$ &  $0.09188$ & $0.08490$
\\ \cline{4-7}
& & & \textcolor{blue}{$(0.00183)$} & \textcolor{blue}{$(0.00163)$} & 
      \textcolor{blue}{$(0.06548)$} & \textcolor{blue}{$(0.05688)$} 
\\ \cline{3-7}                   
          
& & $1.0$  &  $0.00305$ &  $0.00325$ &  $0.09114$ & $0.08385$
\\ \cline{4-7}
& & & \textcolor{blue}{$(0.00184)$} & \textcolor{blue}{$(0.00161)$} & 
      \textcolor{blue}{$(0.06495)$} & \textcolor{blue}{$(0.05604)$} 
\\ \cline{3-7}           
                   
& & Croux  &  $0.00289$ &  $ 0.00306$ &  $0.09302$ & $0.08199$ 
\\ \cline{4-7}
& & & \textcolor{blue}{($0.00185$)} & \textcolor{blue}{$(0.00161)$} & 
      \textcolor{blue}{$(0.06625)$} & \textcolor{blue}{$(0.05269)$} 
\\ \hline 
   \end{tabular}
   \end{adjustbox}
   \label{Table:pure model2 cauchy}
\end{table}

\begin{table}[ht]
\centering
 \caption{Squared bias and MSE of the estimates at \descref{Model 3} with the logit link}
  \begin{adjustbox}{width=0.7\linewidth}
\begin{tabular}{c c c c c c c }
   \hline
Sample size & Link  & Method & $||\hat{\gamma}-\gamma||^{2}$ & $||\hat{\beta}-\beta||^{2}$ & $MSE(\hat{\gamma})$ & $MSE(\hat{\beta})$ 
\\ 
 \hline
                   
$150$ \textcolor{blue}{$(200)$} & logit & MLE  & $0.00169$ &  $0.00041$ &   $0.07006$ &  $0.03587$ \\ \cline{4-7}           
& & & \textcolor{blue}{$(0.00039)$} & \textcolor{blue}{$(0.00020)$} &  
      \textcolor{blue}{$(0.04925)$} & \textcolor{blue}{$(0.02389)$} 
\\ \cline{3-7}         
\rowcolor{yellow}
& & DPD $(\alpha)$ &  &  &  & \\                     
                     
& & $0.1$ &  $0.00247$ & $0.00085$ &  $0.08368$ & $0.03928$
\\ \cline{4-7}
& & & \textcolor{blue}{$(0.00124)$} & \textcolor{blue}{$(0.00068)$} & \textcolor{blue}{$(0.06127)$} & \textcolor{blue}{$(0.02901)$} 
\\ \cline{3-7}                                         
          
& &  $0.2$ &  $0.00380$ &  $0.00132$ &  $0.09504$ & $0.04355$
\\ \cline{4-7}
& & & \textcolor{blue}{$(0.00195)$} & \textcolor{blue}{$(0.00096)$} & 
      \textcolor{blue}{$(0.06969)$} & \textcolor{blue}{$(0.03227)$} 
\\ \cline{3-7}            
                      
& & $0.3$ &  $0.00496$ &  $0.00176$ &  $0.10218$ & $0.04681$
\\ \cline{4-7}
& & & \textcolor{blue}{$(0.00247)$} & \textcolor{blue}{$(0.00118)$} & 
      \textcolor{blue}{$(0.07434)$} & \textcolor{blue}{$(0.03443)$} 
\\ \cline{3-7}  

& & $0.5$ &  $0.00774$ &  $0.00281$ &  $0.11874$ & $0.05477$ 
\\ \cline{4-7}
& & & \textcolor{blue}{$(0.00359)$} & \textcolor{blue}{$(0.00165)$} & 
      \textcolor{blue}{$(0.08417)$} & \textcolor{blue}{$(0.03909)$} 
\\ \cline{3-7}   

& & $0.8$ &  $0.01324$ &  $0.00493$ &  $0.14840$ & $0.06914$ 
\\ \cline{4-7}
& & & \textcolor{blue}{$(0.00571)$} & \textcolor{blue}{$(0.00252)$} & 
      \textcolor{blue}{$(0.10053)$} & \textcolor{blue}{$(0.04693)$} 
\\ \cline{3-7}                                      
                      
& & $1.0$ &  $0.01644$ &  $0.00615$ &  $0.16388$ & $0.07690$ 
\\ \cline{4-7}
& & & \textcolor{blue}{$(0.00692)$} & \textcolor{blue}{$(0.00300)$} & 
      \textcolor{blue}{$(0.10902)$} & \textcolor{blue}{$(0.05129)$} 
\\ \cline{3-7} 
\rowcolor{yellow}
& & Iannario $(c)$ &  &  &  & \\           
                    
& & $0.6$  &  $0.00003$ &  $0.00001$ &  $0.02027$ & $0.01395$
\\ \cline{4-7}
& & & \textcolor{blue}{$(0.00002)$} & \textcolor{blue}{$(0.00002)$} & 
      \textcolor{blue}{$(0.01596)$} & \textcolor{blue}{$(0.01156)$} 
\\ \cline{3-7}                                             
                    
& & $0.8$  &  $0.00001$ &  $0.00000$ &  $0.01770$ & $0.01381$
\\ \cline{4-7}
& & & \textcolor{blue}{$(0.00002)$} & \textcolor{blue}{$(0.00003)$} & 
      \textcolor{blue}{$(0.01294)$} & \textcolor{blue}{$(0.00932)$} 
\\ \cline{3-7}                      
                      
& & $0.9$  &  $0.00002$ &  $0.00004$ &  $0.01857$ & $0.01221$
\\ \cline{4-7}
& & & \textcolor{blue}{$(0.00000)$} & \textcolor{blue}{$(0.00002)$} & 
      \textcolor{blue}{$(0.01248)$} & \textcolor{blue}{$(0.00925)$} 
\\ \cline{3-7}                   
                     
& & $1.0$  &  $0.00003$ &  $0.00001$ &  $0.01763$ & $0.01311$
\\ \cline{4-7}
& & & \textcolor{blue}{$(0.00001)$} & \textcolor{blue}{$(0.00000)$} & 
      \textcolor{blue}{$(0.01210)$} & \textcolor{blue}{$(0.00864)$} 
\\ \cline{3-7}                               
           
& & Croux  &  $0.00406$ &  $0.00138$ &  $0.10083$ & $0.04337$ 
\\ \cline{4-7}
& & & \textcolor{blue}{($0.00233$)} & \textcolor{blue}{$(0.00113)$} & 
      \textcolor{blue}{$(0.07433)$} & \textcolor{blue}{$(0.03232)$} 
\\ \hline 
   \end{tabular}
   \end{adjustbox}
   \label{Table:pure model3 logit}
\end{table}


\begin{table}[ht]
\centering
 \caption{Squared bias and MSE of the estimates at \descref{Model 4} with the probit link}
  \begin{adjustbox}{width=0.7\linewidth}
\begin{tabular}{c c c c c c c }
   \hline
Sample size & Link  & Method & $||\hat{\gamma}-\gamma||^{2}$ & $||\hat{\beta}-\beta||^{2}$ & $MSE(\hat{\gamma})$ & $MSE(\hat{\beta})$ 
\\ 
 \hline         
           
$150$ \textcolor{blue}{$(200)$} & probit & MLE  & $0.01279$ &  $0.00280$ &   $0.19448$ &  $0.06218$ \\ \cline{4-7}           
& & & \textcolor{blue}{$(0.00373)$} & \textcolor{blue}{$(0.00105)$} &  
      \textcolor{blue}{$(0.13881)$} & \textcolor{blue}{$(0.04185)$} 
\\ \cline{3-7}         
\rowcolor{yellow}
& & DPD $(\alpha)$ &  &  &  & \\                     
                    
& & $0.1$ &  $0.08153$ & $0.01777$ &  $0.47914$ & $0.12899$
\\ \cline{4-7}
& & & \textcolor{blue}{$(0.03230)$} & \textcolor{blue}{$(0.00709)$} & \textcolor{blue}{$(0.27740)$} & \textcolor{blue}{$(0.07516)$} 
\\ \cline{3-7}                               
                    
& &  $0.2$ &  $0.09886$ &  $0.02146$ &  $0.46427$ & $0.12885$
\\ \cline{4-7}
& & & \textcolor{blue}{$(0.04529)$} & \textcolor{blue}{$(0.00973)$} & 
      \textcolor{blue}{$(0.30960)$} & \textcolor{blue}{$(0.08144)$} 
\\ \cline{3-7}                      
          
& & $0.3$ &  $0.11786$ &  $0.02552$ &  $0.51792$ & $0.14103$
\\ \cline{4-7}
& & & \textcolor{blue}{$(0.05616)$} & \textcolor{blue}{$(0.01215)$} & 
      \textcolor{blue}{$(0.35480)$} & \textcolor{blue}{$(0.09195)$} 
\\ \cline{3-7}  

& & $0.5$ &  $0.17194$ &  $0.03662$ &  $0.67212$ & $0.17492$ 
\\ \cline{4-7}
& & & \textcolor{blue}{$(0.09193)$} & \textcolor{blue}{$(0.01939)$} & 
      \textcolor{blue}{$(0.52375)$} & \textcolor{blue}{$(0.12404)$} 
\\ \cline{3-7}                                     
                    
& & $0.8$ &  $0.29509$ &  $0.06185$ &  $0.98384$ & $0.24715$ 
\\ \cline{4-7}
& & & \textcolor{blue}{$(0.15115)$} & \textcolor{blue}{$(0.03113)$} & 
      \textcolor{blue}{$(0.65670)$} & \textcolor{blue}{$(0.15294)$} 
\\ \cline{3-7}  

& & $1.0$ &  $0.38689$ &  $0.08023$ &  $1.16475$ & $0.28748$ 
\\ \cline{4-7}
& & & \textcolor{blue}{$(0.19555)$} & \textcolor{blue}{$(0.04004)$} & 
      \textcolor{blue}{$(0.76913)$} & \textcolor{blue}{$(0.17819)$} 
\\ \cline{3-7} 
\rowcolor{yellow}
& & Iannario $(c)$ &  &  &  & \\           
                      
& & $1.1$  &  $0.00216$ &  $0.00190$ &  $0.06028$ & $0.03010$
\\ \cline{4-7}
& & & \textcolor{blue}{$(0.00253)$} & \textcolor{blue}{$(0.00167)$} & 
      \textcolor{blue}{$(0.05822)$} & \textcolor{blue}{$(0.02531)$} 
\\ \cline{3-7}                                             
                   
& & $1.4$  &  $0.00178$ &  $0.00116$ &  $0.05012$ & $0.02313$
\\ \cline{4-7}
& & & \textcolor{blue}{$(0.00112)$} & \textcolor{blue}{$(0.00101)$} & 
      \textcolor{blue}{$(0.03588)$} & \textcolor{blue}{$(0.01929)$} 
\\ \cline{3-7}                      
                    
& & $1.5$  &  $0.00096$ &  $0.00104$ &  $0.04392$ & $0.02456$
\\ \cline{4-7}
& & & \textcolor{blue}{$(0.00191)$} & \textcolor{blue}{$(0.00111)$} & 
      \textcolor{blue}{$(0.03881)$} & \textcolor{blue}{$(0.02191)$} 
\\ \cline{3-7}                   
                    
& & $1.7$  &  $0.00167$ &  $0.00070$ &  $0.51388$ & $0.03080$
\\ \cline{4-7}
& & & \textcolor{blue}{$(0.00180)$} & \textcolor{blue}{$(0.00100)$} & 
      \textcolor{blue}{$(0.07450)$} & \textcolor{blue}{$(0.02285)$} 
\\ \cline{3-7}                     
                    
& & Croux  &  $0.48349$ &  $0.09860$ &  $2.24010$ & $0.50087$ 
\\ \cline{4-7}
& & & \textcolor{blue}{($0.18440$)} & \textcolor{blue}{$(0.03642)$} & 
      \textcolor{blue}{$(0.85851)$} & \textcolor{blue}{$(0.18960)$} 
\\ \hline 
   \end{tabular}
   \end{adjustbox}
   \label{Table:pure model4 probit}
\end{table}

\begin{table}
\centering
 \caption{Squared bias and MSE of the estimates at \descref{Model 5} with the complementary log-log link}
  \begin{adjustbox}{width=0.7\linewidth}
\begin{tabular}{c c c c c c c }
   \hline
Sample size & Link  & Method & $||\hat{\gamma}-\gamma||^{2}$ & $||\hat{\beta}-\beta||^{2}$ & $MSE(\hat{\gamma})$ & $MSE(\hat{\beta})$ 
\\ 
 \hline         
                      
$150$ \textcolor{blue}{$(200)$} & log-log & MLE  & $0.00220$ &     $0.00136$ &   $0.05270$ &  $0.05933$ \\ \cline{4-7}           
& & & \textcolor{blue}{$(0.00104)$} & \textcolor{blue}{$(0.00054)$} &  
      \textcolor{blue}{$(0.03814)$} & \textcolor{blue}{$(0.04114)$} 
\\ \cline{3-7}         
\rowcolor{yellow}
& & DPD $(\alpha)$ &  &  &  & \\                                
           
& & $0.1$ &  $0.00430$ &  $0.00287$ &  $0.06963$ &  $0.07802$ 
\\ \cline{4-7}
& & & \textcolor{blue}{$(0.00237)$} & \textcolor{blue}{$(0.00142)$} & \textcolor{blue}{$(0.05123)$} & \textcolor{blue}{$(0.05546)$} 
\\ \cline{3-7}                                         
          
& &  $0.2$ &  $0.00611$ &   $0.00428$ &   $0.08108$ &  $0.08861$
\\ \cline{4-7}
& & & \textcolor{blue}{$(0.00333)$} & \textcolor{blue}{$(0.00217)$} & 
      \textcolor{blue}{$(0.05868)$} & \textcolor{blue}{$(0.06319)$} 
\\ \cline{3-7}                      
                   
& & $0.3$ &  $0.00749$ &   $0.00539$ &   $0.08978$ &  $0.09701$
\\ \cline{4-7}
& & & \textcolor{blue}{$(0.00397)$} & \textcolor{blue}{$(0.00269)$} & 
      \textcolor{blue}{$(0.06382)$} & \textcolor{blue}{$(0.06859)$} 
\\ \cline{3-7}                     
          
& & $0.5$ &  $0.01250$ &  $0.00930$ &  $0.11349$ &  $0.12154$ 
\\ \cline{4-7}
& & & \textcolor{blue}{$(0.00599)$} & \textcolor{blue}{$(0.00422)$} & 
      \textcolor{blue}{$(0.07664)$} & \textcolor{blue}{$(0.08228)$} 
\\ \cline{3-7}                                              
          
& & $0.8$ &  $0.02330$ &  $0.01753$ &  $0.15353$ &   $0.16217$ 
\\ \cline{4-7}
& & & \textcolor{blue}{$(0.01023)$} & \textcolor{blue}{$(0.00741)$} & 
      \textcolor{blue}{$(0.09707)$} & \textcolor{blue}{$(0.10366)$} 
\\ \cline{3-7}    

& & $1.0$ &  $0.03010$ &  $0.02260$ &  $0.17581$ &  $0.18408$ 
\\ \cline{4-7}
& & & \textcolor{blue}{$(0.01275)$} & \textcolor{blue}{$(0.00929)$} & 
      \textcolor{blue}{$(0.10739)$} & \textcolor{blue}{$(0.11486)$} 
\\ \cline{3-7}  
                   
\rowcolor{yellow}
& & Iannario $(c)$ &  &  &  & \\           
                   
& & $1.1$  &  $0.21656$ & $0.18125$ &   $0.44069$ &   $0.42720$
\\ \cline{4-7}
& & & \textcolor{blue}{$(0.18602)$} & \textcolor{blue}{$(0.15414)$} & 
      \textcolor{blue}{$(0.36320)$} & \textcolor{blue}{$(0.33341)$} 
\\ \cline{3-7}                                             
                     
& & $1.4$  &  $0.12067$ &   $0.10590$ &   $0.26816$ &  $0.27908$
\\ \cline{4-7}
& & & \textcolor{blue}{$(0.09209)$} & \textcolor{blue}{$(0.08140)$} & 
      \textcolor{blue}{$(0.18498)$} & \textcolor{blue}{$(0.19263)$} 
\\ \cline{3-7}                      
                      
& & $1.5$  &  $0.10448$ &   $0.09187$ &  $0.23472$ &   $0.25058$
\\ \cline{4-7}
& & & \textcolor{blue}{$(0.08140)$} & \textcolor{blue}{$(0.07163)$} & 
      \textcolor{blue}{$(0.16667)$} & \textcolor{blue}{$(0.17489)$} 
\\ \cline{3-7}                       
          
& & $1.7$  &  $0.08478$ &  $0.07408$ &  $0.20196$ &  $0.21348$
\\ \cline{4-7}
& & & \textcolor{blue}{$(0.06634)$} & \textcolor{blue}{$(0.05797)$} & 
      \textcolor{blue}{$(0.14491)$} & \textcolor{blue}{$(0.15192)$} 
\\ \cline{3-7}                     
                   
& & Croux  &  $0.00776$ &   $0.00472$ &  $0.09849$ &  $0.09576$ 
\\ \cline{4-7}
& & & \textcolor{blue}{($0.00525$)} & \textcolor{blue}{$(0.00334)$} & 
      \textcolor{blue}{$(0.07474)$} & \textcolor{blue}{$(0.07096)$} 
\\ \hline 
   \end{tabular}
   \end{adjustbox}
   \label{Table:pure model5 loglog}
\end{table}

\clearpage 
\newpage 

\subsection{Simulation Studies: Contaminated Models}
\label{Simulation Studies: Contaminated Models}

Now the above models are contaminated in the following way. 
\begin{description}

    \descitem{Vertical outliers}: In this paper, a data point is said to be a vertical outlier when $Y$ takes the highest categorical value in a manner such that it is inconsistent with the covariates. We add $5\%$ and $10\%$ vertical outliers to the data sets generated by \descref{Model 1} to \descref{Model 5} across all the aforesaid link functions. Numerical values are reported in Tables \ref{Table:model1 probit 0.05 vertical} - \ref{Table:model1 probit-loglog}.

    \descitem{Horizontal outliers}: We have developed the theory where the covariates are assumed to be non-stochastic. However, we might still come across a situation where a small proportion of $x$-values with high magnitudes may destabilise the MLE. We refer to these $x$-values as horizontal outliers because they do not naturally correspond to the ordinal responses. Let the covariate in the data sets, simulated through \descref{Model 2} with the probit link, be contaminated with $5\%-10\%$ horizontal outliers, where the outlying values are chosen to be $5$.  

    \descitem{Misspecification of links}: In this case, data sets are generated through \descref{Model 1} using the probit link. However, the complementary log-log is used in estimation.     
\end{description}
For convenience, we only present the graphs when the sample size is fixed at $n=200$. This is followed throughout the simulation studies only for the contaminated models. However, results corresponding to both these sample sizes are reported in the respective tables. 

Now we present the following general remarks. 
\begin{itemize}
    \item In contaminated models MDPDEs with $\alpha>0$ clearly perform better than the MLE. As $\alpha$ increases its efficiency increases up to a point, then sometimes it decreases slowly. But all the way, it performs much better than the MLE. In Figures \ref{fig:efficiency, contam, model1 and model2} - \ref{fig:efficiency, contam_0.1, model5} we notice that the complementary log-log link performs better than the probit link in \descref{Model 1}. Also in \descref{Model 2}, we find that the probit link yields better efficiency than the logit which in turn holds an edge over the Cauchy link. But, at lower values of the tuning parameter, the Cauchy link dominates the logit link. In both \descref{Model 3} and \descref{Model 4} the probit link performs better than the logit link. In \descref{Model 5} the complementary log-log link performs better than the probit link which performs better than the logit link. 

    \item Let the data sets generated through \descref{Model 2} be contaminated by horizontal outliers as described before. The graphs of the efficiency are plotted for both the $5\%$ and $10\%$ level of contamination in Figure \ref{fig:efficiency, horizontal contam_0.05, model2} and Figure \ref{fig:efficiency, horizontal contam_0.1, model2}.
    Also, we report the squared bias and squared MSE in Table \ref{Table:model2 probit-0.05 horizontal} only for the $5\%$ data contamination. We find that the estimated efficiency improves with the increment of $\alpha$; after some point, it sharply increases and reaches its maximum at some point between $\alpha=0.25$ and $\alpha=0.4$. After that, the estimated efficiency slowly drops, but still stays higher than the MLE. Also, the MDPDE performs better than the M-estimates of Iannario et al. \cite{iannario2017robust}.     

    \item When comparing the MSE, we see that MDPDE performs better than both the Croux et al. \cite{croux2013robust} and the Iannario et al. \cite{iannario2017robust} methods in \descref{Model 1} with the probit link, \descref{Model 2} with the Cauchy and \descref{Model 5} with the complementary log-log link, and also in the case when the probit link is misspecified in \descref{Model 1}. Although the Iannario et al. \cite{iannario2017robust} beats the MDPDE in \descref{Model 3} with the logit link, our method performs better than the Croux et al. \cite{croux2013robust} method.

\end{itemize}

\begin{figure}
\begin{center}
 \includegraphics[scale=0.7]{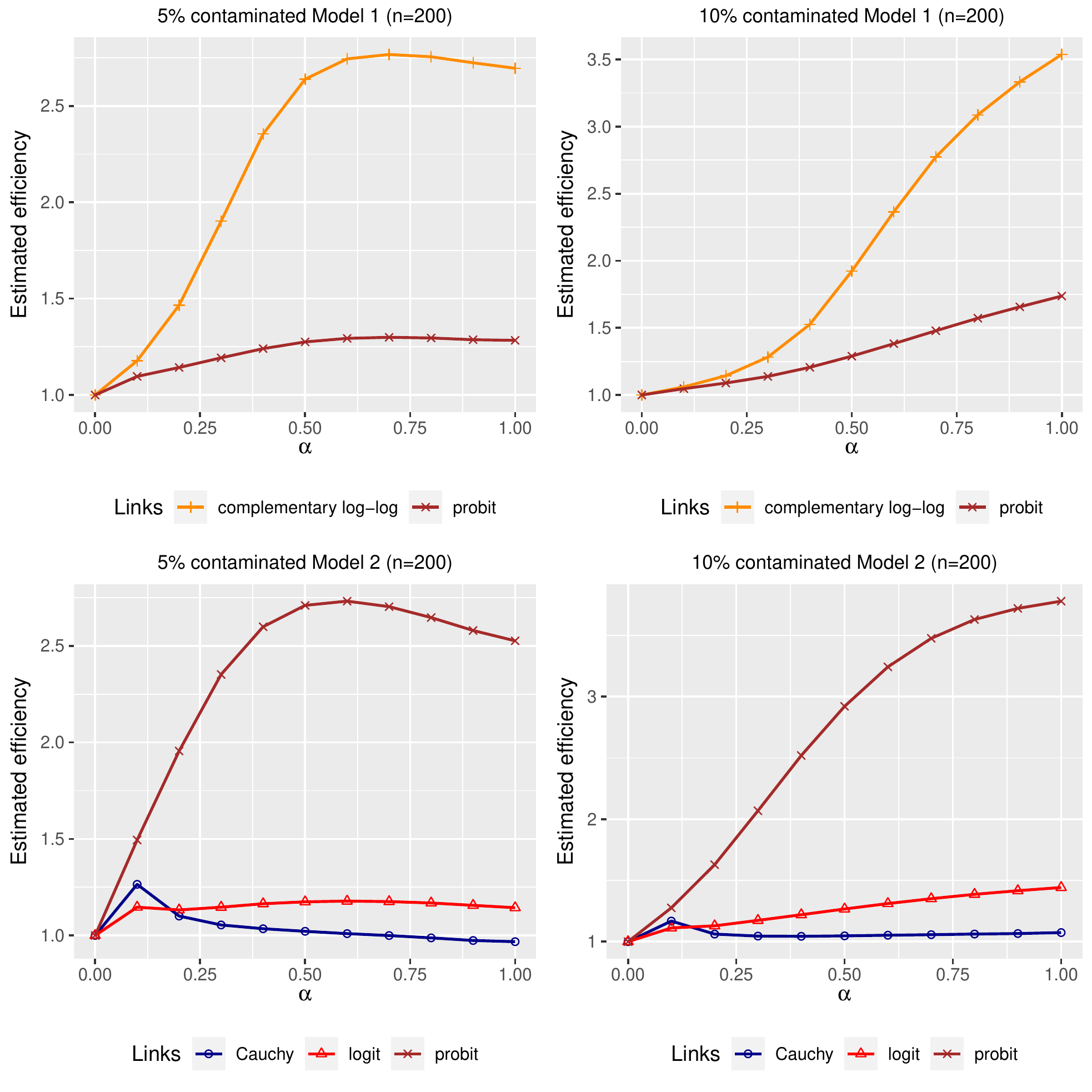} 
\caption{Graphs of efficiency when data generated by \descref{Model 1} and \descref{Model 2} are vertically contaminated at $5\%$ and $10\%$ levels of contamination.} 
\label{fig:efficiency, contam, model1 and model2}
\end{center}
\end{figure} 

\begin{figure}
\begin{center}
 \includegraphics[scale=0.7]{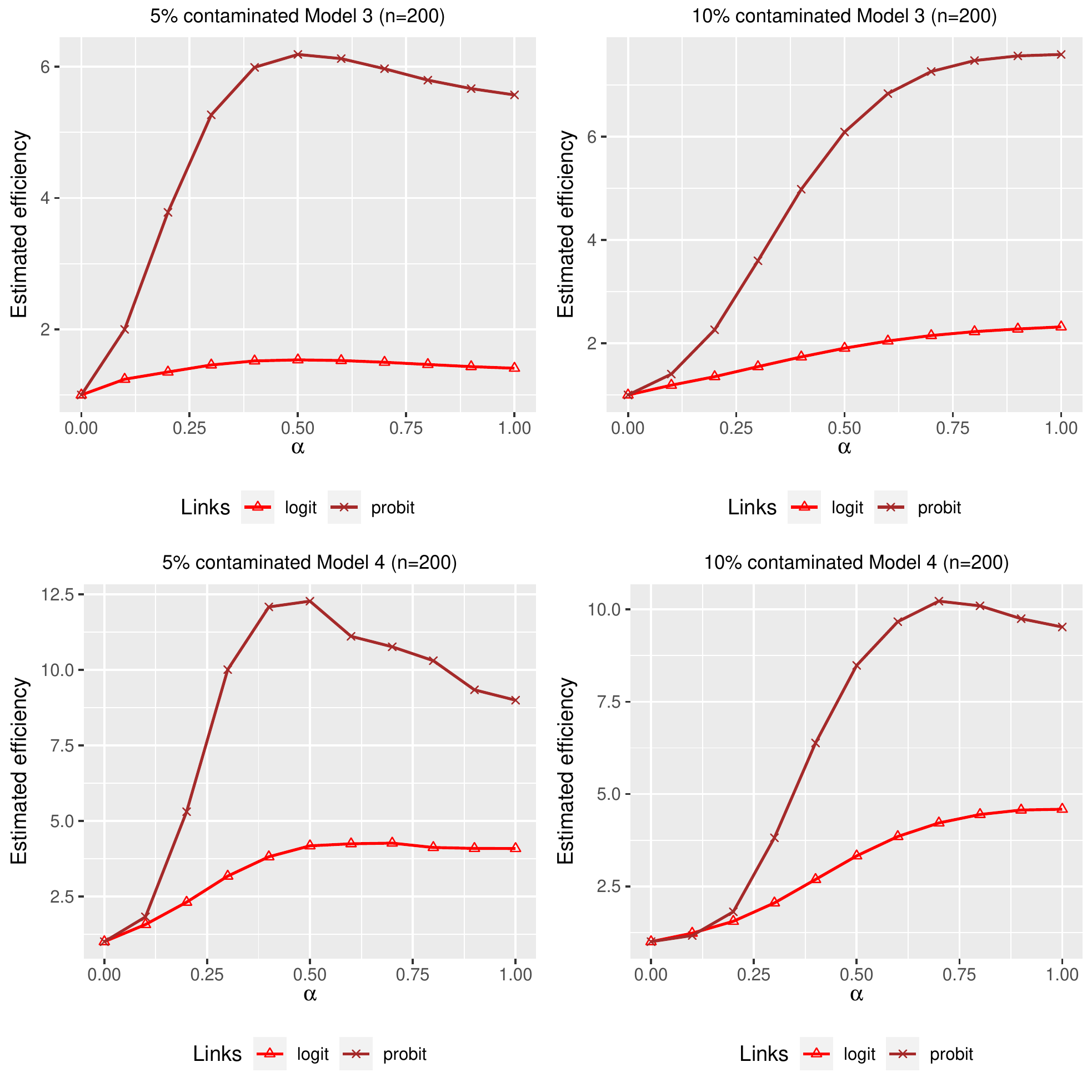} 
\caption{Graphs of efficiency when data generated by \descref{Model 3} and \descref{Model 4} are vertically contaminated at $5\%$ and $10\%$ levels of contamination.}  
\label{fig:efficiency, contam, model3 and model4}
\end{center}
\end{figure} 

\begin{figure}
\begin{multicols}{2}
    \includegraphics[scale=0.4]{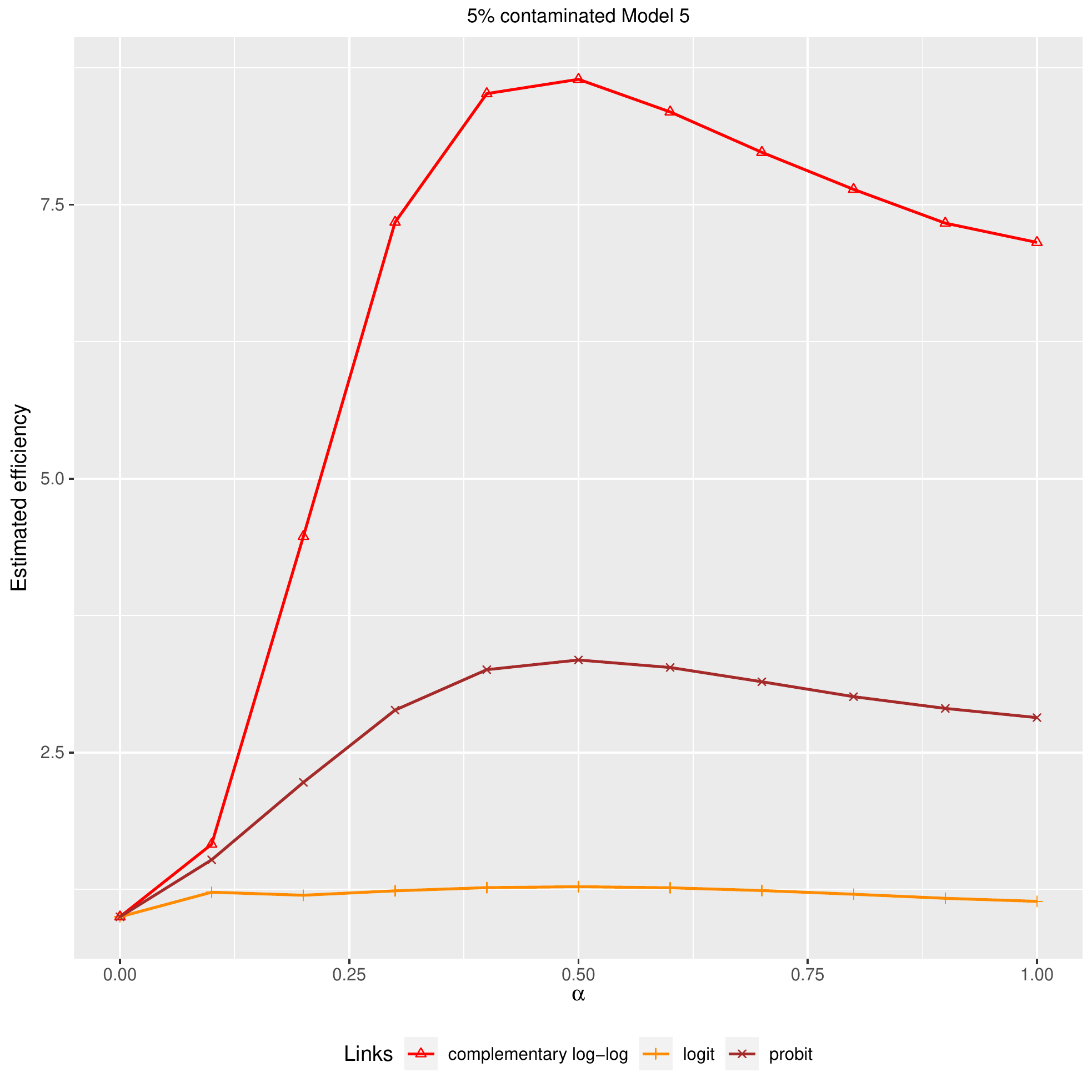} 
\caption{Graphs of efficiency when data generated by \descref{Model 5} is vertically contaminated at $5\%$ level of contamination.}  
\label{fig:efficiency, contam_0.05, model5}
    \par 
    \includegraphics[scale=0.4]{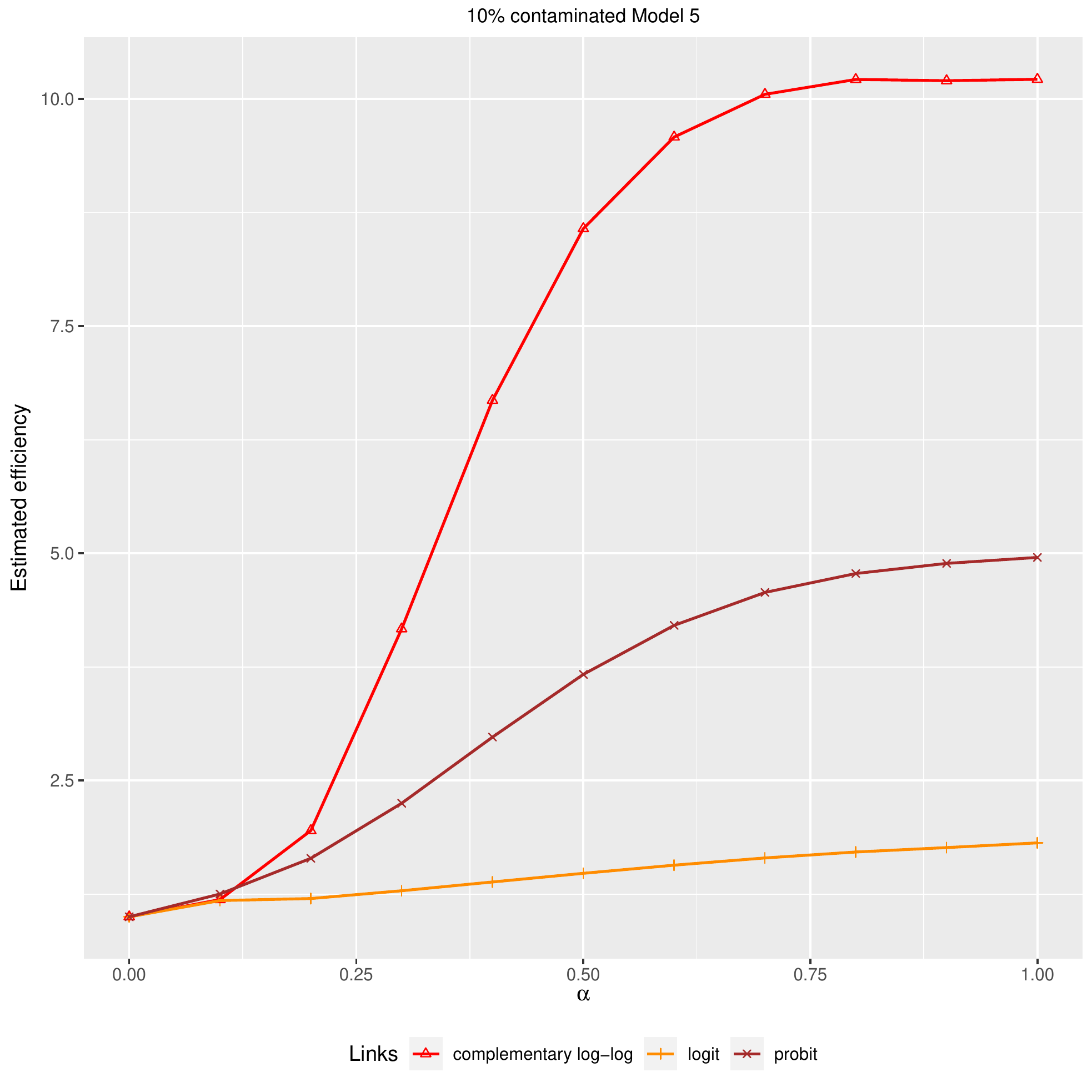} 
\caption{Graphs of efficiency when data generated by \descref{Model 5} is vertically contaminated at $10\%$ level of contamination.}
\label{fig:efficiency, contam_0.1, model5}
    \par 
    \end{multicols}
\end{figure} 

\begin{table}
\centering
 \caption{Squared bias and MSE when $5\%$ vertical outliers are added to data generated by \descref{Model 1} with the probit link}
  \begin{adjustbox}{width=0.7\linewidth}
\begin{tabular}{c c c c c c c }
   \hline
Sample size & Link  & Method & $||\hat{\gamma}-\gamma||^{2}$ & $||\hat{\beta}-\beta||^{2}$ & $MSE(\hat{\gamma})$ & $MSE(\hat{\beta})$ 
\\ 
 \hline         
                      
$150$ \textcolor{blue}{$(200)$} & probit & MLE  & $0.01836$ & $0.01357$ &    $0.03743$ & $0.03508$ \\ \cline{4-7}           
& & & \textcolor{blue}{$(0.02104)$} & \textcolor{blue}{$(0.01564)$} &  
      \textcolor{blue}{$(0.03319)$} & \textcolor{blue}{$(0.03140)$} 
\\ \cline{3-7}         
\rowcolor{yellow}
& & DPD $(\alpha)$ &  &  &  & \\                                
           
& & $0.1$ &  $0.01550$ & $0.01123$ &   $0.03274$ &  $0.03271$
\\ \cline{4-7}
& & & \textcolor{blue}{$(0.01749)$} & \textcolor{blue}{$(0.01306)$} & \textcolor{blue}{$(0.02982)$} & \textcolor{blue}{$(0.02909)$} 
\\ \cline{3-7}                                         
          
& &  $0.2$ &  $0.01353$ &  $0.00971$ &  $0.03193$ &   $0.03210$
\\ \cline{4-7}
& & & \textcolor{blue}{$(0.01550)$} & \textcolor{blue}{$(0.01150)$} & 
      \textcolor{blue}{$(0.02854)$} & \textcolor{blue}{$(0.02798)$} 
\\ \cline{3-7}                      
                   
& & $0.3$ &  $0.01118$ &  $0.00792$ &   $0.03095$ &  $0.03188$
\\ \cline{4-7}
& & & \textcolor{blue}{$(0.01301)$} & \textcolor{blue}{$(0.00961)$} & 
      \textcolor{blue}{$(0.02706)$} & \textcolor{blue}{$(0.02708)$} 
\\ \cline{3-7}                      
          
& & $0.5$ &  $0.00691$ &   $0.00454$ & $0.02969$ &  $0.03237$ 
\\ \cline{4-7}
& & & \textcolor{blue}{$(0.00833)$} & \textcolor{blue}{$(0.00595)$} & 
      \textcolor{blue}{$(0.02474)$} & \textcolor{blue}{$(0.02590)$} 
\\ \cline{3-7}                                              
          
& & $0.8$ &  $0.00312$ &   $0.00159$ &  $0.02941$ &  $0.03549$ 
\\ \cline{4-7}
& & & \textcolor{blue}{$(0.00411)$} & \textcolor{blue}{$(0.00252)$} & 
      \textcolor{blue}{$(0.02334)$} & \textcolor{blue}{$(0.02650)$} 
\\ \cline{3-7} 

& & $1.0$ &  $0.00201$ &  $0.00076$ &  $0.02912$ &  $0.03770$ 
\\ \cline{4-7}
& & & \textcolor{blue}{$(0.00283)$} & \textcolor{blue}{$(0.00140)$} & 
      \textcolor{blue}{$(0.02283)$} & \textcolor{blue}{$(0.02749)$} 
\\ \cline{3-7}  
                   
\rowcolor{yellow}
& & Iannario $(c)$ &  &  &  & \\           
                   
& & $1.1$  &  $0.65004$ &   $0.33193$ &  $0.70072$ &  $0.39141$
\\ \cline{4-7}
& & & \textcolor{blue}{$(0.62722)$} & \textcolor{blue}{$(0.32305)$} & 
      \textcolor{blue}{$(0.66563)$} & \textcolor{blue}{$(0.36653)$} 
\\ \cline{3-7}                                             
                     
& & $1.4$  &  $0.35808$ &  $0.22975$ &   $0.38934$ &  $0.26487$
\\ \cline{4-7}
& & & \textcolor{blue}{$(0.34148)$} & \textcolor{blue}{$(0.22876)$} & 
      \textcolor{blue}{$(0.36280)$} & \textcolor{blue}{$(0.25362)$} 
\\ \cline{3-7}                      
                      
& & $1.5$  &  $0.23591$ & $0.11497$ &   $0.29638$ &   $0.25526$
\\ \cline{4-7}
& & & \textcolor{blue}{$(0.22664)$} & \textcolor{blue}{$(0.11298)$} & 
      \textcolor{blue}{$(0.29271)$} & \textcolor{blue}{$(0.26179)$} 
\\ \cline{3-7}                       
          
& & $1.7$  &  $0.06634$ &   $0.02597$ &  $0.17265$ &   $0.23807$
\\ \cline{4-7}
& & & \textcolor{blue}{$(0.06072)$} & \textcolor{blue}{$(0.02308)$} & 
      \textcolor{blue}{$(0.10544)$} & \textcolor{blue}{$(0.07125)$} 
\\ \cline{3-7}                     
                   
& & Croux  &  $0.01827$ &   $0.01364$ &   $0.03867$ &  $0.03527$ 
\\ \cline{4-7}
& & & \textcolor{blue}{($0.02106$)} & \textcolor{blue}{$(0.01566)$} & 
      \textcolor{blue}{$(0.03327)$} & \textcolor{blue}{$(0.03142)$} 
\\ \hline 
   \end{tabular}
   \end{adjustbox}
   \label{Table:model1 probit 0.05 vertical}
\end{table}


\begin{table}
\centering
 \caption{Squared bias and MSE when $10\%$ vertical outliers are added to data generated by \descref{Model 1} with the probit link}
  \begin{adjustbox}{width=0.7\linewidth}
\begin{tabular}{c c c c c c c }
   \hline
Sample size & Link  & Method & $||\hat{\gamma}-\gamma||^{2}$ & $||\hat{\beta}-\beta||^{2}$ & $MSE(\hat{\gamma})$ & $MSE(\hat{\beta})$ 
\\ 
 \hline         
                      
$150$ \textcolor{blue}{$(200)$} & probit & MLE  & $0.06373$ & $0.04571$ &    $0.08052$ & $0.06505$ \\ \cline{4-7}           
& & & \textcolor{blue}{$(0.06409)$} & \textcolor{blue}{$(0.04689)$} &  
      \textcolor{blue}{$(0.07534)$} & \textcolor{blue}{$(0.06159)$} 
\\ \cline{3-7}         
\rowcolor{yellow}
& & DPD $(\alpha)$ &  &  &  & \\                                
           
& & $0.1$ &  $0.05969$ &   $0.04291$ &  $0.07531$ &   $0.06284$
\\ \cline{4-7}
& & & \textcolor{blue}{$(0.05964)$} & \textcolor{blue}{$(0.04406)$} & \textcolor{blue}{$(0.07144)$} & \textcolor{blue}{$(0.05937)$} 
\\ \cline{3-7}                                         
          
& &  $0.2$ &  $0.05647$ &  $0.04049$ &   $0.07276$ &   $0.06100$
\\ \cline{4-7}
& & & \textcolor{blue}{$(0.05631)$} & \textcolor{blue}{$(0.04159)$} & 
      \textcolor{blue}{$(0.06845)$} & \textcolor{blue}{$(0.05731)$} 
\\ \cline{3-7}                      
                   
& & $0.3$ &  $0.05235$ &   $0.03734$ &  $0.06950$ &   $0.05920$
\\ \cline{4-7}
& & & \textcolor{blue}{$(0.05224)$} & \textcolor{blue}{$(0.03850)$} & 
      \textcolor{blue}{$(0.06508)$} & \textcolor{blue}{$(0.05520)$} 
\\ \cline{3-7}                     
          
& & $0.5$ &  $0.04207$ &   $0.02859$ &  $0.06174$ &  $0.05434$ 
\\ \cline{4-7}
& & & \textcolor{blue}{$(0.04206)$} & \textcolor{blue}{$(0.02989)$} & 
      \textcolor{blue}{$(0.05699)$} & \textcolor{blue}{$(0.04921)$} 
\\ \cline{3-7}                                              
          
& & $0.8$ &  $0.02725$ &    $0.01523$ &  $0.05101$ &  $0.04892$ 
\\ \cline{4-7}
& & & \textcolor{blue}{$(0.02772)$} & \textcolor{blue}{$(0.01682)$} & 
      \textcolor{blue}{$(0.04594)$} & \textcolor{blue}{$(0.04116)$} 
\\ \cline{3-7}   

& & $1.0$ &  $0.02073$ &  $0.00892$ &  $0.04599$ &   $0.04733$ 
\\ \cline{4-7}
& & & \textcolor{blue}{$(0.02155)$} & \textcolor{blue}{$(0.01064)$} & 
      \textcolor{blue}{$(0.04087)$} & \textcolor{blue}{$(0.03795)$} 
\\ \cline{3-7}  
                   
\rowcolor{yellow}
& & Iannario $(c)$ &  &  &  & \\           
                   
& & $1.1$  &  $0.50295$ &    $0.28278$ &  $0.54448$ &  $0.32853$
\\ \cline{4-7}
& & & \textcolor{blue}{$(0.50931)$} & \textcolor{blue}{$(0.27929)$} & 
      \textcolor{blue}{$(0.67603)$} & \textcolor{blue}{$(0.36968)$} 
\\ \cline{3-7}                                             
                     
& & $1.4$  &  $0.25635$ &    $0.17064$ &  $0.28689$ & $0.21103$
\\ \cline{4-7}
& & & \textcolor{blue}{$(0.25041)$} & \textcolor{blue}{$(0.16952)$} & 
      \textcolor{blue}{$(0.27327)$} & \textcolor{blue}{$(0.20127)$} 
\\ \cline{3-7}                      
                      
& & $1.5$  &  $0.15558$ &  $0.08586$ &  $0.19710$ &  $0.14595$
\\ \cline{4-7}
& & & \textcolor{blue}{$(0.15703)$} & \textcolor{blue}{$(0.08411)$} & 
      \textcolor{blue}{$(0.23498)$} & \textcolor{blue}{$(0.14650)$} 
\\ \cline{3-7}                       
          
& & $1.7$  &  $0.04451$ &   $0.02526$ &  $0.09104$ &  $0.07660$
\\ \cline{4-7}
& & & \textcolor{blue}{$(0.04461)$} & \textcolor{blue}{$(0.02489)$} & 
      \textcolor{blue}{$(0.08272)$} & \textcolor{blue}{$(0.06596)$} 
\\ \cline{3-7}                     
                   
& & Croux  &  $0.06351$ &  $0.04584$ &  $0.08197$ &  $0.06534$ 
\\ \cline{4-7}
& & & \textcolor{blue}{($0.06414$)} & \textcolor{blue}{$(0.04693)$} & 
      \textcolor{blue}{$(0.07545)$} & \textcolor{blue}{$(0.06162)$} 
\\ \hline 
   \end{tabular}
   \end{adjustbox}
   \label{Table:model1 probit 0.1 vertical}
\end{table}


\begin{table}
\centering
 \caption{Squared bias and MSE when $5\%$ vertical outliers are added to data generated by \descref{Model 2} with the Cauchy link}
  \begin{adjustbox}{width=0.7\linewidth}
\begin{tabular}{c c c c c c c }
   \hline
Sample size & Link  & Method & $||\hat{\gamma}-\gamma||^{2}$ & $||\hat{\beta}-\beta||^{2}$ & $MSE(\hat{\gamma})$ & $MSE(\hat{\beta})$ 
\\ 
 \hline         
                      
$150$ \textcolor{blue}{$(200)$} & Cauchy & MLE  & $0.01566$ &  $0.00444$ &   $0.08885$ &  $0.07220$ \\ \cline{4-7}           
& & & \textcolor{blue}{$(0.01983)$} & \textcolor{blue}{$(0.00782)$} &  
      \textcolor{blue}{$(0.07032)$} & \textcolor{blue}{$(0.05127)$} 
\\ \cline{3-7}         
\rowcolor{yellow}
& & DPD $(\alpha)$ &  &  &  & \\                                
           
& & $0.1$ &  $0.01319$ &  $0.00327$ &  $0.07317$ &  $0.05511$
\\ \cline{4-7}
& & & \textcolor{blue}{$(0.01572)$} & \textcolor{blue}{$(0.00511)$} & 
      \textcolor{blue}{$(0.05779)$} & \textcolor{blue}{$(0.03834)$} 
\\ \cline{3-7}                                         
          
& &  $0.2$ &  $0.01347$ &  $0.00375$ &  $0.08207$ &  $0.06529$
\\ \cline{4-7}
& & & \textcolor{blue}{$(0.01695)$} & \textcolor{blue}{$(0.00665)$} & 
      \textcolor{blue}{$(0.06449)$} & \textcolor{blue}{$(0.04600)$} 
\\ \cline{3-7}                      
                   
& & $0.3$ &  $0.01262$ &  $0.00349$ &  $0.08502$ &  $0.06942$
\\ \cline{4-7}
& & & \textcolor{blue}{$(0.01638)$} & \textcolor{blue}{$(0.00675)$} & 
      \textcolor{blue}{$(0.06638)$} & \textcolor{blue}{$(0.04893)$} 
\\ \cline{3-7}                     
          
& & $0.5$ &  $0.01056$ &  $0.00251$ &  $0.08754$ &  $0.07341$ 
\\ \cline{4-7}
& & & \textcolor{blue}{$(0.01445)$} & \textcolor{blue}{$(0.00603)$} & 
      \textcolor{blue}{$(0.06761)$} & \textcolor{blue}{$(0.05142)$} 
\\ \cline{3-7}                                               
          
& & $0.8$ &  $0.00788$ &  $0.00125$ &  $0.09167$ &  $0.07902$ 
\\ \cline{4-7}
& & & \textcolor{blue}{$(0.01166)$} & \textcolor{blue}{$(0.00469)$} & 
      \textcolor{blue}{$(0.06930)$} & \textcolor{blue}{$(0.05382)$} 
\\ \cline{3-7}   

& & $1.0$ &  $0.00652$ &  $0.00070$ &  $0.09462$ &  $0.08237$ 
\\ \cline{4-7}
& & & \textcolor{blue}{$(0.01007)$} & \textcolor{blue}{$(0.00388)$} & 
      \textcolor{blue}{$(0.07058)$} & \textcolor{blue}{$(0.05505)$} 
\\ \cline{3-7}  
                   
\rowcolor{yellow}
& & Iannario $(c)$ &  &  &  & \\           
                   
& & $0.6$  &  $0.01963$ &  $0.00362$ &  $0.09957$ &  $0.08104$
\\ \cline{4-7}
& & & \textcolor{blue}{$(0.02450)$} & \textcolor{blue}{$(0.00670)$} & 
      \textcolor{blue}{$(0.08074)$} & \textcolor{blue}{$(0.06024)$} 
\\ \cline{3-7}                                             
                     
& & $0.8$  &  $0.01777$ &  $0.00373$ &  $0.09455$ &  $0.07745$
\\ \cline{4-7}
& & & \textcolor{blue}{$(0.02232)$} & \textcolor{blue}{$(0.00697)$} & 
      \textcolor{blue}{$(0.07586)$} & \textcolor{blue}{$(0.05735)$} 
\\ \cline{3-7}                      
                      
& & $0.9$  &  $0.01715$ &  $0.00386$ &  $0.09305$ &  $0.07618$
\\ \cline{4-7}
& & & \textcolor{blue}{$(0.02156)$} & \textcolor{blue}{$(0.00710)$} & 
      \textcolor{blue}{$(0.07425)$} & \textcolor{blue}{$(0.05621)$} 
\\ \cline{3-7}                       
          
& & $1.0$  &  $0.01668$ &  $0.00397$ &  $0.09193$ &  $0.07512$
\\ \cline{4-7}
& & & \textcolor{blue}{$(0.02100)$} & \textcolor{blue}{$(0.00724)$} & 
      \textcolor{blue}{$(0.07318)$} & \textcolor{blue}{$(0.05538)$} 
\\ \cline{3-7}                     
                   
& & Croux  &  $0.01604$ &  $0.00448$ &  $0.09273$ &  $0.07349$ 
\\ \cline{4-7}
& & & \textcolor{blue}{($0.02022$)} & \textcolor{blue}{$(0.00790)$} & 
      \textcolor{blue}{$(0.07326)$} & \textcolor{blue}{$(0.05207)$} 
\\ \hline 
   \end{tabular}
   \end{adjustbox}
   \label{Table:model2 cauchy 0.05 vertical}
\end{table}


\begin{table}
\centering
 \caption{Squared bias and MSE when $10\%$ vertical outliers are added to data generated by \descref{Model 2} with the Cauchy link}
  \begin{adjustbox}{width=0.7\linewidth}
\begin{tabular}{c c c c c c c }
   \hline
Sample size & Link  & Method & $||\hat{\gamma}-\gamma||^{2}$ & $||\hat{\beta}-\beta||^{2}$ & $MSE(\hat{\gamma})$ & $MSE(\hat{\beta})$ 
\\ 
 \hline         
                      
$150$ \textcolor{blue}{$(200)$} & Cauchy & MLE  & $0.08868$ &   $0.03983$ &    $0.14961$ &   $0.09796$ \\ \cline{4-7}           
& & & \textcolor{blue}{$(0.09109)$} & \textcolor{blue}{$(0.09109)$} &  
      \textcolor{blue}{$(0.13357)$} & \textcolor{blue}{$(0.08222)$} 
\\ \cline{3-7}         
\rowcolor{yellow}
& & DPD $(\alpha)$ &  &  &  & \\                                
           
& & $0.1$ &  $0.07571$ &  $0.02919$ &  $0.13245$ &  $0.08109$
\\ \cline{4-7}
& & & \textcolor{blue}{$(0.07598)$} & \textcolor{blue}{$(0.03074)$} & 
      \textcolor{blue}{$(0.11735)$} & \textcolor{blue}{$(0.06736)$} 
\\ \cline{3-7}                                         
          
& &  $0.2$ &  $0.08032$ &   $0.03608$ &   $0.14006$ &  $0.09224$
\\ \cline{4-7}
& & & \textcolor{blue}{$(0.08214)$} & \textcolor{blue}{$(0.03922)$} & 
      \textcolor{blue}{$(0.12509)$} & \textcolor{blue}{$(0.07813)$} 
\\ \cline{3-7}                      
                   
& & $0.3$ &  $0.07940$ &  $0.03724$ &  $0.14091$ &  $0.09572$
\\ \cline{4-7}
& & & \textcolor{blue}{$(0.08195)$} & \textcolor{blue}{$(0.04126)$} & 
      \textcolor{blue}{$(0.12531)$} & \textcolor{blue}{$(0.08110)$} 
\\ \cline{3-7}                     
          
& & $0.5$ &  $0.07461$ &  $0.03584$ &  $0.13873$ &  $0.09752$ 
\\ \cline{4-7}
& & & \textcolor{blue}{$(0.07816)$} & \textcolor{blue}{$(0.04119)$} & 
      \textcolor{blue}{$(0.12315)$} & \textcolor{blue}{$(0.08288)$} 
\\ \cline{3-7}                                               
          
& & $0.8$ &  $0.06658$ &  $0.03199$ &  $0.13573$ &  $0.09878$ 
\\ \cline{4-7}
& & & \textcolor{blue}{$(0.07139)$} & \textcolor{blue}{$(0.03877)$} & 
      \textcolor{blue}{$(0.11955)$} & \textcolor{blue}{$(0.08359)$} 
\\ \cline{3-7}   

& & $1.0$ &  $0.06146$ &  $0.02922$ &  $0.13415$ &  $0.09925$ 
\\ \cline{4-7}
& & & \textcolor{blue}{$(0.06694)$} & \textcolor{blue}{$(0.03665)$} & 
      \textcolor{blue}{$(0.11743)$} & \textcolor{blue}{$(0.08349)$} 
\\ \cline{3-7}  
                   
\rowcolor{yellow}
& & Iannario $(c)$ &  &  &  & \\           
                   
& & $0.6$  &  $0.09957$ &  $0.03679$ &  $0.16811$ &  $0.10414$
\\ \cline{4-7}
& & & \textcolor{blue}{$(0.10203)$} & \textcolor{blue}{$(0.04039)$} & 
      \textcolor{blue}{$(0.15051)$} & \textcolor{blue}{$(0.08813)$} 
\\ \cline{3-7}                                             
                     
& & $0.8$  &  $0.09401$ &  $0.03748$ &  $0.15930$ &  $0.10102$
\\ \cline{4-7}
& & & \textcolor{blue}{$(0.09652)$} & \textcolor{blue}{$(0.04144)$} & 
      \textcolor{blue}{$(0.14224)$} & \textcolor{blue}{$(0.08605)$} 
\\ \cline{3-7}                      
                      
& & $0.9$  &  $0.09227$ &  $0.03803$ &  $0.15655$ &  $0.10015$
\\ \cline{4-7}
& & & \textcolor{blue}{$(0.09469)$} & \textcolor{blue}{$(0.04199)$} & 
      \textcolor{blue}{$(0.13954)$} & \textcolor{blue}{$(0.08532)$} 
\\ \cline{3-7}                       
          
& & $1.0$  &  $0.09099$ &  $0.03853$ &  $0.15444$ &  $0.09946$
\\ \cline{4-7}
& & & \textcolor{blue}{$(0.09343)$} & \textcolor{blue}{$(0.04252)$} & 
      \textcolor{blue}{$(0.13771)$} & \textcolor{blue}{$(0.08487)$} 
\\ \cline{3-7}                     
                   
& & Croux  &  $0.08982$ &  $0.04025$ &  $0.15346$ &  $0.09912$ 
\\ \cline{4-7}
& & & \textcolor{blue}{($0.09211$)} & \textcolor{blue}{$(0.04424)$} & 
      \textcolor{blue}{$(0.13672)$} & \textcolor{blue}{$(0.08328)$} 
\\ \hline 
   \end{tabular}
   \end{adjustbox}
   \label{Table:model2 cauchy 0.10 vertical}
\end{table}


\begin{table}
\centering
 \caption{Squared bias and MSE when $5\%$ vertical outliers are added to data generated by \descref{Model 3} with the logit link}
  \begin{adjustbox}{width=0.7\linewidth}
\begin{tabular}{c c c c c c c }
   \hline
Sample size & Link  & Method & $||\hat{\gamma}-\gamma||^{2}$ & $||\hat{\beta}-\beta||^{2}$ & $MSE(\hat{\gamma})$ & $MSE(\hat{\beta})$ 
\\ 
 \hline         
                      
$150$ \textcolor{blue}{$(200)$} & logit & MLE  & $0.09349$ &   $0.01687$ &   $0.15705$ &   $0.05238$ \\ \cline{4-7}           
& & & \textcolor{blue}{$(0.11243)$} & \textcolor{blue}{$(0.02058)$} &  
      \textcolor{blue}{$(0.16154)$} & \textcolor{blue}{$(0.04721)$} 
\\ \cline{3-7}         
\rowcolor{yellow}
& & DPD $(\alpha)$ &  &  &  & \\                                
           
& & $0.1$ &  $0.06391$ &  $0.00938$ &  $0.12838$ &  $0.04431$
\\ \cline{4-7}
& & & \textcolor{blue}{$(0.07868)$} & \textcolor{blue}{$(0.01209)$} & \textcolor{blue}{$(0.12957)$} & \textcolor{blue}{$(0.03878)$} 
\\ \cline{3-7}                                         
          
& &  $0.2$ &  $0.04892$ &   $0.00627$ &  $0.12044$ &  $0.04389$
\\ \cline{4-7}
& & & \textcolor{blue}{$(0.06251)$} & \textcolor{blue}{$(0.00869)$} & 
      \textcolor{blue}{$(0.11766)$} & \textcolor{blue}{$(0.03702)$} 
\\ \cline{3-7}                      
                   
& & $0.3$ &  $0.03575$ &   $0.00368$ &  $0.11397$ &  $0.04413$
\\ \cline{4-7}
& & & \textcolor{blue}{$(0.04794)$} & \textcolor{blue}{$(0.00578)$} & 
      \textcolor{blue}{$(0.10756)$} & \textcolor{blue}{$(0.03570)$} 
\\ \cline{3-7}                      
          
& & $0.5$ &  $0.01934$ &   $0.00095$ &  $0.11330$ &  $0.04819$ 
\\ \cline{4-7}
& & & \textcolor{blue}{$(0.02909)$} & \textcolor{blue}{$(0.00245)$} & 
      \textcolor{blue}{$(0.09932)$} & \textcolor{blue}{$(0.03669)$} 
\\ \cline{3-7}                                                 
          
& & $0.8$ &  $0.00908$ &  $0.00001$ &  $0.12563$ &   $0.05788$ 
\\ \cline{4-7}
& & & \textcolor{blue}{$(0.01642)$} & \textcolor{blue}{$(0.00074)$} & 
      \textcolor{blue}{$(0.10112)$} & \textcolor{blue}{$(0.04144)$} 
\\ \cline{3-7} 

& & $1.0$ &  $0.00656$ &  $0.00005$ &  $0.13355$ &  $0.06321$ 
\\ \cline{4-7}
& & & \textcolor{blue}{$(0.01278)$} & \textcolor{blue}{$(0.00041)$} & 
      \textcolor{blue}{$(0.10397)$} & \textcolor{blue}{$(0.04440)$} 
\\ \cline{3-7}  
                   
\rowcolor{yellow}
& & Iannario $(c)$ &  &  &  & \\           
                   
& & $0.6$  &  $0.00097$ &  $0.00014$ &  $0.02866$ &  $0.01818$
\\ \cline{4-7}
& & & \textcolor{blue}{$(0.00136)$} & \textcolor{blue}{$(0.00009)$} & 
      \textcolor{blue}{$(0.02427)$} & \textcolor{blue}{$(0.01593)$} 
\\ \cline{3-7}                                             
                     
& & $0.8$  &  $0.00079$ &  $0.00009$ &  $0.02593$ &  $0.01637$
\\ \cline{4-7}
& & & \textcolor{blue}{$(0.00094)$} & \textcolor{blue}{$(0.00017)$} & 
      \textcolor{blue}{$(0.02294)$} & \textcolor{blue}{$(0.01450)$} 
\\ \cline{3-7}                      
                      
& & $0.9$  &  $0.00086$ &  $0.00012$ &  $0.02390$ &  $0.01663$
\\ \cline{4-7}
& & & \textcolor{blue}{$(0.00129)$} & \textcolor{blue}{$(0.00013)$} & 
      \textcolor{blue}{$(0.02367)$} & \textcolor{blue}{$(0.01551)$} 
\\ \cline{3-7}                       
          
& & $1.0$  &  $0.00078$ &  $0.00013$ &  $0.02415$ &  $0.01577$
\\ \cline{4-7}
& & & \textcolor{blue}{$(0.00090)$} & \textcolor{blue}{$(0.00019)$} & 
      \textcolor{blue}{$(0.02246)$} & \textcolor{blue}{$(0.01476)$} 
\\ \cline{3-7}                     
                   
& & Croux  &  $0.08967$ &  $0.01627$ &  $0.15754$ &  $0.05202$ 
\\ \cline{4-7}
& & & \textcolor{blue}{($0.11268$)} & \textcolor{blue}{$(0.02064)$} & 
      \textcolor{blue}{$(0.16526)$} & \textcolor{blue}{$(0.04740)$} 
\\ \hline 
   \end{tabular}
   \end{adjustbox}
   \label{Table:model3 logit 0.05 vertical}
\end{table}


\begin{table}
\centering
 \caption{Squared bias and MSE when $10\%$ vertical outliers are added to data generated by \descref{Model 3} with the logit link}
  \begin{adjustbox}{width=0.7\linewidth}
\begin{tabular}{c c c c c c c }
   \hline
Sample size & Link  & Method & $||\hat{\gamma}-\gamma||^{2}$ & $||\hat{\beta}-\beta||^{2}$ & $MSE(\hat{\gamma})$ & $MSE(\hat{\beta})$ 
\\ 
 \hline         
                      
$150$ \textcolor{blue}{$(200)$} & logit & MLE  & $0.36860$ &  $0.07404$ &    $0.41935$ &   $0.10829$ \\ \cline{4-7}           
& & & \textcolor{blue}{$(0.37264)$} & \textcolor{blue}{$(0.07511)$} &  
      \textcolor{blue}{$(0.41266)$} & \textcolor{blue}{$(0.10100)$} 
\\ \cline{3-7}         
\rowcolor{yellow}
& & DPD $(\alpha)$ &  &  &  & \\                                
           
& & $0.1$ &  $0.30176$ &  $0.05413$ &  $0.35734$ &  $0.08980$
\\ \cline{4-7}
& & & \textcolor{blue}{$(0.30591)$} & \textcolor{blue}{$(0.05532)$} & \textcolor{blue}{$(0.34978)$} & \textcolor{blue}{$(0.08219)$} 
\\ \cline{3-7}                                         
          
& &  $0.2$ &  $0.25484$ &  $0.04331$ &  $0.31373$ &  $0.08001$
\\ \cline{4-7}
& & & \textcolor{blue}{$(0.26081)$} & \textcolor{blue}{$(0.04507)$} & 
      \textcolor{blue}{$(0.30684)$} & \textcolor{blue}{$(0.07238)$} 
\\ \cline{3-7}                      
                   
& & $0.3$ &  $0.21064$ &  $0.03276$ &   $0.27549$ &  $0.07158$
\\ \cline{4-7}
& & & \textcolor{blue}{$(0.21814)$} & \textcolor{blue}{$(0.03495)$} & 
      \textcolor{blue}{$(0.26821)$} & \textcolor{blue}{$(0.06351)$} 
\\ \cline{3-7}                      
          
& & $0.5$ &  $0.14538$ &  $0.01830$ &  $0.22511$ &  $0.06313$ 
\\ \cline{4-7}
& & & \textcolor{blue}{$(0.15577)$} & \textcolor{blue}{$(0.02121)$} & 
      \textcolor{blue}{$(0.21604)$} & \textcolor{blue}{$(0.05351)$} 
\\ \cline{3-7}                                                 
          
& & $0.8$ &  $0.09268$ &  $0.00860$ &  $0.19391$ &  $0.06302$ 
\\ \cline{4-7}
& & & \textcolor{blue}{$(0.10601)$} & \textcolor{blue}{$(0.01207)$} & 
      \textcolor{blue}{$(0.18052)$} & \textcolor{blue}{$(0.05021)$} 
\\ \cline{3-7}   

& & $1.0$ &  $0.07560$ &  $0.00618$ &  $0.18670$ &  $0.06528$ 
\\ \cline{4-7}
& & & \textcolor{blue}{$(0.08997)$} & \textcolor{blue}{$(0.00980)$} & 
      \textcolor{blue}{$(0.17075)$} & \textcolor{blue}{$(0.05084)$} 
\\ \cline{3-7}  
                   
\rowcolor{yellow}
& & Iannario $(c)$ &  &  &  & \\           
                   
& & $0.6$  &  $0.00551$ &  $0.00069$ &  $0.05755$ &  $0.03125$
\\ \cline{4-7}
& & & \textcolor{blue}{$(0.00700)$} & \textcolor{blue}{$(0.00063)$} & 
      \textcolor{blue}{$(0.05788)$} & \textcolor{blue}{$(0.02746)$} 
\\ \cline{3-7}                                             
                     
& & $0.8$  &  $0.00568$ &   $0.00059$ &  $0.05680$ &  $0.02608$
\\ \cline{4-7}
& & & \textcolor{blue}{$(0.00731)$} & \textcolor{blue}{$(0.00078)$} & 
      \textcolor{blue}{$(0.05857)$} & \textcolor{blue}{$(0.03094)$} 
\\ \cline{3-7}                      
                      
& & $0.9$  &  $0.00719$ &  $0.00123$ &  $0.06402$ &  $0.03791$
\\ \cline{4-7}
& & & \textcolor{blue}{$(0.00639)$} & \textcolor{blue}{$(0.00077)$} & 
      \textcolor{blue}{$(0.05060)$} & \textcolor{blue}{$(0.02869)$} 
\\ \cline{3-7}                       
          
& & $1.0$  &  $0.00673$ &  $0.00073$ &  $0.05699$ &  $0.02974$
\\ \cline{4-7}
& & & \textcolor{blue}{$(0.00645)$} & \textcolor{blue}{$(0.00059)$} & 
      \textcolor{blue}{$(0.05413)$} & \textcolor{blue}{$(0.02740)$} 
\\ \cline{3-7}                     
                   
& & Croux  &  $0.35907$ &  $0.07206$ &  $0.41423$ &  $0.10671$ 
\\ \cline{4-7}
& & & \textcolor{blue}{($0.37283$)} & \textcolor{blue}{$(0.07520)$} & 
      \textcolor{blue}{$(0.41637)$} & \textcolor{blue}{$(0.10117)$} 
\\ \hline 
   \end{tabular}
   \end{adjustbox}
   \label{Table:model3 logit 0.10 vertical}
\end{table}


\begin{table}
\centering
 \caption{Squared bias and MSE when $5\%$ vertical outliers are added to data generated by \descref{Model 5} with the complementary log-log link}
  \begin{adjustbox}{width=0.7\linewidth}
\begin{tabular}{c c c c c c c }
   \hline
Sample size & Link  & Method & $||\hat{\gamma}-\gamma||^{2}$ & $||\hat{\beta}-\beta||^{2}$ & $MSE(\hat{\gamma})$ & $MSE(\hat{\beta})$ 
\\ 
 \hline         
                      
$150$ \textcolor{blue}{$(200)$} & log-log & MLE  & $0.67107$ &   $0.34065$ &   $0.75620$ &   $0.46745$ \\ \cline{4-7}           
& & & \textcolor{blue}{$(0.79469)$} & \textcolor{blue}{$(0.41913)$} &  
      \textcolor{blue}{$(0.84674)$} & \textcolor{blue}{$(0.50749)$} 
\\ \cline{3-7}         
\rowcolor{yellow}
& & DPD $(\alpha)$ &  &  &  & \\                                
           
& & $0.1$ &  $0.35965$ &  $0.16504$ &   $0.45120$ &   $0.27767$
\\ \cline{4-7}
& & & \textcolor{blue}{$(0.44723)$} & \textcolor{blue}{$(0.21468)$} & \textcolor{blue}{$(0.51548)$} & \textcolor{blue}{$(0.29913)$} 
\\ \cline{3-7}                                         
          
& &  $0.2$ &  $0.08589$ &   $0.03262$ &    $0.17626$ &  $0.12863$
\\ \cline{4-7}
& & & \textcolor{blue}{$(0.11621)$} & \textcolor{blue}{$(0.04806)$} & 
      \textcolor{blue}{$(0.18499)$} & \textcolor{blue}{$(0.11780)$} 
\\ \cline{3-7}                      
                   
& & $0.3$ &  $0.02325$ &   $0.00645$ &   $0.11282$ &   $0.10273$
\\ \cline{4-7}
& & & \textcolor{blue}{$(0.03735)$} & \textcolor{blue}{$(0.01254)$} & 
      \textcolor{blue}{$(0.10295)$} & \textcolor{blue}{$(0.08154)$} 
\\ \cline{3-7}                      
          
& & $0.5$ &  $0.00233$ &    $0.00113$ &   $0.10605$ &   $0.11394$ 
\\ \cline{4-7}
& & & \textcolor{blue}{$(0.00784)$} & \textcolor{blue}{$(0.00131)$} & 
      \textcolor{blue}{$(0.07781)$} & \textcolor{blue}{$(0.07884)$} 
\\ \cline{3-7}  

& & $0.8$ &  $0.00112$ &    $0.00454$ & $ 0.12674$ &   $0.14493$ 
\\ \cline{4-7}
& & & \textcolor{blue}{$(0.00167)$} & \textcolor{blue}{$(0.00039)$} & 
      \textcolor{blue}{$(0.08337)$} & \textcolor{blue}{$(0.09388)$} 
\\ \cline{3-7}    

& & $1.0$ &  $0.00202$ &  $0.00619$ &   $0.13590$ &   $0.15810$ 
\\ \cline{4-7}
& & & \textcolor{blue}{$(0.00091)$} & \textcolor{blue}{$(0.00059)$} & 
      \textcolor{blue}{$(0.08772)$} & \textcolor{blue}{$(0.10153)$} 
\\ \cline{3-7}  
                   
\rowcolor{yellow}
& & Iannario $(c)$ &  &  &  & \\           
                   
& & $1.1$  &  $0.03569 $ & $0.07658$ &    $0.28077$ &   $0.31508$
\\ \cline{4-7}
& & & \textcolor{blue}{$(0.02280)$} & \textcolor{blue}{$(0.05765)$} & 
      \textcolor{blue}{$(0.21503)$} & \textcolor{blue}{$(0.23128)$} 
\\ \cline{3-7}                                             
                     
& & $1.4$  &  $0.00845$ &   $0.03774$ &   $ 0.20197$ &   $0.22277$
\\ \cline{4-7}
& & & \textcolor{blue}{$(0.00388)$} & \textcolor{blue}{$(0.02508)$} & 
      \textcolor{blue}{$(0.15595)$} & \textcolor{blue}{$(0.16133)$} 
\\ \cline{3-7}                      
                      
& & $1.5$  &  $0.00433$ &    $0.02996$ &   $0.18551$ &  $0.20015$
\\ \cline{4-7}
& & & \textcolor{blue}{$(0.00220)$} & \textcolor{blue}{$(0.01870)$} & 
      \textcolor{blue}{$(0.14522)$} & \textcolor{blue}{$(0.14665)$} 
\\ \cline{3-7}                       
& & $1.7$  &  $00.00096$ &     $0.01992$ & $0.17543$ &   $0.18171$
\\ \cline{4-7}
& & & \textcolor{blue}{$(0.00175)$} & \textcolor{blue}{$(0.01341)$} & 
      \textcolor{blue}{$(0.15017)$} & \textcolor{blue}{$(0.14232)$} 
\\ \cline{3-7}                     
                   
& & Croux  &  $0.74334$ &  $0.39614$ &   $0.81978$ &   $0.50701$ 
\\ \cline{4-7}
& & & \textcolor{blue}{($0.85310$)} & \textcolor{blue}{$(0.46098)$} & 
      \textcolor{blue}{$(0.89923)$} & \textcolor{blue}{$(0.53467)$} 
\\ \hline 
   \end{tabular}
   \end{adjustbox}
   \label{Table:model5 loglog 0.05 vertical}
\end{table}


\begin{table}
\centering
 \caption{Squared bias and MSE when $10\%$ vertical outliers are added to data generated by \descref{Model 5} with the complementary log-log link}
  \begin{adjustbox}{width=0.7\linewidth}
\begin{tabular}{c c c c c c c }
   \hline
Sample size & Link  & Method & $||\hat{\gamma}-\gamma||^{2}$ & $||\hat{\beta}-\beta||^{2}$ & $MSE(\hat{\gamma})$ & $MSE(\hat{\beta})$ 
\\ 
 \hline         
                      
$150$ \textcolor{blue}{$(200)$} & log-log & MLE  & $1.35901$ &  $0.67394$ &   $1.39573$ &   $0.75649$ \\ \cline{4-7}           
& & & \textcolor{blue}{$(1.39844)$} & \textcolor{blue}{$(0.70918)$} &  
      \textcolor{blue}{$(1.42347)$} & \textcolor{blue}{$(0.76895)$} 
\\ \cline{3-7}         
\rowcolor{yellow}
& & DPD $(\alpha)$ &  &  &  & \\                                
           
& & $0.1$ &  $1.12607$ &  $0.53678$ &  $1.17625$ &  $0.63136$
\\ \cline{4-7}
& & & \textcolor{blue}{$(1.17012)$} & \textcolor{blue}{$(0.56997)$} & \textcolor{blue}{$(1.20455)$} & \textcolor{blue}{$(0.63803)$} 
\\ \cline{3-7}                                         
          
& &  $0.2$ &  $0.64232$ &  $0.27279$ &  $0.73806$ &   $0.39434$
\\ \cline{4-7}
& & & \textcolor{blue}{$(0.67148)$} & \textcolor{blue}{$(0.29143)$} & 
      \textcolor{blue}{$(0.74338)$} & \textcolor{blue}{$(0.38151)$} 
\\ \cline{3-7}                      
                   
& & $0.3$ &  $0.24846$ &  $0.08975$ &  $0.35056$ &   $0.20063$
\\ \cline{4-7}
& & & \textcolor{blue}{$(0.26717)$} & \textcolor{blue}{$(0.10096)$} & 
      \textcolor{blue}{$(0.34647)$} & \textcolor{blue}{$(0.17944)$} 
\\ \cline{3-7}                     
          
& & $0.5$ &  $0.06191$ &  $0.01273$ &   $0.16822$ &   $0.12938$ 
\\ \cline{4-7}
& & & \textcolor{blue}{$(0.07849)$} & \textcolor{blue}{$(0.01925)$} & 
      \textcolor{blue}{$(0.15566)$} & \textcolor{blue}{$(0.10010)$} 
\\ \cline{3-7}                                                
          
& & $0.8$ &  $0.02076$ &  $0.00066$ &  $0.13952$ &   $0.13650$ 
\\ \cline{4-7}
& & & \textcolor{blue}{$(0.03506)$} & \textcolor{blue}{$(0.00358)$} & 
      \textcolor{blue}{$(0.11800)$} & \textcolor{blue}{$(0.09665)$} 
\\ \cline{3-7}  

& & $1.0$ &  $0.01543$ &   $0.00005$ &  $0.13761$ &  $0.14375$ 
\\ \cline{4-7}
& & & \textcolor{blue}{$(0.02927)$} & \textcolor{blue}{$(0.00230)$} & 
      \textcolor{blue}{$(0.11410)$} & \textcolor{blue}{$(0.10052)$} 
\\ \cline{3-7}  
                   
\rowcolor{yellow}
& & Iannario $(c)$ &  &  &  & \\           
                   
& & $1.1$  &  $0.00817$ &   $0.03981$ &  $0.28002$ &   $0.27322$
\\ \cline{4-7}
& & & \textcolor{blue}{$(0.00939)$} & \textcolor{blue}{$(0.03204)$} & 
      \textcolor{blue}{$(0.23622)$} & \textcolor{blue}{$(0.21272)$} 
\\ \cline{3-7}                                             
                     
& & $1.4$  &  $0.02240$ &   $0.02028$ &   $0.26673$ &   $0.22526$
\\ \cline{4-7}
& & & \textcolor{blue}{$(0.02133)$} & \textcolor{blue}{$(0.01424)$} & 
      \textcolor{blue}{$(0.22134)$} & \textcolor{blue}{$(0.16968)$} 
\\ \cline{3-7}                      
                      
& & $1.5$  &  $0.02654$ &   $0.01753$ &   $0.28295$ &   $0.23761$
\\ \cline{4-7}
& & & \textcolor{blue}{$(0.02542)$} & \textcolor{blue}{$(0.01157)$} & 
      \textcolor{blue}{$(0.23144)$} & \textcolor{blue}{$(0.16900)$} 
\\ \cline{3-7}                       
          
& & $1.7$  &  $0.04521$ &   $0.01310$ &   $0.30673$ &   $0.22814$
\\ \cline{4-7}
& & & \textcolor{blue}{$(0.04570)$} & \textcolor{blue}{$(0.01024)$} & 
      \textcolor{blue}{$(0.25892)$} & \textcolor{blue}{$(0.16987)$} 
\\ \cline{3-7}                     
                   
& & Croux  &  $1.35182$ &  $0.67836$ &  $1.39762$ &   $0.75624$ 
\\ \cline{4-7}
& & & \textcolor{blue}{($1.40714$)} & \textcolor{blue}{$(0.71510)$} & 
      \textcolor{blue}{$(1.44104)$} & \textcolor{blue}{$(0.77266)$} 
\\ \hline 
   \end{tabular}
   \end{adjustbox}
   \label{Table:model5 loglog 0.1 vertical}
\end{table}


\begin{figure}
\begin{multicols}{2}
    \includegraphics[scale=0.4]{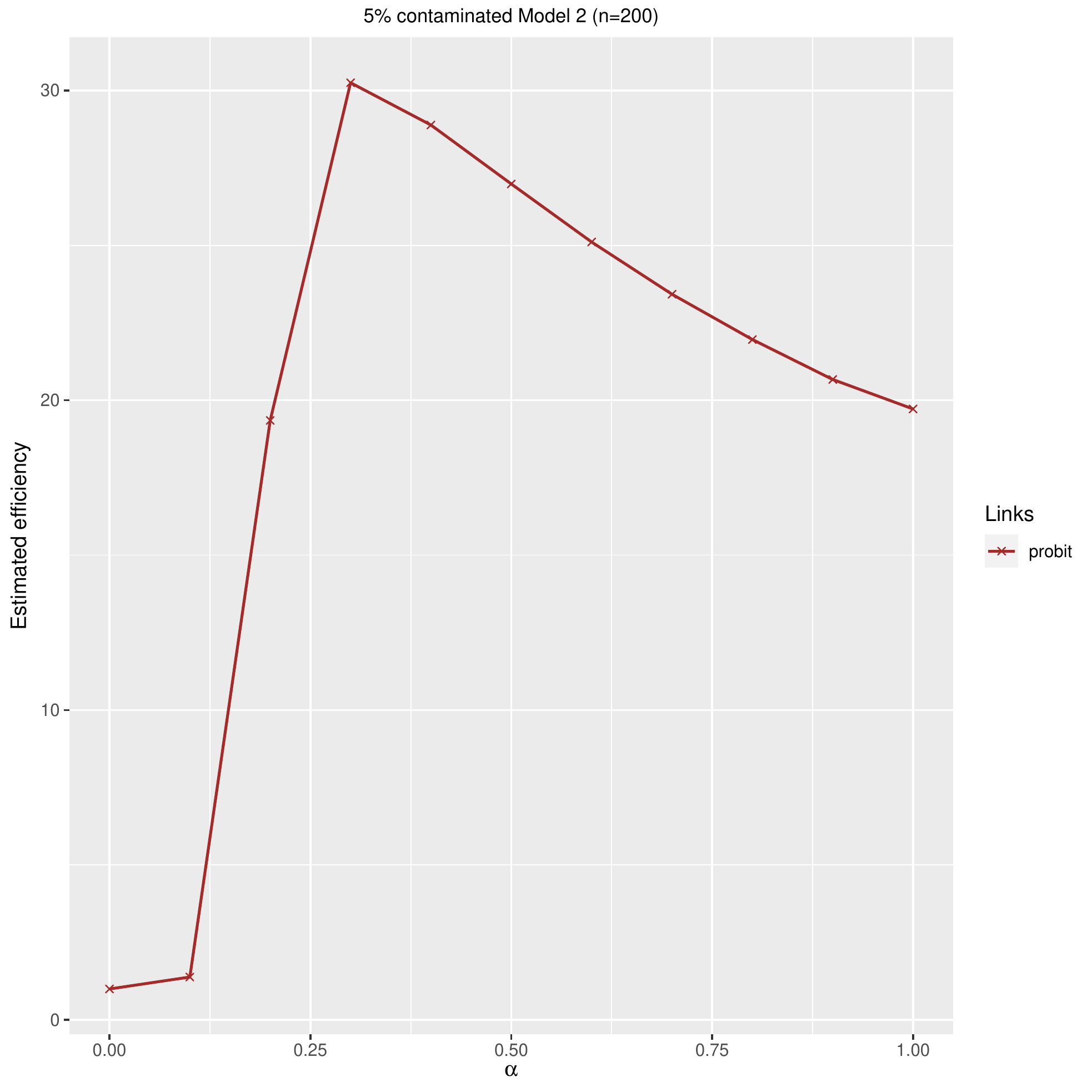} 
\caption{Graphs of efficiency when data generated by \descref{Model 2} is horizontally contaminated at $5\%$ level of contamination.}  
\label{fig:efficiency, horizontal contam_0.05, model2}
    \par 
    \includegraphics[scale=0.4]{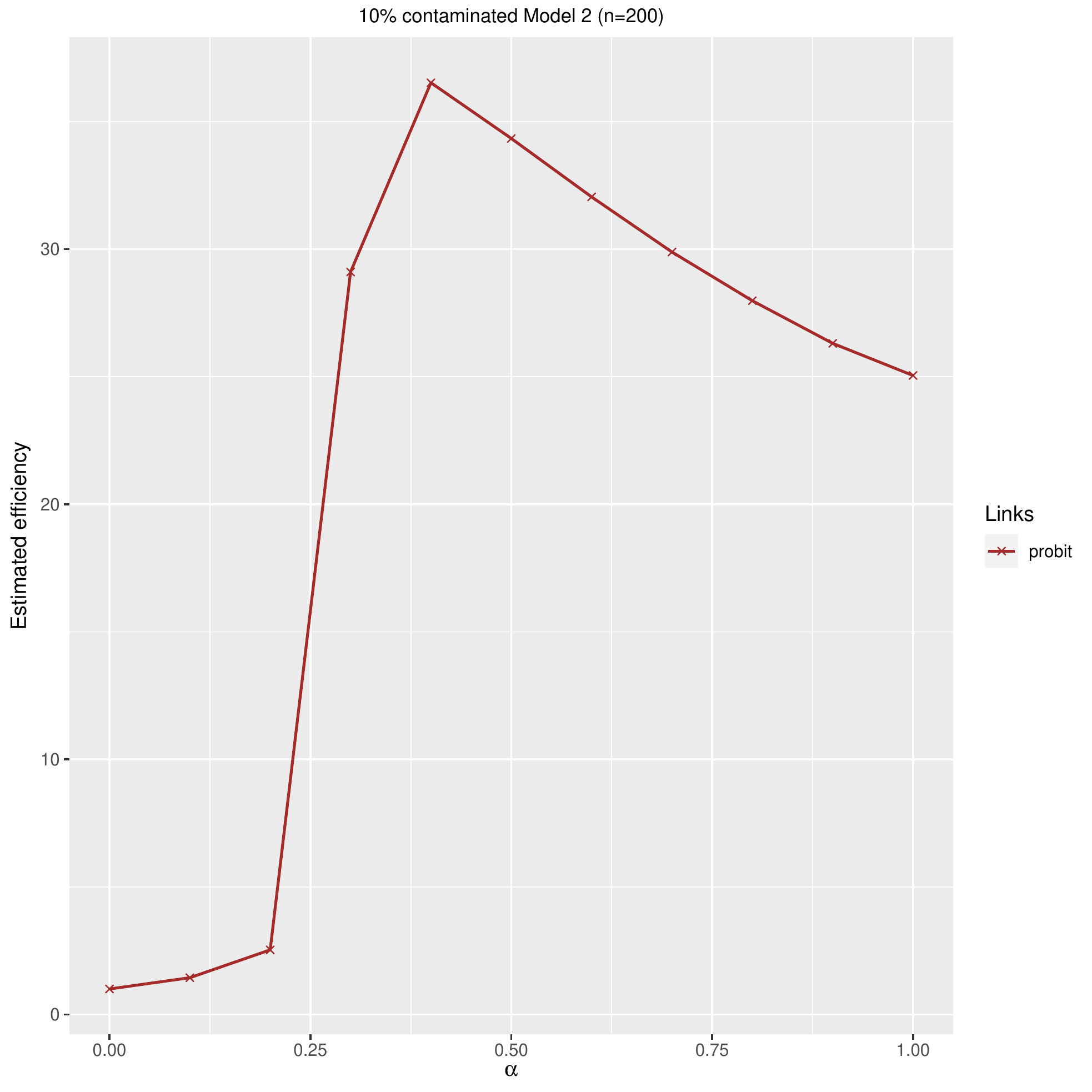} 
\caption{Graphs of efficiency when data generated by \descref{Model 2} is horizontally contaminated at $10\%$ level of contamination.}  
\label{fig:efficiency, horizontal contam_0.1, model2}
    \par 
    \end{multicols}
\end{figure} 

\begin{table}
\centering
\caption{Squared bias and MSE when $5\%$ horizontal outliers are added to data generated by \descref{Model 2} with the probit link}
  \begin{adjustbox}{width=0.7\linewidth}
\begin{tabular}{c c c c c c c }
   \hline
Sample size & Link  & Method & $||\hat{\gamma}-\gamma||^{2}$ & $||\hat{\beta}-\beta||^{2}$ & $MSE(\hat{\gamma})$ & $MSE(\hat{\beta})$ 
\\ 
 \hline         
                      
$150$ \textcolor{blue}{$(200)$} & probit & MLE  & $0.17980$ &   $1.08089$ &    $0.19515$ &  $1.08790$ \\ \cline{4-7}           
& & & \textcolor{blue}{$(0.18975)$} & \textcolor{blue}{$(1.12790)$} &  
      \textcolor{blue}{$(0.20091)$} & \textcolor{blue}{$(1.13275)$} 
\\ \cline{3-7}         
\rowcolor{yellow}
& & DPD $(\alpha)$ &  &  &  & \\                                
           
& & $0.1$ &  $0.10358$ &   $0.62587$ &  $0.13361$ &  $0.71104$
\\ \cline{4-7}
& & & \textcolor{blue}{$(0.12947)$} & \textcolor{blue}{$(0.76588)$} & \textcolor{blue}{$(0.14917)$} & \textcolor{blue}{$(0.81316)$} 
\\ \cline{3-7}                                         
          
& &  $0.2$ &  $0.00011$ & $0.00175$ &   $0.03179$ &   $0.05376$
\\ \cline{4-7}
& & & \textcolor{blue}{$(0.00036)$} & \textcolor{blue}{$(0.00296)$} & 
      \textcolor{blue}{$(0.02494)$} & \textcolor{blue}{$(0.04399)$} 
\\ \cline{3-7}                      
                   
& & $0.3$ &  $0.00028$ &   $0.00049$ &   $0.03098$ &  $0.02797$
\\ \cline{4-7}
& & & \textcolor{blue}{$(0.00010)$} & \textcolor{blue}{$(0.00015)$} & 
      \textcolor{blue}{$(0.02335)$} & \textcolor{blue}{$(0.02075)$} 
\\ \cline{3-7}                     
           
& & $0.5$ &  $0.00063$ &  $0.00141$ &  $0.03461$ &  $0.03072$ 
\\ \cline{4-7}
& & & \textcolor{blue}{$(0.00033)$} & \textcolor{blue}{$(0.00074)$} & 
      \textcolor{blue}{$(0.02626)$} & \textcolor{blue}{$(0.02317)$} 
\\ \cline{3-7}                                                
          
& & $0.8$ &  $0.00122$ &  $0.00266$ &  $0.04123$ &  $0.03939$ 
\\ \cline{4-7}
& & & \textcolor{blue}{$(0.00072)$} & \textcolor{blue}{$(0.00152)$} & 
      \textcolor{blue}{$(0.03122)$} & \textcolor{blue}{$(0.02951)$} 
\\ \cline{3-7}   

& & $1.0$ &  $0.00168$ &  $0.00358$ &  $0.04569$ &   $0.04522$ 
\\ \cline{4-7}
& & & \textcolor{blue}{$(0.00101)$} & \textcolor{blue}{$(0.00206)$} & 
      \textcolor{blue}{$(0.03434)$} & \textcolor{blue}{$(0.03330)$} 
\\ \cline{3-7}  
                   
\rowcolor{yellow}
& & Iannario $(c)$ &  &  &  & \\           
                   
& & $1.1$  &  $0.01437$ &  $0.17591$ &   $0.05819$ &  $0.21230$
\\ \cline{4-7}
& & & \textcolor{blue}{$(0.01111)$} & \textcolor{blue}{$(0.16005)$} & 
      \textcolor{blue}{$(0.04167)$} & \textcolor{blue}{$(0.18579)$} 
\\ \cline{3-7}                                             
                     
& & $1.4$  &  $0.00419$ &  $0.05228$ &  $0.03682$ &   $0.08047$
\\ \cline{4-7}
& & & \textcolor{blue}{$(0.00290)$} & \textcolor{blue}{$(0.04204)$} & 
      \textcolor{blue}{$(0.02671)$} & \textcolor{blue}{$(0.06332)$} 
\\ \cline{3-7}                      
                      
& & $1.5$  &  $0.00267$ &  $0.03273$ &  $0.03350$ &  $0.05880$
\\ \cline{4-7}
& & & \textcolor{blue}{$(0.00179)$} & \textcolor{blue}{$(0.02507)$} & 
      \textcolor{blue}{$(0.02463)$} & \textcolor{blue}{$(0.04524)$} 
\\ \cline{3-7}                       
          
& & $1.7$  &  $0.00091$ &  $0.01047$ &   $0.02978$ &  $0.03424$
\\ \cline{4-7}
& & & \textcolor{blue}{$(0.00059)$} & \textcolor{blue}{$(0.00654)$} & 
      \textcolor{blue}{$(0.02201)$} & \textcolor{blue}{$(0.02488)$} 
\\ \cline{3-7}                     
                   
& & Croux  &  $0.17578$ &  $1.07969$ &   $0.19163$ &  $1.08658$ 
\\ \cline{4-7}
& & & \textcolor{blue}{($0.18979$)} & \textcolor{blue}{$(1.12763)$} & 
      \textcolor{blue}{$(0.20143)$} & \textcolor{blue}{$(1.13247)$} 
\\ \hline 
   \end{tabular}
   \end{adjustbox}
   \label{Table:model2 probit-0.05 horizontal}
\end{table}


\begin{table}
\centering
 \caption{Squared bias and MSE when the probit link is misspecified with the complementary log-log link in \descref{Model 1}}
  \begin{adjustbox}{width=0.7\linewidth}
\begin{tabular}{c c c c c c c }
   \hline
Sample size & Link  & Method & $||\hat{\gamma}-\gamma||^{2}$ & $||\hat{\beta}-\beta||^{2}$ & $MSE(\hat{\gamma})$ & $MSE(\hat{\beta})$ 
\\ 
 \hline         
                      
$150$ \textcolor{blue}{$(200)$} & log-log & MLE  & $0.22796$ &  $0.16405$ &   $0.27993$ &  $0.20407$ \\ \cline{4-7}           
& & & \textcolor{blue}{$(0.21333)$} & \textcolor{blue}{$(0.16156)$} &  
      \textcolor{blue}{$(0.24882)$} & \textcolor{blue}{$(0.18965)$} 
\\ \cline{3-7}         
\rowcolor{yellow}
& & DPD $(\alpha)$ &  &  &  & \\                                
           
& & $0.1$ &  $0.19688$ &  $0.14474$ &   $0.24136$ &  $0.18171$
\\ \cline{4-7}
& & & \textcolor{blue}{$(0.18525)$} & \textcolor{blue}{$(0.14289)$} & \textcolor{blue}{$(0.21630)$} & \textcolor{blue}{$(0.16870)$} 
\\ \cline{3-7}                                         
          
& &  $0.2$ &  $0.20842$ &   $0.14284$ &  $0.25887$ &  $0.18398$
\\ \cline{4-7}
& & & \textcolor{blue}{$(0.19578)$} & \textcolor{blue}{$(0.14061)$} & 
      \textcolor{blue}{$(0.22985)$} & \textcolor{blue}{$(0.16835)$} 
\\ \cline{3-7}                      
                   
& & $0.3$ &  $0.21571$ &  $0.13956$ &  $0.27373$ &  $0.18954$
\\ \cline{4-7}
& & & \textcolor{blue}{$(0.20093)$} & \textcolor{blue}{$(0.13634)$} & 
      \textcolor{blue}{$(0.23818)$} & \textcolor{blue}{$(0.16785)$} 
\\ \cline{3-7}                     
           
& & $0.5$ &  $0.23037$ &  $0.14024$  &  $0.29848$ &   $0.20461$ 
\\ \cline{4-7}
& & & \textcolor{blue}{$(0.21397)$} & \textcolor{blue}{$(0.13506)$} & 
      \textcolor{blue}{$(0.26013)$} & \textcolor{blue}{$(0.17922)$} 
\\ \cline{3-7}                                                
          
& & $0.8$ &  $0.23812$ &  $0.14576$ &  $0.30862$ &  $0.21919$ 
\\ \cline{4-7}
& & & \textcolor{blue}{$(0.22256)$} & \textcolor{blue}{$(0.14158)$} & 
      \textcolor{blue}{$(0.26900)$} & \textcolor{blue}{$(0.19087)$} 
\\ \cline{3-7}   

& & $1.0$ &  $0.23509$ &  $0.14644$ &  $0.30301$ &  $0.22220$ 
\\ \cline{4-7}
& & & \textcolor{blue}{$(0.22104)$} & \textcolor{blue}{$(0.14322)$} & 
      \textcolor{blue}{$(0.26707)$} & \textcolor{blue}{$(0.19585)$} 
\\ \cline{3-7}  
                   
\rowcolor{yellow}
& & Iannario $(c)$ &  &  &  & \\           
                   
& & $1.1$  &  $4.07929$ &  $1.13771$ &  $4.27116$ &  $1.30679$
\\ \cline{4-7}
& & & \textcolor{blue}{$(3.85829)$} & \textcolor{blue}{$(1.05397)$} & 
      \textcolor{blue}{$(4.00773)$} & \textcolor{blue}{$(1.19122)$} 
\\ \cline{3-7}                                             
                     
& & $1.4$  &  $2.22450$ &  $0.25767$ &  $2.45428$ &  $0.46093$
\\ \cline{4-7}
& & & \textcolor{blue}{$(2.09108)$} & \textcolor{blue}{$(0.25334)$} & 
      \textcolor{blue}{$(2.24494)$} & \textcolor{blue}{$(0.40160)$} 
\\ \cline{3-7}                      
                      
& & $1.5$  &  $0.86343$ &  $0.56844$ &  $1.02307$ &  $0.69084$
\\ \cline{4-7}
& & & \textcolor{blue}{$(0.79026)$} & \textcolor{blue}{$(0.54682)$} & 
      \textcolor{blue}{$(0.89390)$} & \textcolor{blue}{$(0.62439)$} 
\\ \cline{3-7}                       
          
& & $1.7$  &  $0.80507$ &  $0.57208$ &  $0.92810$ &  $0.66357$
\\ \cline{4-7}
& & & \textcolor{blue}{$(0.74604)$} & \textcolor{blue}{$(0.55233)$} & 
      \textcolor{blue}{$(0.81903)$} & \textcolor{blue}{$(0.60822)$} 
\\ \cline{3-7}                     
                   
& & Croux  &  $0.22830$ &  $0.16435$ &  $0.28051$ &  $0.20447$ 
\\ \cline{4-7}
& & & \textcolor{blue}{($0.21362$)} & \textcolor{blue}{$(0.16199)$} & 
      \textcolor{blue}{$(0.24918)$} & \textcolor{blue}{$(0.19011)$} 
\\ \hline 
   \end{tabular}
   \end{adjustbox}
   \label{Table:model1 probit-loglog}
\end{table}

\clearpage
\newpage

\section{Data Driven Selection of Tuning Parameter}
\label{tuning parameter selection}

We have seen in simulation studies that small values of $\alpha$ increase the efficiency of an MDPD estimator under the model whereas its robustness increases with $\alpha$. However, an experimenter will not know, a priori, the amount of contamination in a data set. So a data-driven strategy for tuning parameter selection is required to apply this method in real-life data sets. Among different existing methods, we will follow the approach of Warwick and Jones \cite{warwick2005choosing}. They suggest constructing an empirical version of the mean square error, and further minimizing it over the tuning parameter. This method has been generalized by Ghosh and Basu \cite{ghosh2015robust}, and further extended by Basak et al. \cite{basak2021optimal}. Minimizing the empirical MSE has been shown to provide satisfactory performance in selecting an appropriate tuning parameter based on a data set. The empirical version of the asymptotic mean square error, expressed as a function of the tuning parameter and a pilot $\theta_{P}$, is given by 
\begin{equation}
\label{WJ MSE}
    \widehat{MSE}_{\alpha}(\theta_{P})=(\hat{\theta}_{\alpha}-\theta_{P})^{T}(\widehat{\theta}_{\alpha}-\theta_{P})+\frac{1}{n}tr\Big(\widehat{\Psi}_{n}^{-1}(\alpha) \widehat{\Omega}_{n}(\alpha) \hat{\Psi}_{n}^{-1}(\alpha)\Big).
\end{equation}

The optimal $\alpha$ that minimizes (\ref{WJ MSE}) may depend on the choice of the pilot $\theta_{P}$ as well. In Table \ref{Tab: comparison of MSEs with fitted mdpde}, we compare the replication-MSE and the squared bias of the fitted (in the sense of minimized (\ref{WJ MSE})) MDPDEs with that of the MLEs for data sets generated through \descref{Model 3}. In doing so, we have considered both $\hat{\theta}_{0.5}$ and $\hat{\theta}_{1}$ as pilot values. We see that the replication-MSE of the fitted MDPDEs corresponding to the pilot $\hat{\theta}_{0.5}$ gives a much better approximation of the replication-MSE of the MLE (smallest in this case) in pure data sets across two different links. Also, note that the squared biases due to the pilot $\hat{\theta}_{1}$ become almost twice the values generated by the MLE in pure data for both the probit and the logit links.  

Now, we add $10\%$ and $15\%$ vertical outliers in the data sets. As we found, the fitted MDPDEs perform better than the MLE, and the pilot $\hat{\theta}_{1}$ gives the lowest replication-MSE and the squared bias. This gives a clear indication that the more robust we choose a pilot value,  the fitted MDPDEs become more resistant to the outliers, but also they lose their efficiency along the way. Since $\hat{\theta}_{0.5}$ is highly robust (though less robust than $\hat{\theta}_{1}$) and they yield better-fitted MDPDEs in pure data, $\hat{\theta}_{0.5}$ chosen as a pilot would provide a nice trade-off between the efficiency and robustness. This conclusion also validates the empirical evidence of Ghosh and Basu \cite{ghosh2015robust}.    
\begin{table}[!h]
\centering
 \caption{Comparison of the replication-MSE and squared-bias (in bracket) between the MLE (first row for each link) and the fitted MDPDEs with $\hat{\theta}_{0.5}$ and $\hat{\theta}_{1}$ as pilots under different levels ($\epsilon$) of data contamination in vertical direction}
  \begin{adjustbox}{width=0.9\linewidth}
\begin{tabular}{c c c c c c }
   \hline
Model & Link & Pilot & $\epsilon=0$ & $\epsilon=0.10$ & $\epsilon=0.15$ \\ 
 \hline                       
\descref{Model 3} & probit 
&  & 0.18278 (0.01067) & 2.87686 (2.79551) & 4.00026 (3.94187) \\
& & $\hat{\theta}_{0.5}$ & 0.19797 (0.01283) & 0.54000 (0.35334) & 1.07291 (0.89671) \\
& & $\hat{\theta}_{1}$ & 0.24429 (0.02210) & 0.46164 (0.25809) & 0.82525 (0.63395) \\
\cline{2-6}
 & logit 
&  & 0.2782 (0.00514) & 1.47196 (1.30487) & 2.57257 (2.42587)\\
& & $\hat{\theta}_{0.5}$ & 0.29862 (0.00528) & 0.91008 (0.67627) & 1.70114 (1.50296) \\
& & $\hat{\theta}_{1}$ & 0.34358 (0.01027) & 0.77661 (0.50389) & 1.39069 (1.14921) \\
  \hline 
   \end{tabular}
   \end{adjustbox}
   \label{Tab: comparison of MSEs with fitted mdpde}
\end{table}

This demonstrates the effectiveness of the algorithm in picking out a suitable tuning parameter based on a data set. We apply this strategy in the next section to analyze a wine quality data set. Going forward, we will use the $\hat{\theta}_{0.5}$ as our pilot.

\section{Real Data Analysis}
\label{real data analysis}

We analyze a white wine quality data set that is available in the UCI Machine Learning Repository. This data set contains $11$ independent variables and an ordinal response variable. These continuous variables $x_{i}$s determine the wine's quality which is rated on a scale of $3-9$. As we see, the values of the covariates vary a lot. They are standardized using the median ($med$) and the mean absolute deviation (MAD) as 
\begin{equation}
\label{data transformation}
    x_{ij}^{*}=\frac{x_{ij}-med(x_{i})}{1.4828 \times MAD(x_{i})}
    \mbox{ where } x_{i}=(x_{i1}, \ldots , x_{ip})^{T}
    \mbox{ and } j =1, 2, \ldots , p=11
\end{equation}
for each $i$-th data point, to aid better convergence for the optimization algorithms. Having done that, we find the parameter estimates using both the probit and the logit links. In Table \ref{table: data analysis, probit link} and Table \ref{table: data analysis, logit link} the minimum density power divergence estimates are reported along with the $95\%$-trimmed MLE (trim. MLE). The $95\% $-trimmed MLE refers to the computation of the MLE after deletion of the statistical units whose covariates satisfying 
\begin{align}
\label{outlier deletion}
    (x^{*}_{i}-\hat{\mu})^{T}S^{-1}(x^{*}_{i}-\hat{\mu}) \ge \chi^{2}_{0.95, 4897}
    \mbox{ for all } i.
\end{align}
Here $\hat{\mu}$ and $S$ respectively denote the robust location and scale estimates of the data cloud $\{x^{*}_{i}; i=1, \ldots, 4898\}$, and $\chi^{2}_{0.95, 4897}$ being the upper $5\%$ point of central $\chi^{2}$-distribution with $4897$ degrees of freedom. More precisely, $\hat{\mu}$ is obtained by taking component-wise medians, whereas $S$ is constructed from the median absolute deviation for each row of the data matrix. As we find in Table \ref{table: data analysis, probit link} that when $\alpha=0, 0.1, 0.3$, the MDPD estimates of $\beta_{1}, \beta_{2}, \beta_{3}$ differ with the trimmed MLE. This might give an idea that the optimum tuning parameter might occur somewhere above $0.3$ because $\alpha$ close to {\it zero} gives unstable estimates. This speculation is fairly supported in Table \ref{table:optimum alpha selection} which gives the optimum tuning parameter as $\alpha=0.39$ with the lowest value of the MSE for the probit link. When the logit link is used, the optimum tuning parameter turns out $\alpha=0.68$. Here MSE is computed using (\ref{WJ MSE}) with the pilot being chosen as $\hat{\theta}_{0.5}$. Also, we notice that, for the logit link, the matrix $\widehat{\Psi}_{n}(\alpha)$ is singular at $\alpha=0$. Therefore, we report $\alpha=0.01$ in Table \ref{table: data analysis, logit link} in place of $\alpha=0$. We also report the total {\it standard error} (SE) in the second-last row of these two tables. The SE is computed as the sum of asymptotic standard deviations divided by the squared root of the sample size. For these links, SE generally increases with $\alpha$, therefore it is the squared bias term that drives the MSE to have a parabolic shape along with $\alpha$.         

Next, to measure the performance of these estimates we split the entire data set into two parts, namely the training and test data sets. The training data set consisting of $75\%$ data points is used to estimate the parameters that are further used to predict the wine quality levels in the test data set. The proportion of these cases, where the true levels match the predicted values in the test data set, measures the prediction accuracy of a particular method. In Table \ref{table: data analysis, probit link} we notice that the MDPDE with $\alpha>0$ produces better accuracy than both the MLE ($53.6\%$) and the trimmed MLE ($54\%$) for the probit link. However in Table \ref{table: data analysis, logit link} we find that both the trimmed MLE produces slightly better accuracy ($55.4\%$ respectively) than the MDPDE with $\alpha>0$.   


\begin{table}[ht]
\centering
 \caption{Parameters estimates in the wine quality data set with the probit link}
\begin{tabular}{cccccccc}
   \hline
Parameters &  $\alpha=0$ & $\alpha=0.1$ & $\alpha=0.3$ & $\alpha=0.5$ & $\alpha=0.7$ & $\alpha=1$ & $95\%$-trim. MLE\\ 
 \hline
$\hat{\beta}_{1}$ & 0.06268 & 0.07435 & 0.11204 & 0.11696 & 0.11992 & 0.11688 & 0.13085   \\
\hline
$\hat{\beta}_{2}$ & -0.2509 & -0.25329 & -0.25605 & -0.25542 & -0.25602 & -0.25796 & -0.2695   \\
\hline
$\hat{\beta}_{3}$ & 0.00053 & -0.00015 & 0.00791 & 0.01135 & 0.01241 & 0.01230 &
-0.01282   \\
\hline
$\hat{\beta}_{4}$ & 0.61871 & 0.63508 & 0.72930 & 0.73082 & 0.72955 & 0.70955 & 0.6382  \\
\hline
$\hat{\beta}_{5}$ &  -0.00358 & -0.00514 & -0.00374 & -0.0037 & -0.00344 & -0.00282 & -0.07292   \\
\hline
$\hat{\beta}_{6}$ & 0.08815 & 0.11128 & 0.12612 & 0.1368 & 0.14413 & 0.14893 & 0.12094   \\
\hline
$\hat{\beta}_{7}$ & -0.01651 & -0.02224 & -0.03029 & -0.04818 & -0.06126 & -0.07476 & -0.01060   \\
\hline
$\hat{\beta}_{8}$ & -0.66650 &  -0.71737 & -0.90893 & -0.92605 & -0.93696 & -0.91311 & -0.68881  \\
\hline
$\hat{\beta}_{9}$ &  0.13937 & 0.15234 & 0.18989 & 0.20338 & 0.21168 & 0.21468 & 
0.15736 \\
\hline
$\hat{\beta}_{10}$ & 0.09548 & 0.1016 & 0.11539 & 0.12403 & 0.13168 & 0.13694 & 
0.09799
  \\
\hline
$\hat{\beta}_{11}$ & 0.42875 & 0.42235 & 0.34789 & 0.35226 & 0.34956 & 0.36067 &
0.41748  \\
\hline
$\hat{\gamma}_{1}$ & -2.99276 & -3.13133 & -3.36179 & -3.43993 & -3.49516 & -3.47491 & -3.10773   \\
\hline
$\hat{\gamma}_{2}$ & -2.05813 & -2.10963 & -2.18638 & -2.23159 & -2.26262 & -2.27492 & -2.19455  \\
\hline
$\hat{\gamma}_{3}$ & -0.43326 & -0.43169 & -0.43081 & -0.42982 & -0.42819 & -0.42475 & -0.46651   \\
\hline
$\hat{\gamma}_{4}$ &  1.06414 & 1.07661 & 1.10986 & 1.13198 & 1.14808 & 1.15783 & 
1.02281 \\
\hline
$\hat{\gamma}_{5}$ & 2.25888 & 2.29456 & 2.39137 & 2.45673 & 2.50497 & 2.53885 & 2.26259   \\
\hline
$\hat{\gamma}_{6}$ & 24.6605 & 24.6605  & 24.6605 & 24.6605 & 24.6605 & 24.6605 &
24.6605  \\
\hline
SE  & 0.56917 & 0.76282 & 0.75769 & 0.83709 & 0.9544 & 1.12446 & NA     \\
\hline
Accuracy  & 0.53551 & 0.54367 & 0.54449 & 0.54367 & 0.54286 & 0.54041 & 0.54017    \\
\hline
   \end{tabular}
   \label{table: data analysis, probit link}
\end{table}


\begin{table}[!ht]
\centering
 \caption{Parameters estimates in the wine quality data set with the logit link}
\begin{tabular}{cccccccc}
   \hline
Parameters &  $\alpha=0.01$ & $\alpha=0.1$ & $\alpha=0.3$ & $\alpha=0.5$ & $\alpha=0.7$ & $\alpha=1$ & $95\%$-trim. MLE\\ 
 \hline
$\hat{\beta}_{1}$ & 0.29494 & 0.21962 & 0.23269 & 0.22743 & 0.21692 & 0.20681 & 0.29497  \\
\hline
$\hat{\beta}_{2}$ & -0.45306 & -0.47338 & -0.44141 & -0.43249 & -0.42688 & -0.42918 & -0.45306  \\
\hline
$\hat{\beta}_{3}$ & 0.00639 & 0.02054 & 0.01721 & 0.01900 & 0.01979 & 0.01745 & 0.00638  \\
\hline
$\hat{\beta}_{4}$ & 1.40803 & 1.39715 & 1.35650 & 1.32555 & 1.27118 & 1.22729 & 1.40802  \\
\hline
$\hat{\beta}_{5}$ & -0.13639 & -0.00420 & -0.00147 & -0.00008 & -0.00151 & 0.00141 & -0.13660 \\
\hline
$\hat{\beta}_{6}$ & 0.15613 & 0.20760 & 0.21563 & 0.22599 & 0.23991 & 0.23934 & 0.15611  \\
\hline
$\hat{\beta}_{7}$ & 0.01588 & -0.01629 & -0.05074 & -0.07677 & -0.10496 & -0.11556
& 0.01587  \\
\hline
$\hat{\beta}_{8}$ &  -1.76746 & -1.79154 & -1.76090 & -1.73325 & -1.66398 & -1.61093 & -1.76747 \\
\hline
$\hat{\beta}_{9}$ & 0.38776 & 0.35698 & 0.36654 & 0.37110 & 0.36943 & 0.36313 & 0.38776  \\
\hline
$\hat{\beta}_{10}$ & 0.19478 & 0.20459 & 0.21147 & 0.22260 & 0.23045 & 0.23322 & 
0.19478  \\
\hline
$\hat{\beta}_{11}$ & 0.4817 & 0.49613 & 0.48572 & 0.48776 & 0.51220 & 0.52506 &
0.48172  \\
\hline
$\hat{\gamma}_{1}$ & -19.18072 & -19.17419 & -19.17860 & -19.18054 & -19.18059 & 
-19.18051 & -19.18072  \\
\hline
$\hat{\gamma}_{2}$ & -4.07523 & -4.04373 & -3.93328 & -3.91250 & -3.92988 & -3.92934 & -4.07523  \\
\hline
$\hat{\gamma}_{3}$ &  -0.74738 & -0.74259 & -0.72432 & -0.70962 & -0.71042 & -0.70239 & -0.74737 \\
\hline
$\hat{\gamma}_{4}$ & 1.84985 & 1.8959 & 1.89781 & 1.91055 & 1.91435 & 1.91503 &
1.84983 \\
\hline
$\hat{\gamma}_{5}$ & 4.29262 & 4.26090 & 4.24203 & 4.26516 & 4.28086 & 4.30877 &
4.29263  \\
\hline
$\hat{\gamma}_{6}$ &  20.87745 & 20.87584 & 20.87700 & 20.87743 & 20.87745 & 20.87745 & 20.87745 \\
\hline
SE  &   1.33471 & 1.08470 & 1.09194 & 1.1541 & 1.24054 & 1.3822 & NA    \\
\hline
Accuracy & 0.53959 & 0.54122 & 0.53796 & 0.53959 & 0.53878 & 0.53551 & 0.55426 \\
\hline
 \end{tabular}
   \label{table: data analysis, logit link}
\end{table}


\begin{table}
\centering
 \caption{Optimum tuning parameter along with estimated MSE, Accuracy and SE. }
\begin{tabular}{c c c c c c}
   \hline
Method &  Link functions & $\alpha$ & MSE & SE & Accuracy \\ 
 \hline
 \mbox{MDPDE} &  probit & 0.39 & 0.07358  & 0.78485   &  0.54286\\
              &  logit & 0.68 &  0.17656  & 1.23131 & 0.53959   \\
 \hline
 \end{tabular}
 \label{table:optimum alpha selection}
\end{table}

\begin{figure}[!ht]
\begin{multicols}{2}
    \includegraphics[scale=0.5]{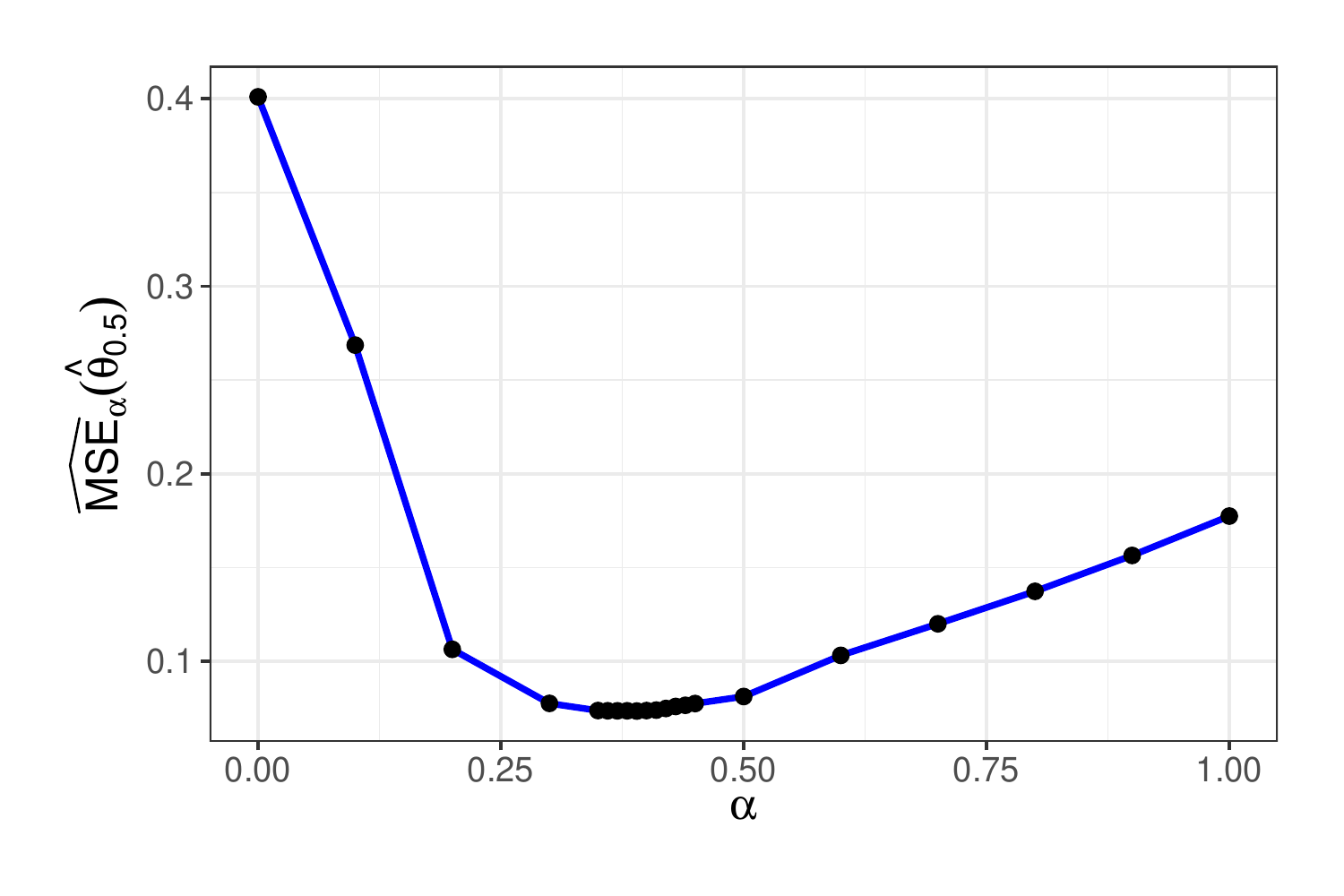} 
    \caption{Graphs of MSE for the probit link}  
     \label{mse:probit}
    \par 
    \includegraphics[scale=0.5]{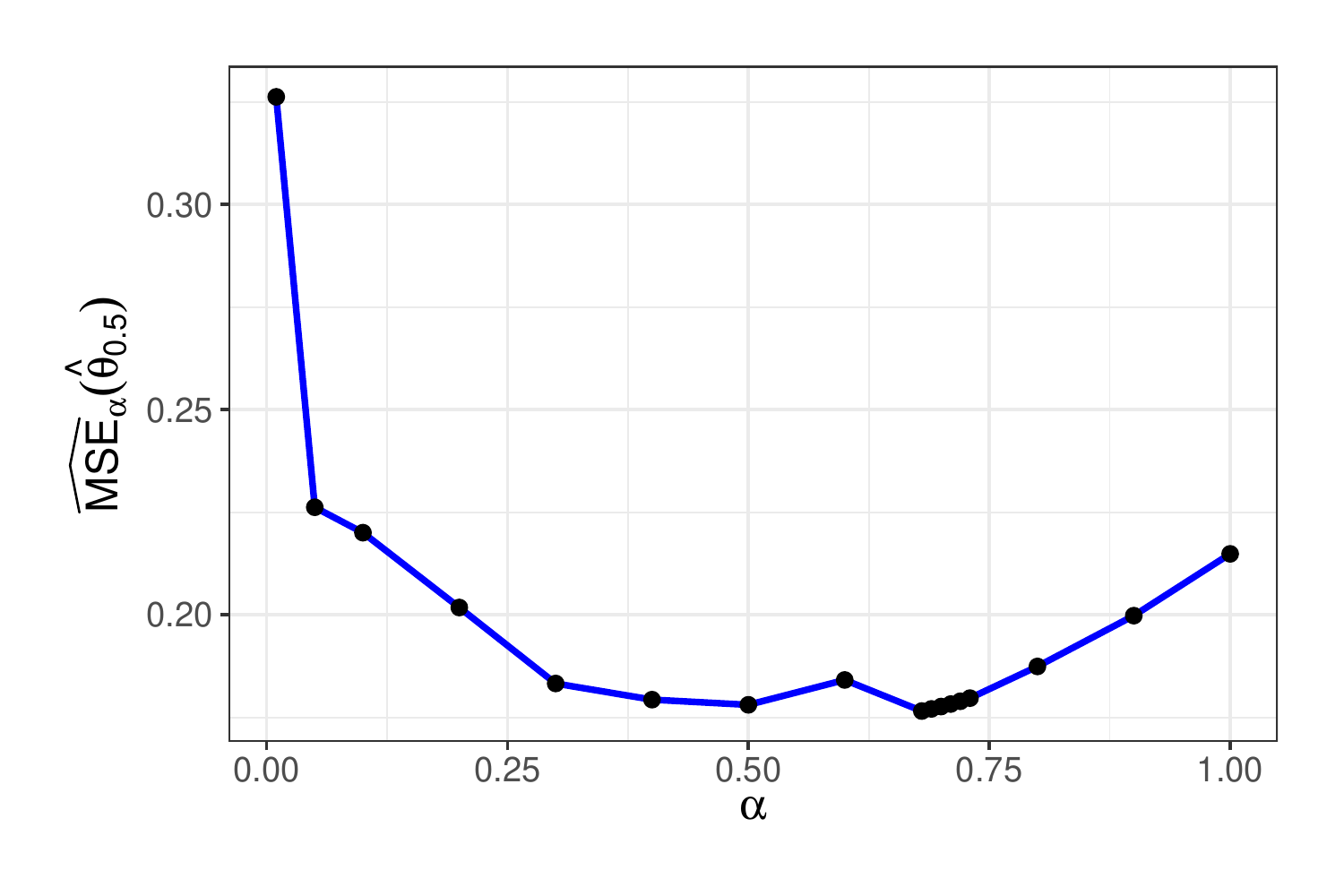} 
    \caption{Graphs of MSE for the logit link}  
    \label{mse:logit}
    \par 
    \end{multicols}
\end{figure}    

Now we proceed to find the optimum value of the tuning parameter using the strategy of Warwick and Jones \cite{warwick2005choosing}. In Figure  \ref{mse:probit} we see that MSE decreases with $\alpha$ roughly up to the point $\alpha=0.39$, after that the curve moves slightly upwards. This implies that MDPDE with $\alpha>0$ performs better than the MLE. This fact is fairly corroborated by the trend of the accuracy values as well. From the discussions of the simulation results, we can fairly conclude that this data set contains outlying observations with respect to the probit link function. Similarly, for the logit link, we observe a similar trend of the MSE values in Figure \ref{mse:logit}. We notice that MSE is minimized when the tuning parameter $\alpha=0.39, 0.68$ respectively for the probit and the logit links. These values are reported in Table \ref{table:optimum alpha selection}. Given this data set, we obtain different optimized MSE values for different link functions. We may take up this as a future research problem in choosing an appropriate link function in this setup.

\section{Conclusions}
\label{conclusions}

The lack of robustness in the likelihood-based inferential procedures poses a major challenge in modelling ordinal response data. Here we explore an alternative robust methodology to estimate the parameters in these statistical models through minimization of the density power divergence. It is based on the theory of an independent but non-homogeneous version of the DPD. We see how the choice of tuning parameters enables the MDPDE to achieve a higher degree of stability against different types of outliers that are inconsistent with a probabilistic model. The robustness of these estimators is discussed through the influence function and breakdown point analysis. Numerically, we have also shown that MDPD estimates also have a high implosive breakdown point at model misspecification. Robustness and asymptotic optimality are generally two competing concepts. The balance between these two is hard to achieve. We have demonstrated through the simulation studies how it is possible to find a suitable trade-off between these two extremities through the proper choice of a tuning parameter. Moreover, our proposed estimates perform better than Croux et al. \cite{croux2013robust}. Also, they are very competitive with Iannario et al. \cite{iannario2017robust}. Factoring in all such possible challenges, we believe that the use of the MDPDE in the ordinal response models provides a useful tool for the applied scientist.   

\section*{Funding Details}
Research of AG is partially supported by the INSPIRE Faculty Research Grant from the Department of Science and Technology, Govt. of India, and an internal research grant from the Indian Statistical Institute, India. The research of AB is supported by the Technology Innovation Hub at the Indian Statistical Institute, Kolkata under Grant NMICPS/006/MD/2020-21 of the Department of Science and Technology, Government of India, dated 16.10.2020.

\section*{Disclosure Statement}
No potential conflict of interest is reported by the authors.

\section*{Data Availability Statement}
The data set that supports the findings of this study is openly available in the UCI Machine Learning Repository at 
\url{https://archive.ics.uci.edu/ml/datasets/wine+quality}. See \cite{cortez2009modeling} to know more about this data set. 

\section*{Acknowledgements}
The authors gratefully acknowledge the comments of the referees and the 
editorial board which helped to significantly improve this manuscript.

\end{document}